\RequirePackage{fix-cm}

\documentclass[twocolumn]{svjour3}          % twocolumn

\smartqed  % flush right qed marks, e.g. at end of proof

\usepackage{natbib}
\usepackage{subcaption,graphicx}                   % figures
\usepackage{algorithm,algorithmic}                 % algorithms
\usepackage{booktabs}                              % table lines
\usepackage{amsmath,amsfonts,bm}                   % math symbols
\usepackage{float,dblfloatfix}
\usepackage{accents}
\usepackage{hyperref}
\usepackage{xcolor}
\hypersetup{
    colorlinks,
    linkcolor={blue!80!black},
    citecolor={blue!80!black},
    urlcolor={blue!80!black}
}

\newcommand*{\dbtilde}[1]{\accentset{\approx}{#1}} % double tilde
\journalname{Structural and Multidisciplinary Optimization}

\begin{document}

\title{Finite Variation Sensitivity Analysis for Discrete Topology Optimization of Continuum Structures\thanks{This work was supported by the São Paulo Research Foundation (FAPESP), grant numbers:  \mbox{2013/08293-7}, \mbox{2019/05393-7}, \mbox{2019/19237-7} and \mbox{2020/07391-9}}}

\author{Daniel Candeloro Cunha         \and
        Breno Vincenzo de Almeida      \and
        Heitor Nigro Lopes             \and
        Renato Pavanello
}

\institute{Daniel Candeloro Cunha \at \email{cunha@fem.unicamp.br}\\
           Department of Computational Mechanics, School of Mechanical Engineering, University of Campinas -- R. Mendeleyev 200, Cidade Universitária, Campinas, Brazil, 13083-860
}

\date{Received: date / Accepted: date}

\maketitle

\begin{abstract}
This paper proposes two novel approaches to perform more suitable sensitivity analyses for discrete topology optimization methods. To properly support them, we introduce a more formal description of the Bi-directional Evolutionary Structural Optimization (BESO) method, in which the sensitivity analysis is based on finite variations of the objective function. The  proposed approaches are compared to a naive strategy; to the conventional strategy, referred to as First-Order Continuous Interpolation (FOCI) approach; and to a strategy previously developed by other researchers, referred to as High-Order Continuous Interpolation (HOCI) approach. The novel Woodbury approach provides exact sensitivity values and is a better alternative to HOCI. Although HOCI and Woodbury approaches may be computationally prohibitive, they provide useful expressions for a better understanding of the problem. The novel Conjugate Gradient Method (CGM) approach provides sensitivity values with arbitrary precision and is computationally viable for a small number of steps. The CGM approach is a better alternative to FOCI since, for appropriate initial conditions, it is always more accurate than the conventional strategy. The standard compliance minimization problem with volume constraint is considered to illustrate the methodology. Numerical examples are presented together with a broad discussion about BESO-type methods.
\end{abstract}

\section{Introduction}

In topology optimization problems, the goal is to obtain the topology of a structure which minimizes (or maximizes) a given objective function. In density methods, the structural domain is discretized in a mesh of finite elements and a density value is assigned to each one of them. It is $0$ for void elements and $1$ for solid elements. Therefore, for a fixed mesh, the vector of density values completely defines the topology of a structure.

Topology optimization of continuum structures is essentially a large-scale non-linear integer programming problem. A common way to solve this binary optimization problem is to perform a continuous relaxation of the discrete variables, then solve the continuous optimization problem with a gradient-based method. A well established approach is to use the Solid Isotropic Material with Penalization (SIMP) interpolation scheme \citep{bendsoe1989optimal,zhou1991coc}, in which a penalization parameter is used to inhibit intermediary density values in the solution. For structural compliance minimization, it can be shown that, if the penalization parameter is sufficiently large, the continuous solution corresponds to a solution of the original binary problem \citep{rietz2001sufficiency,martinez2005note}.

Although this approach is robust and can provide nearly discrete solutions with a proper tuning of the penalization parameter, methods that only consider discrete structures throughout their search algorithm have their own advantages. In discrete methods, no post-processing is necessary to classify remaining elements with intermediary density values; the interfaces between domains are always well defined; multiple solutions can be stored, since a valid candidate for the optimized solution is obtained in each iteration.

The discrete method considered in this work is the Bi-directional Evolutionary Structural Optimization (BESO) \citep{querin1998evolutionary,yang1999bidirectional}, developed after its uni-directional version, Evolutionary Structural Optimization (ESO) \citep{xie1993simple}. This heuristic method produces only discrete topologies, it has been improved over the past years and it now constitutes a well established branch of topology optimization methods \citep{huang2007convergent,xia2018bi}. It has been extended to a wide range of problems, among them: optimization considering multiple materials \citep{huang2009bi}; problems with displacement constraints \citep{huang2010further}; problems with topology-dependent fluid pressure loads \citep{picelli2015bi,cunha2017evolutionary}; multi-objective and multi-scale optimization \citep{yan2015two}; problems with multi-scale non-linear structures \citep{xia2017recent}; maximum stress minimization \citep{xia2018stress}; frequency responses minimization \citep{vicente2016concurrent}; frequency gaps maximization \citep{lopes2021high}; design of acoustic mufflers \citep{azevedo2018topology}; design of piezoelectric energy harvesters with topology-dependent constraints \citep{de2019topology}.

There are alternative strategies to BESO-type methods, e.g., algorithms based on discrete mathematical programming \citep{beckers1999topology}. By linearizing the functions, integer Linear Programming (LP) can be used to solve the discrete optimization problems. Such approach, together with a strategy to easily include constraints, has been referred to as Topology Optimization of Binary Structures (TOBS) \citep{sivapuram2018topology,sivapuram2018topology2}. Moreover, for specific settings, it has been shown that the BESO method can be equivalent to a sequential LP approach \citep{tanskanen2002evolutionary}.

All optimization methods mentioned thus far are based on sensitivity analysis, which determines how the objective function is affected by a minimal variation of each design variable. In continuous relaxation approaches, such as SIMP, the sensitivity analysis consists in computing the gradient vector of the objective function. While in discrete methods, such as BESO, it should ideally consist in computing the objective function after switching the state of each discrete design variable. Since this is usually an excessively expensive procedure, estimations are used instead. The sensitivity analysis for discrete methods, based on finite variations, is henceforth referred to as Finite Variation Sensitivity Analysis (FVSA).

Another branch of topology optimization methods uses the concept of topological derivative to perform the sensitivity analysis \citep{cea2000shape,novotny2003topological}. It is commonly used together with level set methods to obtain optimal designs \citep{norato2007topological,van2013level}. In level set methods, interpolation schemes are unnecessary and topologies with no intermediary density values are obtained. Similarly to discrete density methods, they are advantageous in problems where interfaces must be well delineated, e.g., the design of fluid flow channels \citep{sa2016topological}. Despite these similarities, the development of the present paper is not directly applicable to this class of optimization methods.

The paper focuses on BESO-type methods. In order to develop the FVSA procedures, a more formal description of this class of methods is presented, in which the sensitivity analysis is properly defined to consider finite variations. To narrow down the focus of the study, the specific case of structural compliance minimization subject to a volume constraint is taken. It should be noted, however, that the proposed procedures can be extended to different objective functions and they can be adjusted to produce improved linearizations for LP approaches.

BESO-type methods, also referred to as Sequential Element Rejections and Admissions (SERA) \citep{rozvany2002combining,rozvany2002theoretical}, have two main hypotheses. If they are reasonably satisfied, these methods should be able to provide efficient optimized structures. It is assumed that the variation in the objective function is approximately equal to the sum of the variations which would occur if only one element were switched at a time (i.e., the objective function can be well approximated by an additively separable function). The second assumption is that the objective function is approximately linear with respect to the relaxed continuum density values (i.e., the objective function can be well approximated by a sum of linear continuous functions).

If the second hypothesis is satisfied for a given continuous interpolation function, the sensitivity analysis in these methods becomes the same as the one from continuous relaxation approaches: it is performed by computing the gradient vector of the objective function, which is usually a fairly simple operation.

In \citet{rozvany2009critical}, SERA and SIMP approaches were compared and some shortcomings of SERA methods were discussed, e.g., the lack of rigorous proof of efficacy due to its heuristic nature. As shown in \citet{zhou2001validity}, SERA methods may produce topologies far from the global optimum. This means that this class of methods cannot be blindly applied, it is necessary to discuss what are the conditions in which it should be used, and how reasonable are the hypotheses under such conditions.

The BESO algorithm cannot circumvent the requirement of the first hypothesis. Even if the combined effects of switching multiple elements simultaneously could be predicted, a new algorithm would have to be developed to take this information into account. However, the second hypothesis may be discarded if the variation of the objective function after switching the state of an element can be accurately predicted. Such an improvement may overcome some of the current limitations of this class of methods.

The problem of predicting the effects of finite variations has been studied by some researchers in the field of topology optimization. In \citet{mroz2003finite}, finite variations in topological parameters were considered to optimize truss, beam and frame structures. In \citet{bojczuk2009topological}, topological derivatives together with finite topology modifications were used in a heuristic algorithm to optimize bending plates. In algorithms based on topological derivatives, high-order terms of the topological asymptotic expansion can be considered to improve predictions \citep{de2007second,hassine2016higher}. Likewise, quadratic approximations can be produced from second-order derivatives of the objective function \citep{groenwold2010quadratic}, which can be used in integer Quadratic Programming (QP) approaches \citep{liang2019topology}. For well behaved functions, QP should perform better than LP when dealing with finite variations. For continuum structures, specifically for compliance minimization ESO, \citet{ghabraie2015eso} proposed an accurate sensitivity analysis, using high-order terms of the Taylor series expansion of the objective function.

In this paper, three existing FVSA procedures are shown and two novel ones are proposed and tested. All sensitivity formulations are developed for the structural compliance minimization problem, with volume constraint. The finite element method is used to solve the equilibrium equation, which corresponds to a static linear-elastic problem, with homogeneous isotropic material, under constant load, constrained by homogeneous Dirichlet boundary conditions.

In Section~\ref{sec:density}, the stiffness matrix, the displacements vector, the volume of material and the structural compliance are described as functions of the density vector. In Section~\ref{sec:discrete}, a variation-based representation for functions of discrete variables is presented. Using such representation for the objective function, Section~\ref{sec:heuristic} presents the considered heuristic optimization method, that is just a more formal description of the BESO method that considers finite variations of the objective function in the sensitivity analysis. In Section~\ref{sec:fvsens}, the five sensitivity analyses are presented: a naive approach, used as reference; the First-Order Continuous Interpolation (FOCI) approach, which corresponds to the standard procedure in literature; the High-Order Continuous Interpolation (HOCI) approach, which corresponds to the one presented in \citet{ghabraie2015eso}; the Woodbury approach, proposed in this work; and the Conjugate Gradient Method (CGM) approach, also proposed in this work. Then, some considerations are made about the error of the developed sensitivity expressions. In Section~\ref{sec:results}, numerical examples are discussed. And, in Section~\ref{sec:conclusions}, the main conclusions are summarized.

In \ref{ap_eigenvalues_A}, there is a proof of convergence of the Taylor series used in the HOCI sensitivity analysis for solid elements. In \ref{ap_counterexample}, there is a counterexample to prove that the corresponding series may be divergent for void elements. In \ref{ap_selective}, there is a procedure to update the selective inverse of the system matrix, which can be useful for both HOCI and Woodbury approaches. In \ref{ap_cgm_explicit}, there are some explicit sensitivity formulations for the CGM approach, considering different initial conditions, for $1$ and $2$ CGM steps.

\section{Density-based topology}
\label{sec:density}

For a fixed finite element mesh of $N$ elements, the topology of a structure can be described by a density vector $\mathbf{x} \in \{0,1\}^N$. If $x_i = 0$, the $i$th element is a void element with no stiffness; if $x_i = 1$, the $i$th element is a solid element with the material stiffness. The global matrix is given by

\begin{equation}
 \mathbf{K}(\mathbf{x}) = \sum\limits_{i=1}^{N} x_i\,\mathbf{K^{[0]}_i} \,,
 \label{eq:stiffness0}
\end{equation}

\noindent
where $\mathbf{K^{[0]}_i}$ is the stiffness matrix of the $i$th element, when it is solid. Since only homogeneous Dirichlet boundary conditions are considered, the constraints can be applied by simply removing from the stiffness matrices the rows and columns corresponding to restricted degrees of freedom. The matrices $\mathbf{K_i^{[0]}}$ are already the constrained ones. Furthermore, they are all symmetric matrices of dimensions $G\times G$, where $G$ is the number of unconstrained degrees of freedom of the discretized structure. Each one of them is positive semi-definite and assumes zero values everywhere outside a small submatrix of dimensions $g_i \times g_i$, where $g_i$ is the number of unconstrained degrees of freedom of the $i$th element.

To prevent $\mathbf{K}$ from becoming singular as solid elements are removed (turned into voids), a small stiffness is assigned to void elements. For a given soft-kill parameter $\varepsilon_k$, the stiffness matrix assigned to the $i$th element, when it is void, is $\varepsilon_k\,\mathbf{K^{[0]}_i}$. Thus, the following base stiffness is assigned to the whole structure:

\begin{equation}
 \mathbf{K_0} = \varepsilon_k \sum\limits_{i=1}^{N} \mathbf{K^{[0]}_i} \,.
 \label{eq:small_stiffness}
\end{equation}

The elemental variation matrix can be defined as

\begin{equation}
\mathbf{K_i} = (1-\varepsilon_k)\,\mathbf{K^{[0]}_i} \,,
\label{eq:var_stiffness}
\end{equation}

\noindent
then, the global matrix can be redefined as

\begin{equation}
 \mathbf{K}(\mathbf{x}) = \mathbf{K_0} + \sum\limits_{i=1}^{N} x_i\,\mathbf{K_i} \,.
 \label{eq:stiffness}
\end{equation}

The matrices $\mathbf{K_i}$ are all symmetric positive semi-definite; and the matrices $\mathbf{K_0}$ and $\mathbf{K}$ are both symmetric positive definite.

The relation between the stiffness matrix $\mathbf{K}$, the displacements vector $\mathbf{u}$ and the load vector $\mathbf{f}$ is given by the equilibrium equation 

\begin{equation}
 \mathbf{K}\,\mathbf{u} = \mathbf{f} \,.
 \label{eq:equilibrium}
\end{equation}

Since constant load is considered, the dependence of $\mathbf{u}$ on $\mathbf{x}$ is expressed as

\begin{equation}
 \mathbf{u}(\mathbf{x}) = \left[\mathbf{K}(\mathbf{x})\right]^{-1} \mathbf{f} \,.
 \label{eq:displacements}
\end{equation}

The total volume of material in the topology $\mathbf{x}$ is given by

\begin{equation}
 V(\mathbf{x}) = \sum\limits_{i=1}^N x_i\,V_i \,,
 \label{eq:volume}
\end{equation}

\noindent
where $V_i$ is the volume of the $i$th element.

By only considering meshes with elements of the same volume, the number of solid elements can be used as a measure of volume, so $V(\mathbf{x})$ can be redefined as

\begin{equation}
 V(\mathbf{x}) = \sum\limits_{i=1}^N x_i \,.
 \label{eq:volume_unitary}
\end{equation}

The structural compliance $C$ is defined as

\begin{equation}
 C = \frac{1}{2}\,\mathbf{u}^T \, \mathbf{K} \, \mathbf{u}
 \label{eq:compliance}
\end{equation}

\noindent
and its dependence on $\mathbf{x}$ is given by

\begin{equation}
 C(\mathbf{x}) = \frac{1}{2}\,\mathbf{f}^T \left[\mathbf{K}(\mathbf{x})\right]^{-1} \mathbf{f} \,.
 \label{eq:compliance_of_x}
\end{equation}

\section{Discrete function representation}
\label{sec:discrete}

The domain $\{0,1\}^N$, in which the design variables are defined, has $2^N$ elements. Given an ordering rule for the possible entries, any discrete scalar function $h(\mathbf{x})$ of $N$ binary variables can be defined by a finite ordered set of output values.

\begin{figure}[h!]
\centering
\includegraphics[width=0.45\textwidth]{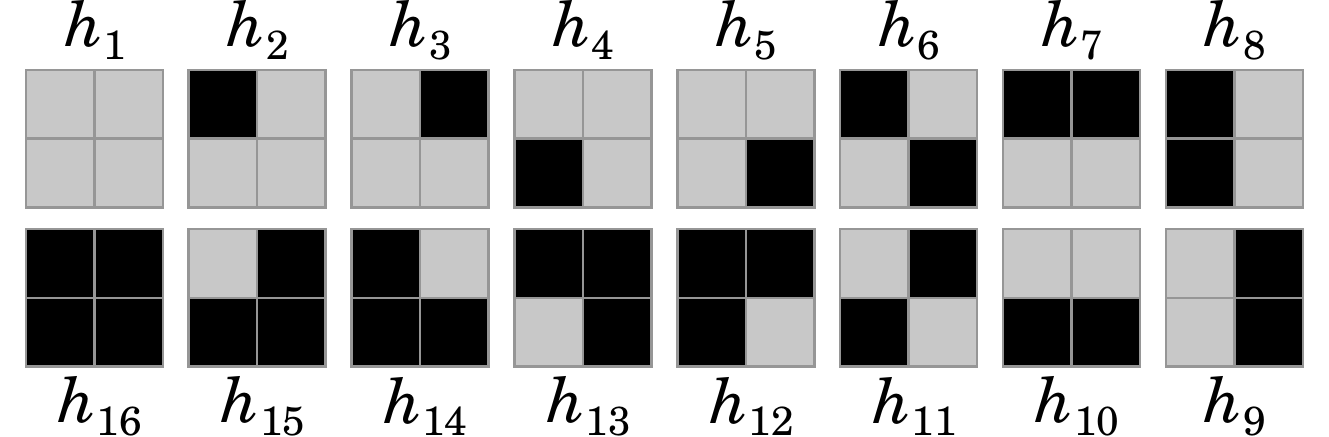}
\caption{Discrete function of binary variables for a $2\times2$ mesh ($N=4$).}
\label{fig:discrete_function}
\end{figure}

In Fig.~\ref{fig:discrete_function}, this is illustrated for a particular case, with $N=4$. The $16$ possible entries are presented as topologies in a $2\times2$ mesh, solid elements are represented in black and void elements are represented in gray. Each entry is paired with a corresponding output value, so the set $\{h_1, h_2, \hdots, h_{16}\}$ completely defines the discrete scalar function $h(\mathbf{x})$.

Another way of representing such a function is by variations around a point $\mathbf{\bar{x}}$. Denoting $N_v$ the number of $0$-valued terms in $\mathbf{\bar{x}}$, and $N_s$ the number of $1$-valued terms in it, a variation vector can be defined in $\{0,1\}^{N_v} \times \{0,-1\}^{N_s}$ as

\begin{equation}
 \mathbf{y} = \mathbf{x} - \mathbf{\bar{x}} \,.
 \label{eq:variation}
\end{equation}

Based on $\mathbf{\bar{x}}$, if $y_i = 0$, there is no variation in the state of the $i$th element, if $y_i = \pm 1$, the state is switched (from void to solid, or from solid to void). Thus, for a given $\mathbf{\bar{x}}$, a new function can be defined as

\begin{equation}
 \tilde{h}(\mathbf{y}) = \tilde{h}(\mathbf{y}(\mathbf{x})) = h(\mathbf{x}) \,.
 \label{eq:variation_function}
\end{equation}

It can be parameterized by a scalar $\alpha^{\langle 0 \rangle}$ and $N$ tensors of ascending order $\bm{\alpha}^{\langle k \rangle}$, as shown below:

\begin{equation}
 \tilde{h}(\mathbf{y}) = \alpha^{\langle 0 \rangle} + \sum\limits_{k=1}^{N} \bm{\alpha}^{\langle k \rangle} \,(\cdot)^k\, \mathbf{y}^k \,.
 \label{eq:parametrized_function}
\end{equation}

The $(\cdot)^k$-product represents the operation given by

\begin{equation}
 \bm{\alpha}^{\langle k \rangle} \,(\cdot)^k\, \mathbf{y}^k = \sum\limits_{i_1=1}^{N} \sum\limits_{i_2=1}^{N} \hdots \sum\limits_{i_k=1}^{N} \alpha^{\langle k \rangle}_{i_1 i_2 \hdots i_k} y_{i_1} y_{i_2} \hdots y_{i_k} \,.
 \label{eq:notation_dot}
\end{equation}

Except for $\bm{\alpha}^{\langle 1 \rangle}$, which is a vector, each tensor $\bm{\alpha}^{\langle k \rangle}$ is strictly upper triangular, so $\alpha^{\langle k \rangle}_{i_1 i_2 \hdots i_k}$ only assumes non-zero values when $i_1 < i_2 < \hdots < i_k$. This means that $\bm{\alpha}^{\langle k \rangle}$ is defined by $\binom{N}{k}$ parameters. Together with the scalar $\alpha^{\langle 0 \rangle}$, they all sum up to $2^N$.

The scalar $\alpha^{\langle 0 \rangle}$ corresponds to the value of the function without variation, $\alpha^{\langle 0 \rangle} = \tilde{h}(\mathbf{0}) = h(\bar{\mathbf{x}})$. The $i$th term of the $1$st-order tensor, $\alpha^{\langle 1 \rangle}_i$, is related to the variation of $h$ when only $x_i$ is switched. The $i_1i_2$th term of the $2$nd-order tensor, $\alpha^{\langle 2 \rangle}_{i_1i_2}$ with $i_1<i_2$, takes into account the coupled effect of switching both $x_{i_1}$ and $x_{i_2}$. It somewhat predicts how $\alpha^{\langle 1 \rangle}_{i_1}$ (or $\alpha^{\langle 1 \rangle}_{i_2}$) would change after switching $x_{i_2}$ (or $x_{i_1}$) and rewriting $\tilde{h}$ around the new point. As a general interpretation, the $k$th-order tensor $\bm{\alpha}^{\langle k \rangle}$ is related to the combined effect of switching $k$ variables simultaneously.

For a given non-negative integer $d \leq N$, let the $d$-neighborhood of $\mathbf{\bar{x}}$ be defined as the set $\mathbb{B}_d(\mathbf{\bar{x}})$ of vectors $\mathbf{x}$ that differ from $\mathbf{\bar{x}}$ in at most $d$ terms, which can be expressed as

\begin{equation}
 \mathbb{B}_d(\mathbf{\bar{x}}) = \left\{\;\mathbf{x} \in \{0,1\}^N \;\; \Bigg| \;\; \sum\limits_{i=1}^{N} |x_i - \bar{x}_i | \leq d \right\} \,,
 \label{eq:n_neighborhood}
\end{equation}

\noindent
so, for any $\mathbf{\bar{x}}$, $\{\mathbf{\bar{x}}\} = \mathbb{B}_0(\mathbf{\bar{x}}) \subset \mathbb{B}_1(\mathbf{\bar{x}}) \subset \hdots \subset \mathbb{B}_N(\mathbf{\bar{x}}) = \{0,1\}^N$.

A reduced domain may be considered to ignore topologies that are too far away from the current one. For a specified parameter $n$, if only points in $\mathbb{B}_n(\mathbf{\bar{x}})$ are considered, $\mathbf{y}$ can assume at most $n$ non-zero values, i.e., $\mathbf{y}^T \mathbf{y} \leq n$. Under this restriction, the tensors $\bm{\alpha}^{\langle k \rangle}$ become irrelevant for $k>n$, so the function can be redefined with only $\sum_{k=0}^{n} \binom{N}{k}$ parameters as

\begin{equation}
 \dbtilde{h}(\mathbf{y}) = \alpha^{\langle 0 \rangle} + \sum\limits_{k=1}^{n} \bm{\alpha}^{\langle k \rangle} \,(\cdot)^k\, \mathbf{y}^k \,.
 \label{eq:parametrized_function_reduced}
\end{equation}

\begin{figure}[h!]
\centering
\includegraphics[width=0.45\textwidth]{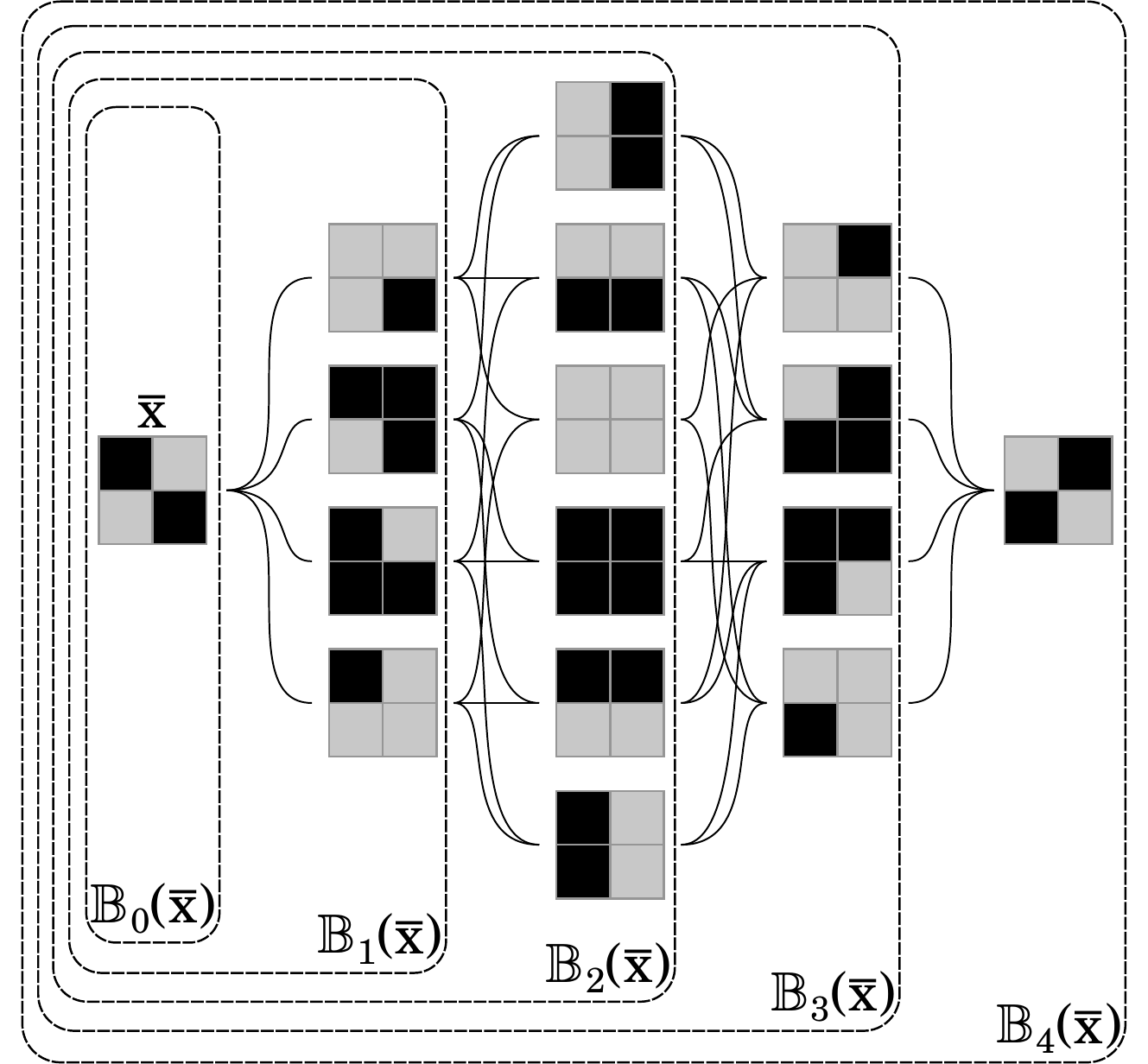}
\caption{Neighborhood sets for a $2\times2$ mesh ($N=4$).}
\label{fig:neighborhood}
\end{figure}

Although Eq.~(\ref{eq:parametrized_function}) presents a very compact algebraic form for the function $\tilde{h}$, it may be challenging to readily understand it. In Fig.~\ref{fig:neighborhood}, for a particular case, $\mathbf{\bar{x}}$ is defined and all the possible $d$-neighborhood sets are presented. The topologies are linked to their $N$ immediate neighbors, composing a graph. By assigning proper values to the edges, this graph provides an equivalent representation of $\tilde{h}$, a less compact but clearer one.

The function value in the leftmost node of the graph is given by $h(\mathbf{\bar{x}}) = \tilde{h}(\mathbf{0}) = \alpha^{\langle 0\rangle}$. Any topological variation corresponds to a path in the graph and produces a variation $\Delta h$ in the function. Thus, the edge-values are defined so their sum over a path results in $\Delta h$. To use an undirected graph, it can be established that steps from left to right add up the values, and steps from right to left subtract them. There are $\binom{4}{1}$ topologies in the second layer, so $4$ parameters (related to $\bm{\alpha}^{\langle 1 \rangle}$) must be defined to assign the $4$ edge-values between the first and second layers. There are $\binom{4}{2}$ topologies in the third layer, so $6$ parameters (related to $\bm{\alpha}^{\langle 2 \rangle}$) must be defined to assign the $12$ edge-values between the second and third layers. There are $\binom{4}{3}$ topologies in the fourth layer, so $4$ parameters (related to $\bm{\alpha}^{\langle 3 \rangle}$) must be defined to assign the $12$ edge-values between the third and fourth layers. Finally, since there is $\binom{4}{4}$ topology in the last layer, $1$ parameter (related to $\bm{\alpha}^{\langle 4 \rangle}$) must be defined to assign the $4$ edge-values between the fourth and last layers.

When considering $\mathbf{y}^T \mathbf{y} \leq n$, the tensors $\bm{\alpha}^{\langle k \rangle}$ become irrelevant for $k>n$ because the information they carry is about unreachable topologies, i.e., topologies outside the $n$-neighborhood of $\mathbf{\bar{x}}$.

Furthermore, the reduced function $\dbtilde{h}$ can also be interpreted as a truncated approximation of $\tilde{h}$. It estimates $\tilde{h}$ values outside $\mathbb{B}_n(\mathbf{\bar{x}})$ by disregarding some of the combined effects of simultaneously switching more than $n$ elements.

\section{Heuristic optimization}
\label{sec:heuristic}

It is desired to solve the discrete optimization problem

\begin{equation}
\begin{array}{c}
 \mathbf{x}^* = \underset{\mathbf{x}}{\text{arg min}}\; C(\mathbf{x})\\
 \text{subject to}\\
 V(\mathbf{x}) = V^* \,,
\end{array}
\label{eq:optimization}
\end{equation}

\noindent
where $V^*$ corresponds to a specified target volume for the structure.

Considering a point $\mathbf{\bar{x}}$ of volume $V(\mathbf{\bar{x}})=\bar{V}$, the optimization problem can be written in its variation-based form as

\begin{equation}
\begin{array}{c}
 \mathbf{y}^* = \underset{\mathbf{y}}{\text{arg min}}\; \tilde{C}(\mathbf{y})\\
 \text{subject to}\\
 \sum\limits_{i=1}^{N} y_i = V^* - \bar{V} \,.
\end{array}
\label{eq:optimization_y}
\end{equation}

The strategy to solve this problem is to start from a point $\mathbf{\bar{x}}^{(0)}$ and progress iteratively towards a local minimum, defined as follows. For a given non-negative integer $d \leq N$, a $d$-local minimum of a function $h(\mathbf{x})$ is any point $\mathbf{x}^*$ such that $h(\mathbf{x}^*) \leq h(\mathbf{x}), \forall \mathbf{x} \in \mathbb{B}_d(\mathbf{x}^*)$.

For a specified parameter $n$, the procedure consists in performing, in each iteration, an optimal move in the $n$-neighborhood of the current point. Therefore, $\mathbf{\bar{x}}^{(j+1)} \in \mathbb{B}_n(\mathbf{\bar{x}}^{(j)})$ is obtained by solving the subproblem

\begin{equation}
\begin{array}{c}
 \mathbf{y}^{*} = \underset{\mathbf{y}}{\text{arg min}}\; \dbtilde{C}(\mathbf{y})\\
 \text{subject to}\\
 \sum\limits_{i=1}^{N} y_i = \bar{V}^{(j+1)} - \bar{V}^{(j)}\\
 \mathbf{y}^T\,\mathbf{y} \leq n \,,
\end{array}
\label{eq:optimization_y_red}
\end{equation}

\noindent
where the variation vector $\mathbf{y}$ and the function $\dbtilde{C}(\mathbf{y})$ are defined around $\mathbf{\bar{x}}^{(j)}$. As in Eq.~)\ref{eq:parametrized_function_reduced}), $\dbtilde{C}$ corresponds to the reduced objective function given by

\begin{equation}
 \dbtilde{C}(\mathbf{y}) = \alpha^{\langle 0 \rangle} + \sum\limits_{k=1}^{n} \bm{\alpha}^{\langle k \rangle} \,(\cdot)^k\, \mathbf{y}^k \,.
 \label{eq:subproblem_objective}
\end{equation}

Since the volume change is limited by $n$, a volume value $\bar{V}^{(j)}$ must be assigned for each iteration, assuring a proper progression from $\bar{V}^{(0)}$ to $V^*$. If it can be performed, this optimization procedure is guaranteed to converge to an $n$-local minimum of $C$.

However, the $\sum_{k=0}^{n} \binom{N}{k}$ parameters of $\dbtilde{C}$ are not easily obtained and the subproblem is not much simpler than the original one. Thus, a heuristic approach is adopted. Instead of optimizing $\dbtilde{C}$, the subproblem is redefined for the additively separable function

\begin{equation}
 \hat{C}(\mathbf{y}) = \alpha^{\langle 0 \rangle} + \bm{\alpha}^{\langle 1 \rangle} \cdot\mathbf{y} \,.
 \label{eq:subproblem_linearized}
\end{equation}

This means that the combined effects, given by $\bm{\alpha}^{\langle k \rangle}$ for $k$ from $2$ to $n$, are disregarded. The scalar $\alpha^{\langle 0 \rangle}$ corresponds to $C(\mathbf{\bar{x}}^{(j)})$ and the vector $\bm{\alpha}^{\langle 1 \rangle}$ is given by

\begin{equation}
 \alpha^{\langle 1 \rangle}_i = C(\mathbf{\bar{x}}^{(j)},x_i=1) - C(\mathbf{\bar{x}}^{(j)},x_i=0) \,,
 \label{eq:alpha}
\end{equation}

\noindent
where the arguments $(\mathbf{\bar{x}}^{(j)},x_i=1)$ and $(\mathbf{\bar{x}}^{(j)},x_i=0)$ denote vectors that are equal to $\mathbf{\bar{x}}^{(j)}$ except at their $i$th term, which assumes the explicitly defined value. In order to obtain $\alpha^{\langle 0 \rangle}$ and $\bm{\alpha}^{\langle 1 \rangle}$, $C(\mathbf{x})$ must be known for all $\mathbf{x} \in \mathbb{B}_1(\mathbf{\bar{x}}^{(j)})$.

The vector $\bm{\alpha}^{\langle 1 \rangle}$ is the sensitivity of $\hat{C}$ to any change on $\mathbf{x}$. The class of optimization algorithms considered in this work is based on the estimation of this sensitivity vector.

Functions that measure the volume variation, $V\!V(\mathbf{y})$, and the topological variation, $T\!V(\mathbf{y})$, can be defined as

\begin{equation}
V\!V(\mathbf{y}) = \sum\limits_{i=1}^{N} y_i
\label{eq:variations_parameters_1}
\end{equation}

\noindent
and

\begin{equation}
T\!V(\mathbf{y}) = \sum\limits_{i=1}^{N} y_i^2 \,.
\label{eq:variations_parameters_2}
\end{equation}

The constraints over these variations are expressed by specified values for the volume variation $\bar{V}^{(j+1)}~-~\bar{V}^{(j)}$, which will be denoted by $V\!V^{(j)}$, and for the maximal topological variation $n$, which will be denoted by $T\!V_{\max}^{(j)}$. Both must be assigned for each iteration of the optimization procedure. It should be noted that $V\!V^{(j)}$ corresponds to the Evolutionary Rate ($E\!R$) and $T\!V_{\max}^{(j)}$ is related to the maximal Addition Ratio ($A\!R_{\max}$) from the BESO method, as shown below:

\begin{equation}
V\!V^{(j)} = N\,\cdot\,E\!R \;;
\label{eq:variation_beso_1}
\end{equation}

\begin{equation}
T\!V_{\max}^{(j)} = N\,\cdot\,\left(E\!R + 2\,A\!R_{\max}\right) \,.
\label{eq:variation_beso_2}
\end{equation}

Finally, the simplified subproblem

\begin{equation}
\begin{array}{c}
 \mathbf{y}^* = \underset{\mathbf{y}}{\text{arg min}}\; \hat{C}(\mathbf{y})\\
 \text{subject to}\\
 V\!V(\mathbf{y}) = V\!V^{(j)}\\
 T\!V(\mathbf{y}) \leq T\!V_{\max}^{(j)}
\end{array}
\label{eq:optimization_y_lin}
\end{equation}

\noindent
can be solved as follows: to obtain $\mathbf{\bar{x}}^{(j+1)}$, the elements are ordered by their sensitivity values, then the elements with higher values are turned into voids while the elements with lower values are turned into solids, ensuring that the volume variation and topological variation constraints are both satisfied.

It is assumed that, starting from $\mathbf{\bar{x}}^{(0)}$, successive solutions of the simplified subproblems move towards an $n$-local minimum of the original problem $\mathbf{x}^*$. If such assumption is realistic, after a finite number of iterations, the solutions of the simplified subproblems will oscillate around $\mathbf{x}^*$, i.e., $\exists \,p,J \;|\; \mathbf{\bar{x}}^{(j)} \in \mathbb{B}_p(\mathbf{x}^*), \forall j > J$. The smallest value for $p$ and its corresponding $J$ represent, respectively, the accuracy of the method and the number of iterations that characterizes its convergence.

After reaching the target volume $V^*$, the topology vector should be stored whenever the objective function assumes a value lower than the smallest one obtained thus far. A patience parameter defines the maximum number of iterations that the algorithm can perform with no improvements in the objective function. If this number is reached, the best topology obtained thus far is returned as the optimized result. This is a coherent stop criterion for this algorithm, it is similar to the one used in \citet{xia2018stress}.

The main hypothesis here is that by making small alterations in the topology, only local effects, related to $\bm{\alpha}^{\langle 1 \rangle}$ and $\mathbb{B}_1$, are relevant. Most topology updates only change the shape of the existing solid boundaries, if a small number of elements is switched, it is reasonable to consider that the global behavior of the structure will be nearly the same between iterations. However, if all of the switched elements are clustered, the combined effects in that disturbed region would not be negligible anymore. Such case should not spoil the method though, since, if there is a critical region, many of the elements in it should indeed be switched. The most problematic scenario is when new boundaries are created or old ones are removed, which can produce unpredictable changes and undesirable moves may be made. Nevertheless, practice shows that, when applied methodically, this heuristic approach consistently provides satisfactory results \citep{xia2018bi}.

As in other density methods, sensitivity filters can be used to deal with the checkerboard problem and mesh dependency \citep{sigmund1998numerical,sigmund2012sensitivity}. Also, to improve stability, momentum strategies can be included. A popular approach is to average the sensitivity vector with its previous values throughout the iterations, properly weighting each term \citep{huang2007convergent}.

In this work, a simple conic filter was used to smoothen the sensitivity map and equal weights were used for the momentum: the filtered sensitivity number of the current iteration is added to the final sensitivity number of the last one (which contain previous momentum effects), then the sum is divided by $2$.

\section{Finite Variation Sensitivity Analysis (FVSA)}
\label{sec:fvsens}

The linear system of Eq.~(\ref{eq:equilibrium}) must be solved to obtain the displacements vector and compute the objective function. Other than that, the only potentially costly task in the optimization algorithm is to perform the sensitivity analysis. This section presents some ways to obtain or estimate the sensitivity vector $\bm{\alpha}^{\langle 1 \rangle}$, defined in Eq.~(\ref{eq:alpha}).

The usual approach is to define a continuous interpolation function to describe the material behavior for intermediary density values, then obtain the sensitivity vector from the truncated Taylor series of the objective function. In the proposed alternative approaches, the sensitivity vector is obtained without continuous interpolations, by considering finite variations of the elemental stiffness matrices.

\begin{figure}[h!]
\centering
\includegraphics[width=0.45\textwidth]{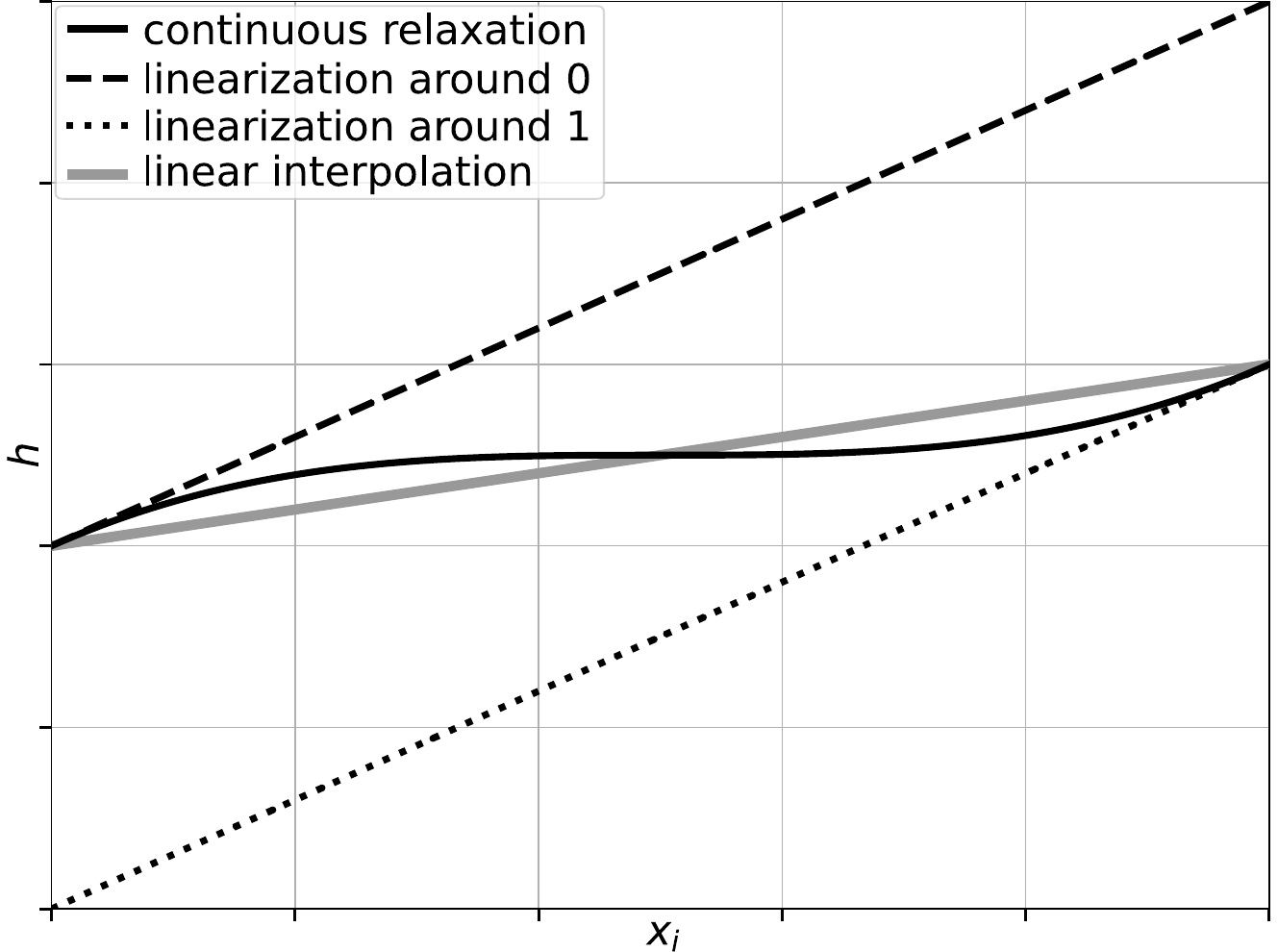}
\caption{Different linearizations of a function.}
\label{fig:linearizations}
\end{figure}

The proposed FVSA can also be interpreted as an improved linearization for integer LP approaches. The usual linearization takes into account only local information (function and gradient values in an extreme point), however, since the variables are binary, the local behavior is unimportant, the relevant information is only at the extreme points. Therefore, a more appropriate linearization can be produced through FVSA, which is a linear interpolation of extreme values. This is illustrated in Fig.~\ref{fig:linearizations}, in which the continuous relaxation of $h(x_i)$ corresponds to a simple cubic polynomial.

Even if is not possible to obtain the exact function variation between extreme points, estimated values can be enough to produce better results than local linearizations.

\subsection{Naive approach}

The most obvious way to perform the FVSA is through exhaustive effort. For a given topology, the displacements vector after switching the state of the $i$th element can be computed as

\begin{equation}
 \mathbf{\bar{\bar{u}}} = \left\{
 \begin{aligned}
  & \left[\mathbf{\bar{K}} + \mathbf{K_i}\right]^{-1} \mathbf{f}\;, & \text{if } x_i = 0 \,,\\
  & \left[\mathbf{\bar{K}} - \mathbf{K_i}\right]^{-1} \mathbf{f}\;, & \text{if } x_i = 1 \,,
 \end{aligned}\right.
 \label{eq:displacements_update_brute}
\end{equation}

\noindent
where $\mathbf{\bar{K}}$ is the stiffness matrix for the current topology. Thus, each term of the sensitivity vector can be obtained by

\begin{equation}
 \boxed{\alpha^{\langle 1 \rangle}_i = \left\{
 \begin{aligned}
  & -\frac{1}{2}\,\mathbf{f}^T \left[\mathbf{\bar{u}} - \mathbf{\bar{\bar{u}}}\right]\;, & \text{if } x_i = 0 \,,\\
  & -\frac{1}{2}\,\mathbf{f}^T \left[\mathbf{\bar{\bar{u}}} - \mathbf{\bar{u}}\right]\;, & \text{if } x_i = 1 \,,
 \end{aligned}\right.}
 \label{eq:sensitivity_brute}
\end{equation}

\noindent
where $\mathbf{\bar{u}}$ is the displacements vector for the current topology.

Evidently, this is not a viable strategy, since $N$ linear systems must be solved before each topology update. Yet, it can be used as reference when evaluating other strategies.

\subsection{First-Order Continuous Interpolation (FOCI) approach}

By considering a continuous density vector, $\mathbf{x} \in [0,1]^N$, an interpolation function $\gamma(x_i)$ can be defined with the following properties: $\gamma(0) = 0$; $\gamma(1) = 1$; $\gamma$ is differentiable and monotonic in $[0,1]$. Thus, the global stiffness matrix can be redefined as a differentiable function of $\mathbf{x}$, given by

\begin{equation}
 \mathbf{K}(\mathbf{x}) = \mathbf{K_0} + \sum\limits_{i=1}^{N} \gamma(x_i)\,\mathbf{K_i} \,.
 \label{eq:stiffness_continuous}
\end{equation}

By using such $\mathbf{K}(\mathbf{x})$ in Eqs.~(\ref{eq:displacements}) and (\ref{eq:compliance_of_x}), $\mathbf{u}$ and $C$ can also be described as differentiable functions of $\mathbf{x}$. Then, a first-order approximation of $C(x_i)$ can be obtained by disregarding the high-order terms of its Taylor expansion centered at $\mathbf{\bar{x}}$. From this approximation, the sensitivity vector can be calculated as

\begin{equation}
 \alpha_i^{\langle 1 \rangle} = -\frac{1}{2}\,\frac{\partial \gamma}{\partial x_i}\,\mathbf{\bar{u}}^T\,\mathbf{K_i}\,\mathbf{\bar{u}} \,,
 \label{eq:sensitivity_first_order}
\end{equation}

\noindent
which is easily obtained since $\mathbf{\bar{u}}$ is the same for every element and all the matrices $\mathbf{K_i}$ are known.

Since the sensitivity values are only used to compare elements, they can all be multiplied by a positive value without affecting the optimization process. Therefore, the sensitivity vector can be redefined as

\begin{equation}
 \boxed{\alpha^{\langle 1 \rangle}_i =
 \left\{\begin{aligned}
  & -\frac{\varepsilon_v}{2}\,\mathbf{\bar{u}}^T\,\mathbf{K_i}\,\mathbf{\bar{u}}\;, & \text{if } x_i = 0 \,,\\
  & -\frac{1}{2}\,\mathbf{\bar{u}}^T\,\mathbf{K_i}\,\mathbf{\bar{u}}\;, & \text{if } x_i = 1 \,,
 \end{aligned}\right.}
 \label{eq:sensitivity_first_order_penalized}
\end{equation}

\noindent
where a penalization factor is defined for void elements as $\varepsilon_v = \frac{\partial \gamma}{\partial x_i}(0) / \frac{\partial \gamma}{\partial x_i}(1)$.

The interpolation function is only relevant to define the derivatives at the two extreme points, so there is no reason to define it differently from the linear interpolation, while using a specified penalization parameter for void elements. As long as the same interpolation function is considered for every element, the presence of the parameter $\varepsilon_v$ covers all possible definitions for such function.

\subsection{High-Order Continuous Interpolation (HOCI) approach}

If the whole Taylor expansion of $C(x_i)$ is considered, the sensitivity vector can be obtained as the series

\begin{equation}
 \alpha^{\langle 1 \rangle}_i =
 \left\{\begin{aligned}
  & \sum\limits_{a=1}^{\infty}\frac{1}{a!}\,\frac{\partial^a C}{\partial x_i^a}(\mathbf{\bar{x}}) \phantom{(-)^{a+1}}, & \text{if } x_i = 0 \,,\\
  & \sum\limits_{a=1}^{\infty}\frac{(-1)^{a+1}}{a!}\,\frac{\partial^a C}{\partial x_i^a}(\mathbf{\bar{x}}) \;, & \text{if } x_i = 1 \,.
 \end{aligned}\right.
 \label{eq:sensitivity_taylor}
\end{equation}

For a general interpolation function $\gamma$, this series becomes a quite complex expression. In \citet{ghabraie2015eso}, it was simplified by considering $\gamma(x_i) = x_i$. In such case, the sensitivity vector is given by

\begin{equation}
 \alpha^{\langle 1 \rangle}_i =
 \left\{\begin{aligned}
  & \frac{1}{2} \sum\limits_{a=1}^{\infty} \mathbf{f}^T \left[-\mathbf{\bar{K}}^{-1} \, \mathbf{K_i}\right]^{a} \mathbf{\bar{K}}^{-1} \mathbf{f}\;, & \text{if } x_i = 0 \,,\\
  & \frac{1}{2} \sum\limits_{a=1}^{\infty} -\mathbf{f}^T \left[\mathbf{\bar{K}}^{-1} \, \mathbf{K_i}\right]^{a} \mathbf{\bar{K}}^{-1} \mathbf{f}\;, & \text{if } x_i = 1 \,.
 \end{aligned}\right.
 \label{eq:sensitivity_series}
\end{equation}

There is an alternate, more natural way to obtain this same expression. Since the displacements field $\mathbf{u}$ is a function of $\mathbf{K}(\mathbf{x})$, with no explicit dependence on $\mathbf{x}$, its variation can be evaluated with respect to variations of $\mathbf{K}$. Considering $\mathbf{K}$ an independent variable, Eq.~(\ref{eq:displacements}) can be rewritten as the Taylor series

\begin{equation}
 \mathbf{u}(\mathbf{\bar{K}} + \mathbf{\Delta K}) = \sum\limits_{a=0}^{\infty} \left[-\mathbf{\bar{K}}^{-1} \mathbf{\Delta K}\right]^{a} \mathbf{u}(\mathbf{\bar{K}}) \,,
 \label{eq:displacements_taylor}
\end{equation}

\noindent
from which Eq.~(\ref{eq:sensitivity_series}) can be obtained as the $i$th sensitivity value.

There is a more convenient form for the sensitivity expression, that will shorten the discussion about the series convergence. It can be written in terms of the symmetric positive semi-definite matrix

\begin{equation}
 \mathbf{\bar{A}_i} = \sqrt{\mathbf{K_i}}\,\mathbf{\bar{K}}^{-1} \sqrt{\mathbf{K_i}}
 \label{eq:matrix_A}
\end{equation}

\noindent
and the vector

\begin{equation}
 \mathbf{\bar{v}_i} = \sqrt{\mathbf{K_i}}\,\mathbf{\bar{u}}
 \label{eq:vector_v} \,.
\end{equation}

The resultant sensitivity expression is given by

\begin{equation}
 \boxed{\alpha^{\langle 1 \rangle}_i =
 \left\{\begin{aligned}
  & -\frac{1}{2} \sum\limits_{a=1}^{\infty} \mathbf{\bar{v}_i}^T \left[-\mathbf{\bar{A}_i}\right]^{a-1} \mathbf{\bar{v}_i}\;, & \text{if } x_i = 0 \,,\\
  & -\frac{1}{2} \sum\limits_{a=1}^{\infty} \mathbf{\bar{v}_i}^T \left[\mathbf{\bar{A}_i}\right]^{a-1} \mathbf{\bar{v}_i}\phantom{-}\;, & \text{if } x_i = 1 \,.
 \end{aligned}\right.}
 \label{eq:sensitivity_series_sym}
\end{equation}

The convergence of such a series depends on the eigenvalues of $\mathbf{\bar{A}_i}$. Let $\mathbf{\bar{\bm{\Lambda}}_i}$ be the diagonal eigenvalues matrix and $\mathbf{\bar{\Phi}_i}$ be the orthogonal eigenvectors matrix, so that

\begin{equation}
 \mathbf{\bar{A}_i} = \mathbf{\bar{\Phi}_i} \, \mathbf{\bar{\bm{\Lambda}}_i} \, \mathbf{\bar{\Phi}_i}^T \,,
 \label{eq:eigenproblem_A}
\end{equation}

\noindent
by defining the vector $\mathbf{\bar{w}_i}$ as

\begin{equation}
 \mathbf{\bar{w}_i} = \mathbf{\bar{\Phi}_i}^T \, \mathbf{\bar{v}_i} \,,
 \label{eq:vector_w}
\end{equation}

\noindent
the sensitivity vector can be written as

\begin{equation}
 \boxed{\alpha^{\langle 1 \rangle}_i = \left\{\begin{aligned}
  & -\frac{1}{2} \sum\limits_{a=1}^{\infty} \mathbf{\bar{w}_i}^T \left[- \mathbf{\bar{\bm{\Lambda}}_i}\right]^{a-1} \mathbf{\bar{w}_i}\;, & \text{if } x_i = 0 \,,\\
  & -\frac{1}{2} \sum\limits_{a=1}^{\infty} \mathbf{\bar{w}_i}^T \left[ \mathbf{\bar{\bm{\Lambda}}_i}\right]^{a-1} \mathbf{\bar{w}_i}\phantom{-}\;, & \text{if } x_i = 1 \,.
 \end{aligned}\right.}
 \label{eq:sensitivity_series_eig}
\end{equation}

For any vector $\mathbf{\bar{w}_i}$, those series are convergent if and only if $\|\mathbf{\bar{A}_i}\|_2 = \max(\mathbf{\bar{\bm{\Lambda}}_i}) < 1$. In \ref{ap_eigenvalues_A}, it is shown that this condition is always satisfied when $x_i = 1$. And in \ref{ap_counterexample}, it is shown that it may not be satisfied when $x_i = 0$.

Therefore, through the truncated series from Eqs.~(\ref{eq:sensitivity_series_sym}) or (\ref{eq:sensitivity_series_eig}), this approach can estimate, with arbitrary precision, the sensitivity values for solid elements. Since it may fail to provide sensitivity values for void elements, it requires further considerations. A possible strategy is to assign all void sensitivity numbers as $0$. In such case, the solid part of the topology would guide the optimization process, and void elements could only be turned into solid ones through the effects of a sensitivity filter.

\subsection{Woodbury approach}

This approach is named after Woodbury because the Woodbury matrix identity is used to obtain the sensitivity expressions. For a given perturbation term added to a matrix, the Woodbury formula provides an expression for the inverse of the perturbed matrix. Its general formulation and some applications can be seen in \citet{hager1989updating}. Although it is named after Woodbury's report \citep{woodbury1950inverting}, published in 1950, it should be noted that it had already appeared in previous papers of other authors, e.g., \citet{guttman1946enlargement}.

Without the need of an interpolation function, after switching an element state, the inverse of the updated global matrix can be obtained by the Woodbury identity as

\begin{equation}
\left[\mathbf{\bar{K}} \pm \mathbf{K_i}\right]^{-1} = \mathbf{\bar{K}}^{-1} \mp \mathbf{\bar{K}}^{-1} \sqrt{\mathbf{K_i}} \left[\mathbf{I} \pm \mathbf{\bar{A}_i}\right]^{-1} \sqrt{\mathbf{K_i}}\,\mathbf{\bar{K}}^{-1} \,.
 \label{eq:woodbury}
\end{equation}

It can be used to compute $\mathbf{\bar{\bar{u}}}$ from Eq.~(\ref{eq:displacements_update_brute}), resulting in the following expression for the sensitivity vector:

\begin{equation}
 \boxed{\alpha^{\langle 1 \rangle}_i =
 \left\{\begin{aligned}
  & -\frac{1}{2}\,\mathbf{\bar{v}_i}^T \left[\mathbf{I} + \mathbf{\bar{A}_i}\right]^{-1}\mathbf{\bar{v}_i}\;, & \text{if } x_i = 0 \,,\\
  & -\frac{1}{2}\,\mathbf{\bar{v}_i}^T \left[\mathbf{I} - \mathbf{\bar{A}_i}\right]^{-1} \mathbf{\bar{v}_i}\;, & \text{if } x_i = 1 \,.
 \end{aligned}\right.}
 \label{eq:sensitivity_woodbury}
\end{equation}

Which can be written in terms of $\mathbf{\bar{\bm{\Lambda}}_i}$ and $\mathbf{\bar{w}_i}$ as

\begin{equation}
 \boxed{\alpha^{\langle 1 \rangle}_i =
 \left\{\begin{aligned}
  & -\frac{1}{2}\,\mathbf{\bar{w}_i}^T\left[\mathbf{I} + \mathbf{\bar{\bm{\Lambda}}_i}\right]^{-1} \mathbf{\bar{w}_i}\;, & \text{if } x_i = 0 \,,\\
  & -\frac{1}{2}\,\mathbf{\bar{w}_i}^T\left[\mathbf{I} - \mathbf{\bar{\bm{\Lambda}}_i}\right]^{-1} \mathbf{\bar{w}_i}\;, & \text{if } x_i = 1 \,.
 \end{aligned}\right.}
 \label{eq:sensitivity_woodbury_eig}
\end{equation}

It should be noted that, when $\|\mathbf{\bar{A}_i}\|_2 < 1$, those expressions correspond to the sums of the power series from Eqs.~(\ref{eq:sensitivity_series_sym}) and (\ref{eq:sensitivity_series_eig}). Moreover, this approach can provide sensitivity values even for void elements such that $\|\mathbf{\bar{A}_i}\|_2 \geq 1$.

From Eq.~(\ref{eq:sensitivity_woodbury_eig}), it should be clear that the FOCI approach, with linear interpolation, always overestimates $\big|\alpha_i^{\langle 1 \rangle}\big|$ for void elements and always underestimates it for solid elements. Thus, the void penalization factor $\varepsilon_v$, from Eq.~(\ref{eq:sensitivity_first_order_penalized}), should always be defined in $[0,1[$ for more accurate comparisons between elements in different states.

In both HOCI and Woodbury approaches, each matrix $\mathbf{\bar{A}_i} = \sqrt{\mathbf{K_i}}\,\mathbf{\bar{K}}^{-1}\sqrt{\mathbf{K_i}}$ must be known to obtain the sensitivity vector. This means that it is necessary to know a selective inverse of $\mathbf{\bar{K}}$, more specifically: all the terms $\bar{K}^{-1}_{i_1i_2}$ with indexes $(i_1,i_2)$ corresponding to valued entries of the sparse matrix $\mathbf{\bar{K}}$.

The calculation of this selective inverse is fairly expensive and may take these expressions out of practical use. However, this statement should not be taken as conclusive. Since it was out of the scope of this work, no specific algorithm for selective inversion was implemented to test how prohibitive this task would actually be \citep{lin2011selinv,jacquelin2016pselinv}. Nonetheless, if the selective inverse is known in the first iteration, after each topology change, it can be updated by the procedure presented in \ref{ap_selective}. In this work, the initial selective inverses were computed through exhaustive effort.

\subsection{Conjugate Gradient Method (CGM) approach}

In this section, an iterative strategy that does not require interpolation functions or selective inversions is proposed.

A variation $\mathbf{\Delta K}$ in $\mathbf{\bar{K}}$,

\begin{equation}
 \mathbf{\bar{\bar{K}}} = \mathbf{\bar{K}} + \mathbf{\Delta K} \,,
 \label{eq:variation_K}
\end{equation}

\noindent
results in a variation $\mathbf{\Delta u}$ in $\mathbf{\bar{u}}$, 

\begin{equation}
 \mathbf{\bar{\bar{u}}} = \mathbf{\bar{u}} + \mathbf{\Delta u} \,.
 \label{eq:variation_u}
\end{equation}

Since the load vector is constant, the following equation must be satisfied:

\begin{equation}
 \mathbf{\bar{\bar{K}}}\,\mathbf{\bar{\bar{u}}} = \mathbf{\bar{K}}\,\mathbf{\bar{u}} = \mathbf{f} \,.
 \label{eq:variation_identity}
\end{equation}

Thus, two expressions can be obtained for the corresponding variation of structural compliance

\begin{equation}
 \Delta C = -\frac{1}{2}\,\mathbf{\bar{u}}^T \mathbf{\Delta K}\,\mathbf{\bar{\bar{u}}} = \frac{1}{2}\,\mathbf{f}^T\,\mathbf{\Delta u} \,.
 \label{eq:variation_C}
\end{equation}

In order to obtain an approximated value for $\Delta C$, $\mathbf{\bar{\bar{u}}}$ must be estimated. It can be done by using a preconditioned Conjugate Gradient Method (CGM) to solve the linear system presented in Eq.~(\ref{eq:variation_identity}).

For given initial guess $\mathbf{u_0}$, initial direction $\mathbf{d_0}$ and preconditioner matrix $\mathbf{M}$, the CGM provides a set of $\mathbf{\bar{\bar{K}}}$-orthogonal directions $\{\mathbf{d_0},\mathbf{d_1},\hdots,\mathbf{d_{G-1}}\}$ and a corresponding set of coefficients $\{\mu_0,\mu_1,\hdots,\mu_{G-1}\}$, from which $\mathbf{\bar{\bar{u}}}$ can be obtained as

\begin{equation}
  \mathbf{\bar{\bar{u}}} = \mathbf{u_0} + \sum\limits_{k=0}^{G-1} \mu_k\,\mathbf{d_k} \,.
  \label{eq:solution_cgm}
\end{equation}

If $m$ directions and coefficients are calculated, with $m < G$, Eq.~(\ref{eq:solution_cgm}) can be rewritten as

\begin{equation}
 \mathbf{\bar{\bar{u}}} = \mathbf{u_0} + \bm{\delta}_\mathbf{u}^{(m)} + \bm{\varepsilon}_{\mathbf{u}}^{(m)} \,,
 \label{eq:solution_cgm_error}
\end{equation}

\noindent
where $\bm{\delta}_\mathbf{u}^{(m)}$ is a known term, given by

\begin{equation}
 \bm{\delta}_\mathbf{u}^{(m)} = \sum\limits_{k=0}^{m-1} \mu_k\,\mathbf{d_k} \,,
 \label{eq:known_term}
\end{equation}

\noindent
and $\bm{\varepsilon}_\mathbf{u}^{(m)}$ is the unknown error, given by

\begin{equation}
 \bm{\varepsilon}_\mathbf{u}^{(m)} = \sum\limits_{k=m}^{G-1} \mu_k\,\mathbf{d_k} \,.
 \label{eq:unknown_error}
\end{equation}

Therefore, the $m$-steps estimation of $\mathbf{\bar{\bar{u}}}$ can be expressed as

\begin{equation}
 \mathbf{u_m} = \mathbf{u_0} + \bm{\delta}_\mathbf{u}^{(m)} \,.
 \label{eq:u_approx}
\end{equation}

From Eq.~(\ref{eq:variation_C}), the variation $\Delta C$ can be written as

\begin{equation}
 \Delta C = D_{\mathbf{u}}^{(m)} + E_{\mathbf{u}}^{(m)} = D_{\mathbf{f}}^{(m)} + E_{\mathbf{f}}^{(m)} \,,
 \label{eq:variation_C_cgm}
\end{equation}

\noindent
where the known terms $D_{\mathbf{u}}^{(m)}$ and $D_{\mathbf{f}}^{(m)}$ are given by

\begin{equation}
 D_{\mathbf{u}}^{(m)} = -\frac{1}{2}\,\mathbf{\bar{u}}^T\,\mathbf{\Delta K}\, \mathbf{u_m}
 \label{eq:Du}
\end{equation}

\noindent
and

\begin{equation}
 D_{\mathbf{f}}^{(m)} = \frac{1}{2}\,\mathbf{f}^T \left[\mathbf{u_m} - \mathbf{\bar{u}}\right] \,;
 \label{eq:Df}
\end{equation}

\noindent
and the unknown errors $E_{\mathbf{u}}^{(m)}$ and $E_{\mathbf{f}}^{(m)}$ are given by

\begin{equation}
\begin{aligned}
 E_{\mathbf{u}}^{(m)} {}& = \frac{1}{2} \left[\mathbf{u_0} - \mathbf{\bar{u}} + \bm{\varepsilon}_{\mathbf{u}}^{(m)}\right]^T \mathbf{\bar{\bar{K}}}\,\bm{\varepsilon}_{\mathbf{u}}^{(m)} \\& = \frac{1}{2} \left\langle \mathbf{u_0}-\mathbf{\bar{u}} \,,\, \bm{\varepsilon}_{\mathbf{u}}^{(m)} \right\rangle_{\mathbf{\bar{\bar{K}}}} + \frac{1}{2} \left\|\bm{\varepsilon}_{\mathbf{u}}^{(m)}\right\|_{\mathbf{\bar{\bar{K}}}}^2
\end{aligned}
 \label{eq:Eu}
\end{equation}

\noindent
and

\begin{equation}
\begin{aligned}
 E_{\mathbf{f}}^{(m)} {}& = \frac{1}{2} \left[\mathbf{u_0} + \bm{\varepsilon}_{\mathbf{u}}^{(m)}\right]^T \mathbf{\bar{\bar{K}}}\,\bm{\varepsilon}_{\mathbf{u}}^{(m)} \\& = \frac{1}{2} \left\langle \mathbf{u_0} \,,\, \bm{\varepsilon}_{\mathbf{u}}^{(m)} \right\rangle_{\mathbf{\bar{\bar{K}}}} + \frac{1}{2} \left\|\bm{\varepsilon}_{\mathbf{u}}^{(m)}\right\|_{\mathbf{\bar{\bar{K}}}}^2 \,.
\end{aligned}
 \label{eq:Ef}
\end{equation}

The preconditioned CGM procedure is presented in Algorithm~\ref{alg:precond_cgm}. For a given problem (defined by the pair $\mathbf{\bar{\bar{K}}}$ and $\mathbf{f}$), an initial guess $\mathbf{u_0}$, an initial direction $\mathbf{d_0}$, a preconditioner matrix $\mathbf{M}$, the maximum number of iterations $m$ and an early stop criteria $\tau$ must be provided. This procedure returns $\mathbf{u_m}$, used to compute the estimations of $\Delta C$: $D_\mathbf{u}^{(m)}$ and $D_\mathbf{f}^{(m)}$. The direction $\mathbf{d_m}$ is also returned so that, in case the $m$ iterations were not enough to obtain the desired precision, more iterations can be performed, using $\mathbf{u_m}$ and $\mathbf{d_m}$ as input. 

\begin{algorithm}
\caption{Preconditioned Conjugate Gradient Method.}
\label{alg:precond_cgm}
\begin{algorithmic}
\STATE Given Problem: $\mathbf{\bar{\bar{K}}},\,\mathbf{f}$
\STATE Input: $\mathbf{u_0},\,\mathbf{d_0},\,\mathbf{M},\,m,\,\tau$
\STATE $\mathbf{g_0} \;\leftarrow\; \mathbf{\bar{\bar{K}}}\,\mathbf{u_0} - \mathbf{f}$
\FOR{$k \in \{0,1,\hdots,m-1\}$}
\IF{$\|\mathbf{g_{k}}\| < \tau$}
\RETURN $\mathbf{u_{k}},\,\mathbf{d_{k}},\,k$
\ENDIF
\STATE $\mathbf{e_k}\phantom{_{+1}} \;\leftarrow\; \phantom{-}\mathbf{\bar{\bar{K}}}\,\mathbf{d_k}$
\STATE $\mu_k\phantom{_{+1}} \;\leftarrow\; -\frac{\mathbf{d_k}^T \mathbf{g_k}}{\mathbf{d_k}^T \mathbf{e_k}}$
\STATE $\mathbf{u_{k+1}} \;\leftarrow\; \phantom{-}\mathbf{u_k} + \mu_k\,\mathbf{d_k}$
\STATE $\mathbf{g_{k+1}} \;\leftarrow\; \phantom{-}\mathbf{g_k} + \mu_k\,\mathbf{e_k}$
\STATE $\mathbf{q_k}\phantom{_{+1}} \;\leftarrow\; \phantom{-}\mathbf{M}^{-1}\,\mathbf{g_{k+1}}$
\STATE $\beta_k\phantom{_{+1}} \;\leftarrow\; \phantom{-}\frac{\mathbf{e_k}^T\mathbf{q_k}}{\mathbf{d_k}^T\mathbf{e_k}}$
\STATE $\mathbf{d_{k+1}} \;\leftarrow\; -\mathbf{q_k} + \beta_k\,\mathbf{d_k}$
\ENDFOR
\RETURN $\mathbf{u_m},\,\mathbf{d_m},\,m$
\end{algorithmic}
\end{algorithm}

The stiffness variation can be defined so that the variation of structural compliance corresponds to the sensitivity number of an element: if $x_i = 0$ and $\mathbf{\Delta K} = \mathbf{K_i}$, $\alpha_i^{\langle 1 \rangle} = \Delta C$; if $x_i = 1$ and $\mathbf{\Delta K} = -\mathbf{K_i}$, $\alpha_i^{\langle 1 \rangle} = -\Delta C$.

Considering $\mathbf{u_0} = \mathbf{0}$ and $\mathbf{d_0} = \mathbf{M}^{-1} \mathbf{f}$, $E_{\mathbf{f}}^{(m)}$ decreases monotonically as more CGM iterations are performed. Thus, the sensitivity values can be estimated, with arbitrary precision, as

\begin{equation}
 \boxed{\alpha^{\langle 1 \rangle}_i =
 \left\{\begin{aligned}
  & -\frac{1}{2}\,\mathbf{f}^T \left[\mathbf{\bar{u}} - \mathbf{u_m}\right]  \;, & \text{if } x_i = 0 \,,\\
  &  -\frac{1}{2}\,\mathbf{f}^T \left[\mathbf{u_m} - \mathbf{\bar{u}} \right]  \;, & \text{if } x_i = 1 \,.
 \end{aligned}\right.}
 \label{eq:sensitivity_cgm_u0df}
\end{equation}

Considering $\mathbf{u_0} = \mathbf{\bar{u}}$ and $\mathbf{d_0} = -\mathbf{M}^{-1} \mathbf{\Delta K}\,\mathbf{\bar{u}}$, $E_{\mathbf{u}}^{(m)}$ decreases monotonically as more CGM iterations are performed. Thus, the sensitivity values can be estimated, with arbitrary precision, as

\begin{equation}
 \boxed{\alpha^{\langle 1 \rangle}_i = -\frac{1}{2}\,\mathbf{\bar{u}}^T\,\mathbf{K_i}\,\mathbf{u_m} \;,\;\; x_i \in \{0,1\} \,.}
 \label{eq:sensitivity_cgm_uudg}
\end{equation}

Considering $\mathbf{u_0} = \mathbf{0}$ and $\mathbf{d_0} = \mathbf{\bar{u}}$, both $E_{\mathbf{f}}^{(m)}$ and  $E_{\mathbf{u}}^{(m)}$ decrease monotonically as more iterations are performed. Thus, either Eq.~(\ref{eq:sensitivity_cgm_u0df}) or Eq.~(\ref{eq:sensitivity_cgm_uudg}) can be used to estimate the sensitivity values.

In \ref{ap_cgm_explicit}, sensitivity expressions are presented for each considered initial condition $(\mathbf{u_0} \,,\, \mathbf{d_0})$, for $1$ and $2$ CGM steps.

For $\mathbf{u_0} = \mathbf{\bar{u}}$ and $\mathbf{d_0} = -\mathbf{M}^{-1} \mathbf{\Delta K}\,\mathbf{\bar{u}}$, Eq.~(\ref{eq:sensitivity_cgm_uudg}) always provides a better result than the FOCI approach with linear interpolation. For this case, considering no preconditioning, i.e., $\mathbf{M} = \mathbf{I}$, the following sensitivity values are obtained after performing $1$ CGM step:

\begin{equation}
\boxed{\alpha^{\langle 1 \rangle}_i = \left\{\begin{aligned}
&\begin{aligned}
-{}&\frac{1}{2} \Bigg[\mathbf{\bar{u}}^T\,\mathbf{K_i}\,\mathbf{\bar{u}}-\frac{\left[\mathbf{\bar{u}}^T \left[\mathbf{K_i}^2\right] \mathbf{\bar{u}}\right]^2}{\mathbf{\bar{u}}^T \left[\mathbf{K_i}\,\mathbf{\bar{K}}\,\mathbf{K_i}\right] \mathbf{\bar{u}} + \mathbf{\bar{u}}^T \left[\mathbf{K_i}^3\right] \mathbf{\bar{u}}}\Bigg]
\\& \text{if } x_i = 0 \,,
\end{aligned}\\
&\begin{aligned}
-{}&\frac{1}{2} \Bigg[\mathbf{\bar{u}}^T\,\mathbf{K_i}\,\mathbf{\bar{u}}+\frac{\left[\mathbf{\bar{u}}^T \left[\mathbf{K_i}^2\right] \mathbf{\bar{u}}\right]^2}{\mathbf{\bar{u}}^T \left[\mathbf{K_i}\,\mathbf{\bar{K}}\,\mathbf{K_i}\right] \mathbf{\bar{u}} - \mathbf{\bar{u}}^T \left[\mathbf{K_i}^3\right] \mathbf{\bar{u}}}\Bigg]
\\& \text{if } x_i = 1 \,.
\end{aligned}
\end{aligned}\right.}
\label{eq:cgm_simple}
\end{equation}

\subsection{Sensitivity error}

Both the naive and Woodbury expressions provide the exact sensitivity values, thus, there is no error in such approaches. The element-wise sensitivity error for FOCI and HOCI expressions can be obtained by subtracting the truncated series of Eq.~(\ref{eq:sensitivity_series_eig}) from Eq.~(\ref{eq:sensitivity_woodbury_eig}), then taking the absolute value of the result.

For void elements, only the first-order approximation should be considered, since the series can diverge. For the $i$th element, such that $x_i = 0$, the error $\varepsilon_{\alpha}$ of FOCI sensitivity expression is bounded, as given below:

\begin{equation}
\varepsilon_{\alpha} \leq \left[\frac{1}{2}\,\mathbf{\bar{u}}^T\,\mathbf{K_i}\,\mathbf{\bar{u}}\right] \left[\frac{\|\mathbf{\bar{A}_i}\|_2}{1+\|\mathbf{\bar{A}_i}\|_2}\right] = [C_i] \, [B_r] \,.
\label{eq:error_bound_void}
\end{equation}

\begin{figure}[h!]
\centering
\includegraphics[width=0.45\textwidth]{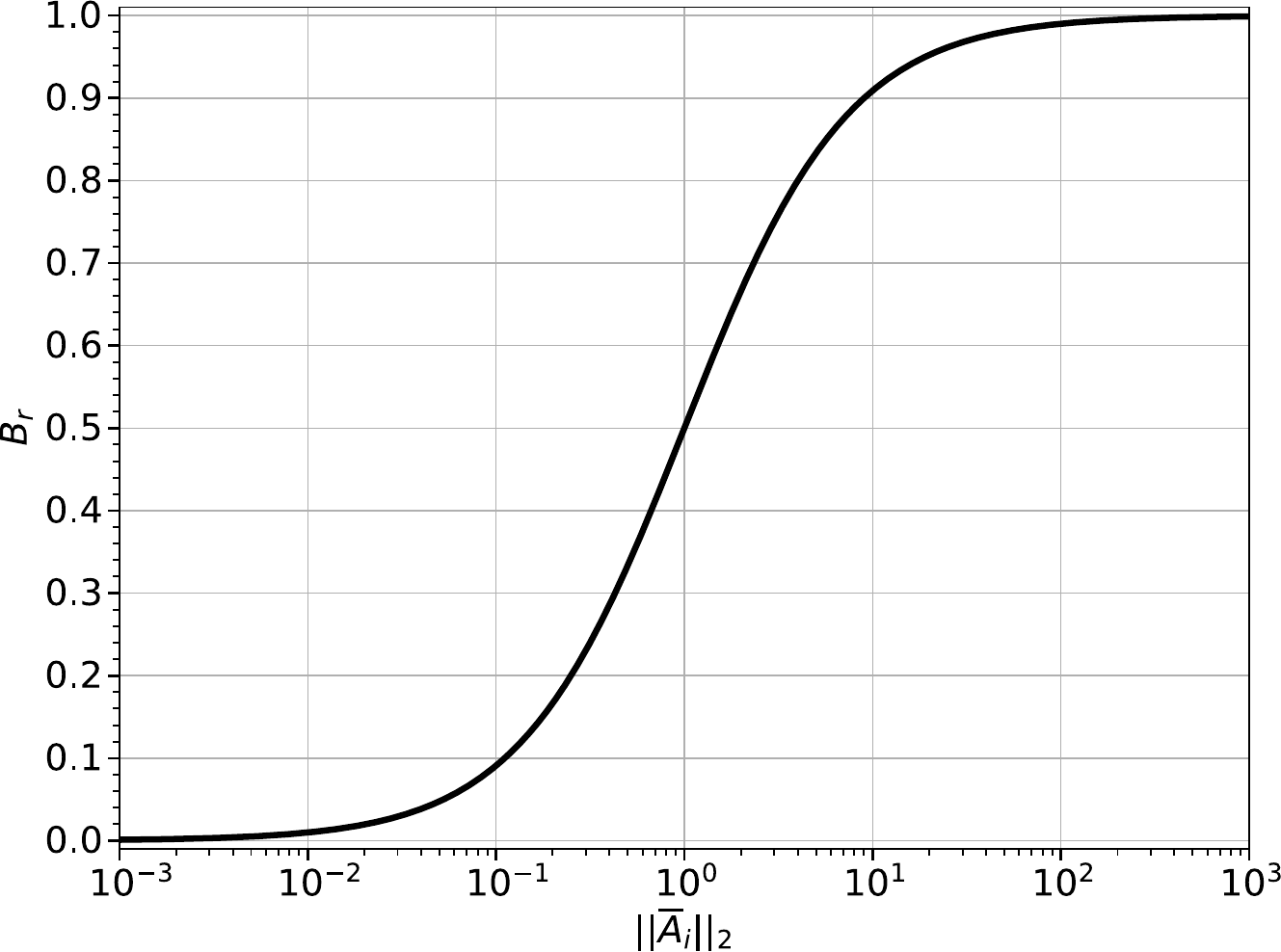}
\caption{Upper bound for the sensitivity relative error of a void element.}
\label{fig:sens_error_void}
\end{figure}

The term on the left, denoted by $C_i$, is the first-order sensitivity number itself. The term on the right, denoted by $B_r$, is an upper bound for the elemental sensitivity relative error. Its behavior with respect to $\|\mathbf{\bar{A}_i}\|_2$ is presented in Fig.~\ref{fig:sens_error_void}.

For a sufficiently small $\|\mathbf{\bar{A}_i}\|_2$, the linear approximation is accurate. As it increases, the relative error bound goes towards its maximal value of $100\%$.

The series will be considered up to its $q$th order term for solid elements. For the $i$th element, such that $x_i = 1$, the error $\varepsilon_{\alpha}$ of the HOCI sensitivity expression is bounded, as given below:

\begin{equation}
\varepsilon_{\alpha} \leq \left[\frac{1}{2}\,\mathbf{\bar{u}}^T\,\mathbf{K_i}\,\mathbf{\bar{u}}\right] \left[\frac{{\|\mathbf{\bar{A}_i}\|_2}^q}{1-\|\mathbf{\bar{A}_i}\|_2}\right] = [C_i]\,[B_r] \,.
\label{eq:error_bound_solid}
\end{equation}

\begin{figure}[h!]
\centering
\includegraphics[width=0.45\textwidth]{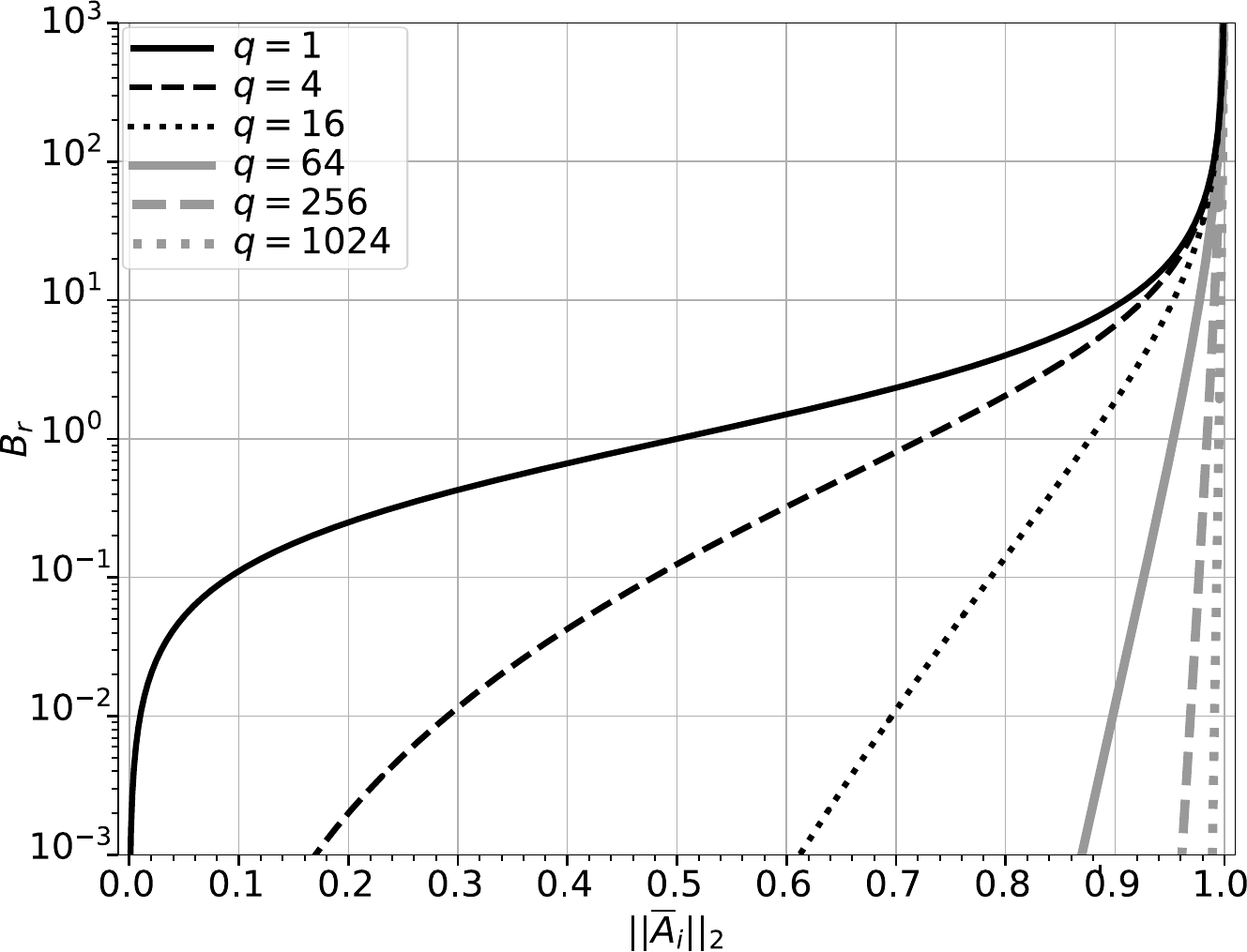}
\caption{Upper bound for the sensitivity relative error of a solid element.}
\label{fig:sens_error_solid}
\end{figure}

Again, the term on the left is the first-order sensitivity number; and the term on the right is an upper bound for the elemental sensitivity relative error. Its behavior with respect to $\|\mathbf{\bar{A}_i}\|_2$ is presented in Fig.~\ref{fig:sens_error_solid} for different values of $q$.

For a sufficiently small $\|\mathbf{\bar{A}_i}\|_2$, the linear approximation is accurate. As it increases, the relative error bound goes towards indefinitely large values. It should be noted that even for very high values of $q$, the error bound rapidly increases as $\|\mathbf{\bar{A}_i}\|_2$ gets near to $1$.

Therefore, when considering linear continuous interpolations for the density vector, the maximal eigenvalue of $\mathbf{\bar{A}_i}$ can be used as a measure of potential non-linearity of the objective function with respect to $x_i$.

When using $\mathbf{u_0} = \mathbf{\bar{u}}$, for any number of steps, the error for the CGM sensitivity values is always smaller than the one for the FOCI expression. So the FOCI error bound is also a bound for the error of CGM approach.

An upper bound for $\|\mathbf{\bar{A}_i}\|_2$ can be obtained as

\begin{equation}
 \|\mathbf{\bar{A}_i}\|_2 \leq \|\mathbf{K_i}\|_2\,\|\mathbf{\bar{K}}^{-1}\|_2 = \frac{\|\mathbf{K_i}\|_2}{ \|\mathbf{\bar{K}}\|_2}\,\kappa(\mathbf{\bar{K}}) \,,
\label{eq:bound_eigenA}
\end{equation}

\noindent
where $\kappa(\mathbf{\bar{K}})$ denotes the condition number of $\mathbf{\bar{K}}$.

It can be shown that $\|\mathbf{\bar{K}}\|_2 \geq \|\mathbf{K_i}\|_2$ \citep{fried1973bounds}. This means that $\kappa(\mathbf{\bar{K}})$ is itself an upper bound for $\|\mathbf{\bar{A}_i}\|_2$:

\begin{equation}
 \|\mathbf{\bar{A}_i}\|_2 \leq \kappa(\mathbf{\bar{K}}) \,.
 \label{eq:bound_eigenA2}
\end{equation}

This is uninformative for solid elements, since a lower upper bound has already been obtained in \ref{ap_eigenvalues_A}, however, no bound had been presented for void elements until now. This expression shows that $\|\mathbf{\bar{A}_i}\|_2$ cannot be arbitrarily high for a fixed non-singular system.

Although expressions for void elements were presented, the relevance of assigning sensitivity numbers to void elements should be discussed. In the compliance minimization problem, any element with no solid elements in its immediate neighborhood must have a sensitivity value of $0$, since any change on this disconnected element would have no effect on the structural compliance. This means that, for a refined mesh, the sensitivity values of most of the void elements will be $0$ and the optimization process will be guided only by the solid part of the topology, through the effects of the sensitivity filter. So it would hardly be beneficial to compute the sensitivity numbers of void elements in a refined mesh. It is a reasonable action to simply assign all void sensitivity numbers as $0$ before the filtering procedure.

Besides disconnected elements, there is another configuration for an element in which the sensitivity number is known beforehand. An element which is the sole solid element connecting an imposed load to a constrained part of the structure will have an arbitrarily high sensitivity absolute value, according to the soft-kill parameter $\varepsilon_k$. In practical applications with refined meshes, this should only happen in elements that are directly loaded by external forces. Since any element within the sensitivity filter radius from these connective solid elements would be artificially removed from the design domain, it is a reasonable action to simply disregard their sensitivity values in the filtering procedure.

\section{Numerical examples}
\label{sec:results}

In the following numerical examples, the structural compliance minimization was performed through successive solutions of the optimization subproblem from Eq.~(\ref{eq:optimization_y_lin}), which corresponds to the BESO method. In all problems, a linear-elastic, homogeneous and isotropic material was considered; four-nodes bilinear square elements in plane stress state were used; and the soft-kill parameter was $\varepsilon_k = 10^{-9}$.

The exact sensitivity analyses were performed through the Woodbury approach. Such analyses were duplicated through the naive approach in order to validate the implementation.

Although the implemented algorithm works with the presented volume measure $V(\mathbf{x})$, the information commonly used to define constraints and compare results is the fraction of solid material in the design domain, defined as

\begin{equation}
 V_f(\mathbf{x}) = \frac{V(\mathbf{x})}{N} \,.
 \label{eq:volume_fraction}
\end{equation}

Therefore, in the following numerical examples, $V_f(\mathbf{x})$ and $V_f^*$ were used instead of $V(\mathbf{x})$ and $V^*$.

\subsection{Cantilever tie-beam}

\begin{figure}[h!]
\centering
\includegraphics[width=0.45\textwidth]{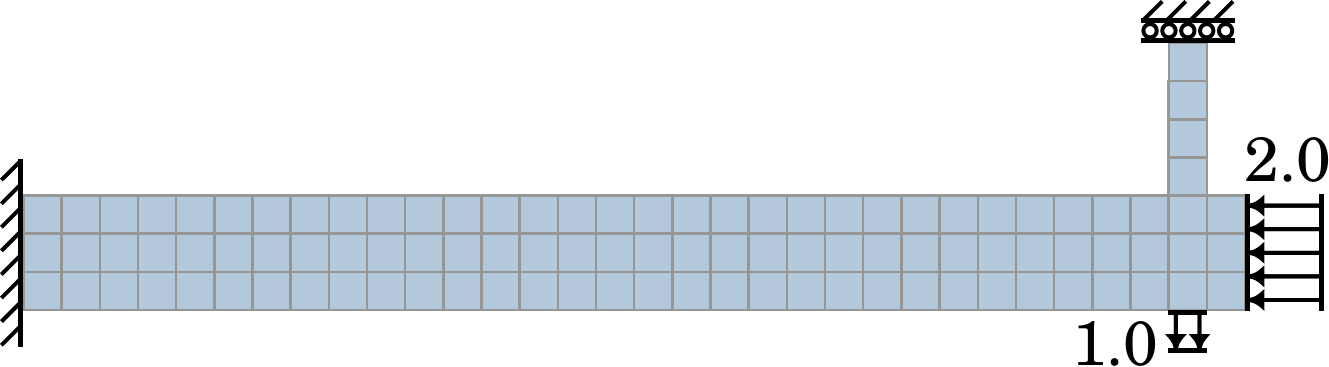}
\caption{Cantilever tie-beam.}
\label{fig:tie_beam}
\end{figure}

Fig.~\ref{fig:tie_beam} presents a $100$ elements mesh for a cantilever tie-beam. The material has a Young's modulus of $1.0$ and a Poisson's ratio of $0.0$; the dimensions of the elements are $1.0 \times 1.0$; over the rightmost edge, the intensity of the horizontal load per unit length is $2.0$; on the bottom edge, below the vertical tie, the intensity of the vertical load per unit length is $1.0$.

In \citet{zhou2001validity} this problem was presented as a simple example in which the standard ESO method results in a highly non-optimal design. For this mesh, the FOCI approach is not enough to correctly predict the variation of the objective function when the state of an element in the vertical tie is switched. Such imprecision results in the removal of all the elements from the vertical tie, which are critical elements. Their removal qualitatively changes the behavior of the structure since a mechanical restriction is artificially removed from the problem. It was stated that this problem would still happen for fine meshes, when certain values of rejection ratio are used \citep{zhou2001validity}. However, it has already been shown that by using fine meshes with appropriate parameters, the ESO method is able to obtain satisfactory optimized solutions \citep{edwards2007evaluative,huang2008new}.

% [ Figure 8 -- repositioned so it appears in the right place ]
\setcounter{figure}{7}
\begin{figure*}[b]
\centering
  \begin{subfigure}{0.40\textwidth}
    \includegraphics[width=\linewidth]{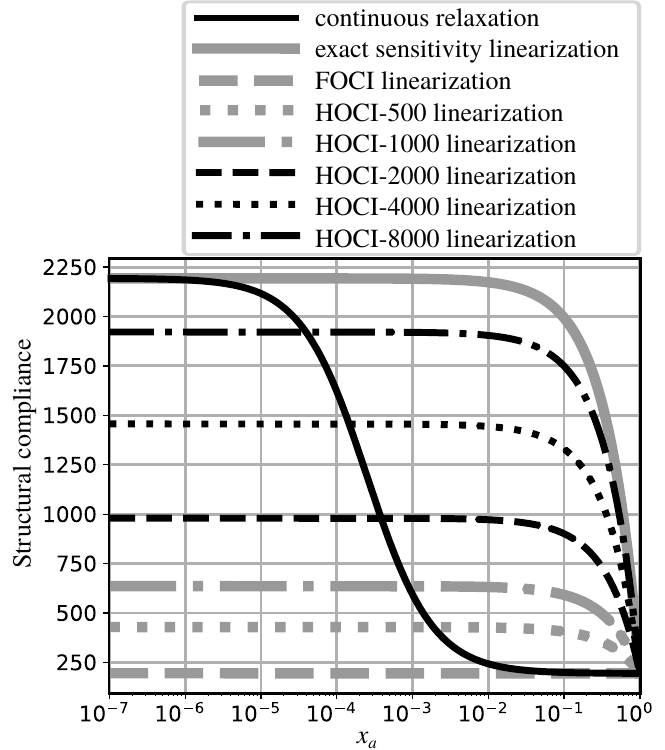}
    \caption{HOCI linearizations.}
  \end{subfigure}
  \hspace*{2em}
  \begin{subfigure}{0.40\textwidth}
    \includegraphics[width=\linewidth]{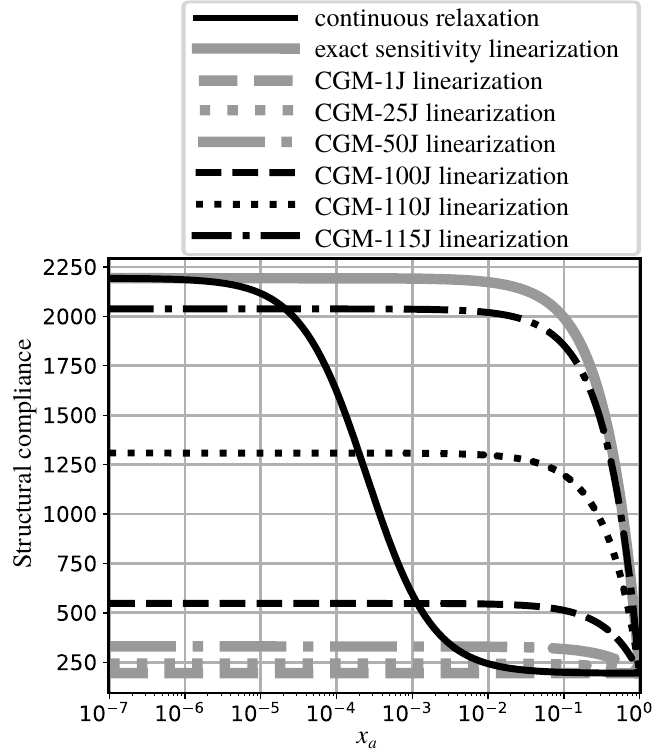}
    \caption{CGM linearizations.}
  \end{subfigure}
\caption{FVSA linearizations of the structural compliance with respect to $x_a$.}
\label{fig:linA_ros100}
\end{figure*}

To discourage misleading discussions over this problem, some complementary remarks should be made. Unlike problems in which there are trivial solutions that correspond to disconnected structures \citep{wang2009analysis}, this problem does not require any further considerations, it is solved simply by proper usage of the method. When density methods are used to optimize continuum structures, the mesh should be fine enough to produce complex (or at least non-trivial) topologies for the structures. Moreover, when discrete methods are used, the mesh should be fine enough so any  disconnection between structural components happens gradually, which may avoid highly non-optimal topology alterations. Lastly, although $1\%$ rejection ratio would usually be a reasonable value, because the mesh is too coarse, it forces the method to remove all of the material from the same spot, which should be enough to highlight how poorly defined is the setting of the presented problem.

Nonetheless, an example with extreme properties can be useful to evaluate potential improvements on this kind of method. In \citet{ghabraie2015eso}, it was shown that the removal of the vertical tie can be prevented by using accurate sensitivity values. In this section, this problem was further explored considering the proposed approaches.

By considering $V\!V^{(j)} = -1$ and $T\!V_{\max}^{(j)} = 1$ in all iterations, the proposed optimization procedure becomes the ESO method with rejection ratio of $1\%$. Firstly, a fully solid initial topology was considered. It has a volume fraction of $V_f(\mathbf{\bar{x}}^{(0)}) = 100\%$ and a structural compliance of $C(\mathbf{\bar{x}}^{(0)}) = 194.4$. 

\setcounter{figure}{6}
\begin{figure}[h!]
\centering
\begin{subfigure}{0.45\textwidth}
\includegraphics[width=\linewidth]{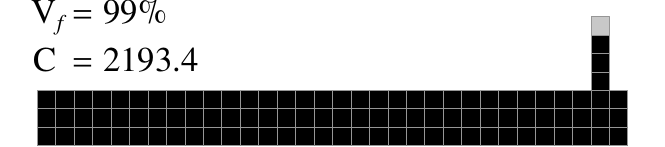}
\caption{Optimized with FOCI sensitivity analysis.}
\end{subfigure}\vspace*{1em}\\
\begin{subfigure}{0.45\textwidth}
\includegraphics[width=\linewidth]{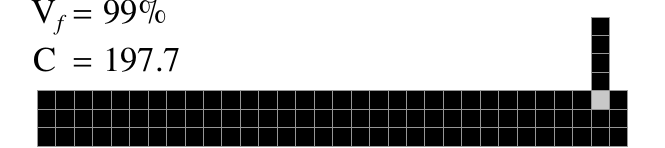}
\caption{Optimized with exact sensitivity analysis.}
\end{subfigure}
\caption{Cantilever tie-beams optimized for $V_f^*=99\%$.}
\label{fig:ergraphs_ros100}
\end{figure}

Fig.~\ref{fig:ergraphs_ros100} shows the solutions for $V_f^* = 99\%$ when FOCI approach is adopted and when the exact sensitivity vector is used. The solid elements are represented in black; the void elements are represented in gray; the index of the void element from Fig.~\ref{fig:ergraphs_ros100}(a) is denoted by $a$; and the index of the void element from Fig.~\ref{fig:ergraphs_ros100}(b) is denoted by $b$. It can be seen that the FOCI prediction, based only on small density variations, results in a very inefficient structure.

In Figs.~\ref{fig:linA_ros100} and \ref{fig:linB_ros100}, the FVSA linearizations produced from HOCI and CGM approaches are presented for the elements $a$ and $b$. The behavior of the relaxed continuum function for a linear interpolation is also presented. On the legend, the number next to HOCI indicates its order (HOCI-$1$ is referred to as FOCI); the number next to CGM indicates how many steps were performed; and the letter ``J'' after CGM indicates that Jacobi preconditioning were used. For the CGM approach, $\mathbf{u_0} = \mathbf{\bar{u}}$ and $\mathbf{d_0} = \pm\mathbf{M}^{-1}\,\mathbf{K_i}\,\mathbf{\bar{u}}$ were considered. In Fig.~\ref{fig:linA_ros100}, the linearizations are not straight lines, since the horizontal axis is in a logarithmic scale.

% [ Figure 8 -- original position ]
\setcounter{figure}{8}

\begin{figure*}[h!]
\centering
  \begin{subfigure}{0.40\textwidth}
    \includegraphics[width=\linewidth]{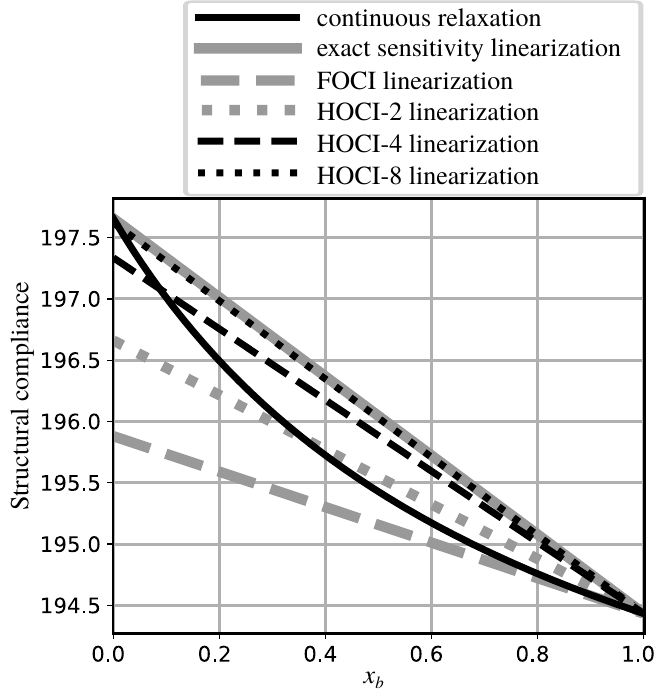}
    \caption{HOCI linearizations.}
  \end{subfigure}
  \hspace*{2em}
  \begin{subfigure}{0.40\textwidth}
    \includegraphics[width=\linewidth]{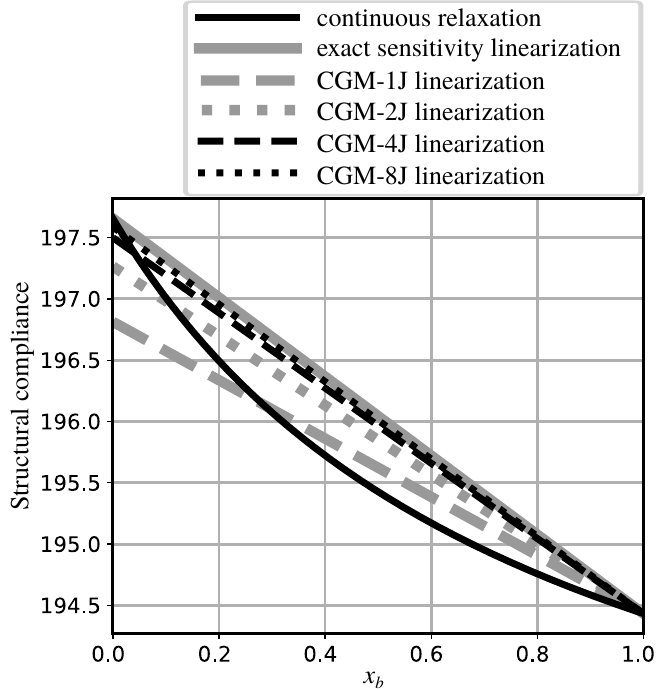}
    \caption{CGM linearizations.}
  \end{subfigure}
\caption{FVSA linearizations of the structural compliance with respect to $x_b$.}
\label{fig:linB_ros100}
\end{figure*}

The continuous relaxation of $C(x_b)$ is fairly well behaved. The FOCI approach produces a reasonable result and, for slightly higher orders, the HOCI approach can provide accurate sensitivity values. The CGM approach can also provide accurate sensitivity values in a small number of steps.

On the other hand, $C(x_a)$ has very distinct local and global behaviors. When $x_a$ is near $1$, it has little influence in the structural compliance. When $x_a$ goes below $0.01$, the effect of disconnecting the vertical tie appears as an abrupt compliance variation, which cannot be properly predicted by a local sensitivity analysis. The HOCI approach provides reasonable values only for orders higher than $1000$. And, to produce reasonable values, the CGM approach needs nearly half of the maximal number of CGM steps, which would provide the exact result.

\begin{figure}[h!]
\centering
\includegraphics[width=0.45\textwidth]{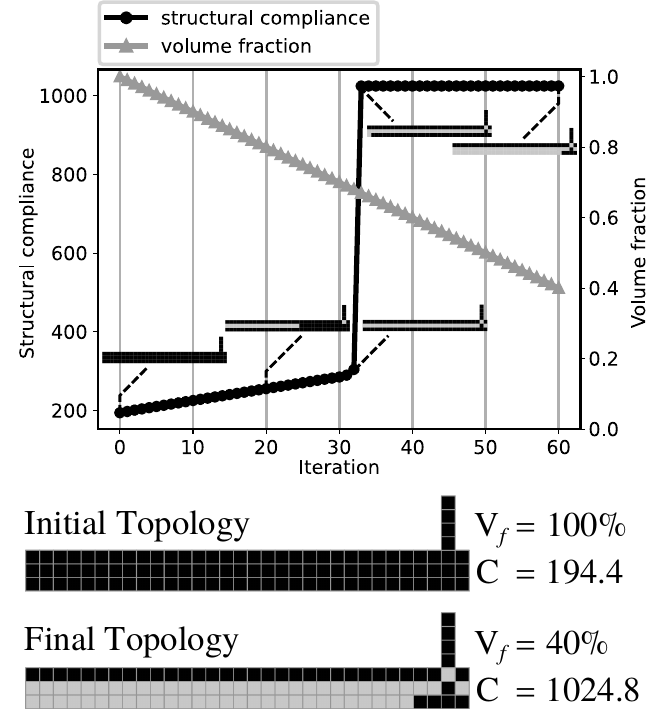}
\caption{Cantilever tie-beam optimization with $V_f(\mathbf{\bar{x}}^{(0)})=100\%$ and $V_f^*=40\%$.}
\label{fig:opt_ros100}
\end{figure}

By performing the optimization with exact sensitivity values, for a target volume fraction of $40\%$, the result shown in Fig.~\ref{fig:opt_ros100} is obtained. Since exact predictions are used, each iteration produces a topology with minimal compliance gain. It successfully prevented the removal of the vertical tie. It should be noted that some non-physical effects are present due to the type of finite element used: elements connected only by their nodes do not negatively affect the structural stiffness; the three rightmost elements have a distributed load on their edges, even so, the one in the middle can be removed without producing a singularity.

After removing all but one of the middle elements of the horizontal beam, the method is forced to remove a critical element. It chooses to break the bottom horizontal beam, because it would result in a smaller compliance gain than to break the top horizontal beam or the vertical tie. The breaking point can be easily identified in the plot of the objective function over the iterations, it is the point where there is an abrupt growth of the structural compliance.

% [ Figure 12 -- repositioned so it appears in the right place ]
\setcounter{figure}{11}
\begin{figure*}[b]
\centering
  \begin{subfigure}{0.9\textwidth}
    \includegraphics[width=\textwidth]{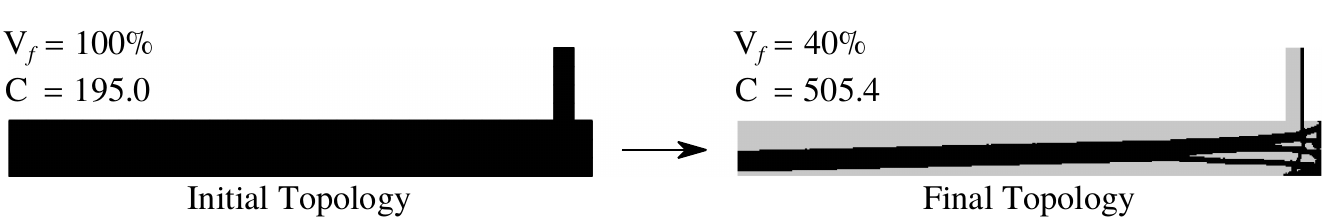}
    \caption{Optimization with $V_f(\mathbf{\bar{x}}^{(0)})=100\%$ and $V_f^*=40\%$.}
  \end{subfigure}\\
  \begin{subfigure}{0.9\textwidth}
    \includegraphics[width=\textwidth]{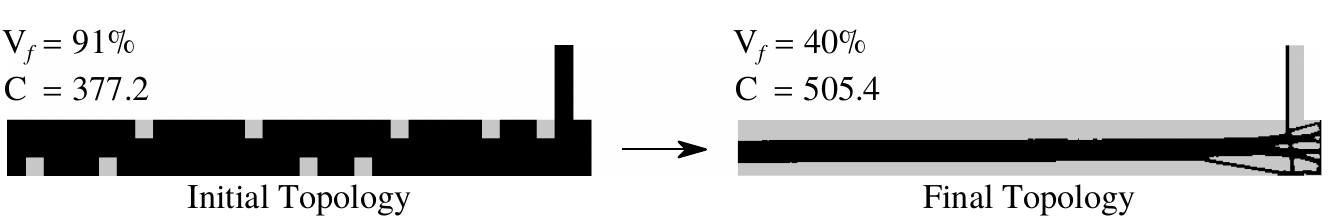}
    \caption{Optimization with $V_f(\mathbf{\bar{x}}^{(0)})=91\%$ and $V_f^*=40\%$.}
  \end{subfigure}
\caption{Cantilever tie-beam optimizations with refined meshes.}
\label{fig:opt_ref_ros}
\end{figure*}

The obtained result is worse than the solution presented in \citet{zhou2001validity}. In most iterative optimization methods, to obtain the best possible solution, the initial guess must be within the basins of attraction of the global optimum. In a sense, this is also true for this discrete method, a path of successive topologies is generated and its final point will always depend on the first one. It was shown that, when the optimization process starts from the fully solid topology, the succession of topological variations with minimal compliance gain does not lead to the desired optimum. However, when the optimization with exact sensitivity values starts from the initial topology of Fig.~\ref{fig:opt_ros91}, with a volume fraction of $91\%$, the reference solution is obtained.

\setcounter{figure}{10}
\begin{figure}[h!]
\centering
\includegraphics[width=0.45\textwidth]{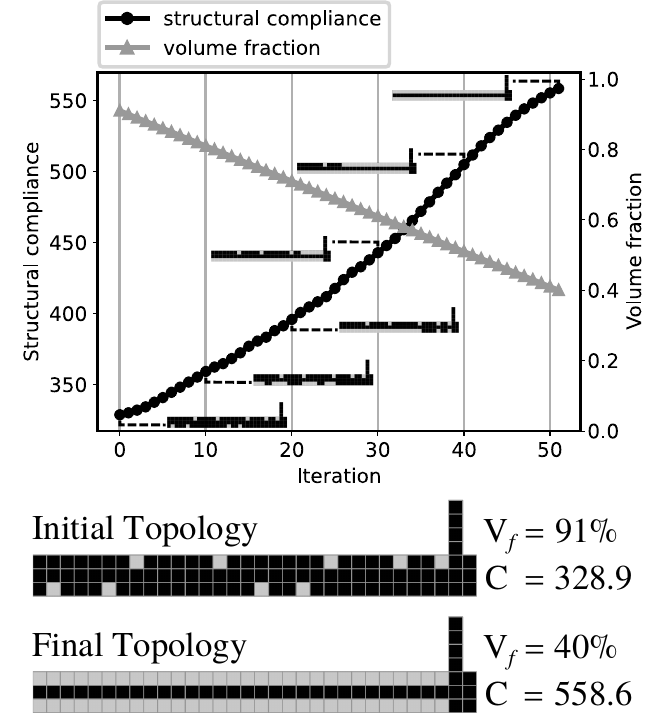}
\caption{Cantilever tie-beam optimization with $V_f(\mathbf{\bar{x}}^{(0)})=91\%$ and $V_f^*=40\%$.}
\label{fig:opt_ros91}
\end{figure}

As discussed, to approach this problem systematically, a finer mesh should be used with appropriate optimization parameters. A mesh of $25600$ elements was considered, obtained by remeshing each coarse element as $16\times 16$ fine elements of dimensions $0.0625 \times 0.0625$. Starting from a fully solid topology and from the initial topology of Fig.~\ref{fig:opt_ros91}, the optimization was performed for a target volume fraction of $40\%$. A sensitivity filter with radius of $0.2$ was used, as well as the presented momentum strategy. In iterations with changing volume, the constraints were given by $V\!V^{(j)} = -128$ ($0.5\%$) and $T\!V_{\max}^{(j)} = 1152$ ($4.5 \%$); in iterations with constant volume, they were given by $V\!V^{(j)} = 0$ ($0\%$) and $T\!V_{\max}^{(j)} = 1024$ ($4 \%$). Which corresponds to the BESO method with $E\!R = 0.5\%$ and $A\!R_{\max}=2\%$. The FOCI approach was used, with void sensitivity penalization of $\varepsilon_v = 10^{-6}$. The results are presented in Fig.~\ref{fig:opt_ref_ros}.

% [ Figure 12 -- original position ]
\setcounter{figure}{12}

Two different solutions with the same structural compliance were obtained, both are more efficient than the solution presented in \citet{zhou2001validity}, whose structural compliance is $562.9$ when calculated with this mesh. Therefore, even using the simplest sensitivity estimation, when a proper setting is defined, the method successfully produces optimized structures for this problem.

The mesh refinement not only allows the method to distribute the topological variations over the domain, but it also improves the general behavior of the objective function with respect to each design variable. A way to observe this effect is through maps of $\|\mathbf{\bar{A}_i}\|_2$ for different meshes, as shown in Fig.~\ref{fig:refining_eigenvalues} for fully solid topologies. As presented in Eqs.~(\ref{eq:error_bound_void}) and (\ref{eq:error_bound_solid}), these norms define upper bounds for the error of the sensitivity estimations.

% [ Figure 14 -- repositioned so it appears in the right place ]
\setcounter{figure}{13}
\begin{figure*}[b]
\centering
\begin{subfigure}{0.40\textwidth}
    \includegraphics[width=\linewidth]{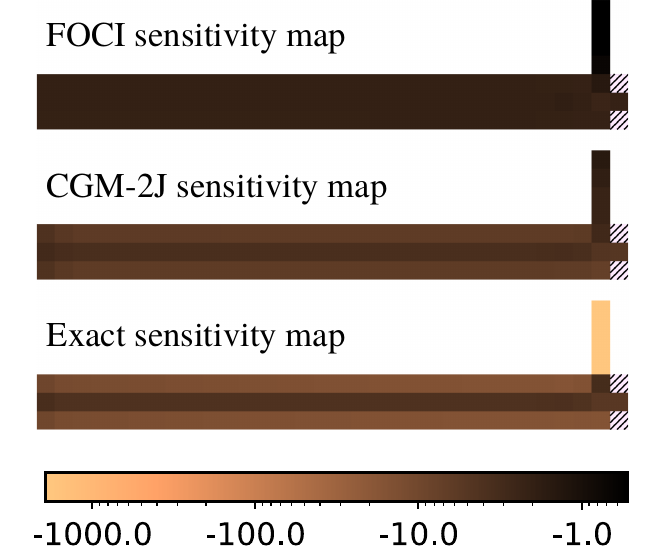}
    \caption{Sensitivity maps for the $100$ elements mesh.}
\end{subfigure}
\hspace*{2em}
\begin{subfigure}{0.40\textwidth}
    \includegraphics[width=\linewidth]{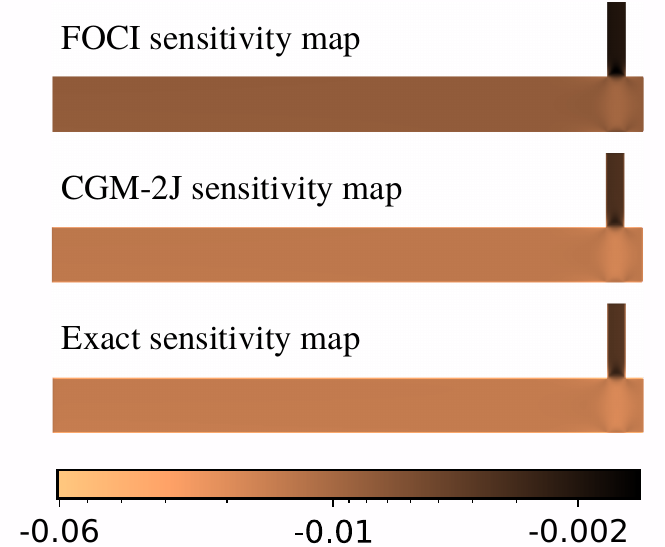}
    \caption{Sensitivity maps for the $25600$ elements mesh.}
\end{subfigure}
\caption{FOCI, CGM and exact sensitivity maps for fully solid topologies.}
\label{fig:refining_sens}
\end{figure*}

\setcounter{figure}{12}
\begin{figure}[h!]
\centering
\includegraphics[width=0.45\textwidth]{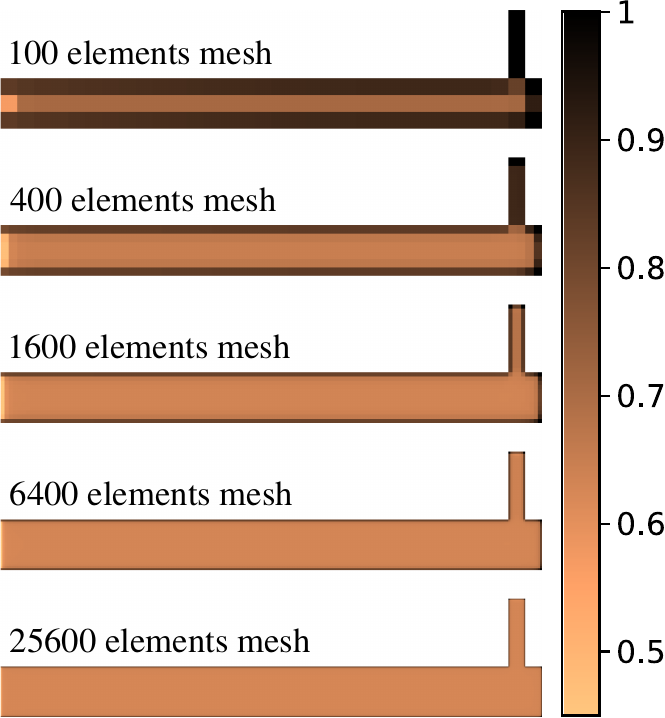}
\caption{Maps of $\|\mathbf{\bar{A}_i}\|_2$ for the fully solid topology in different meshes.}
\label{fig:refining_eigenvalues}
\end{figure}

The average of all $\|\mathbf{\bar{A}_i}\|_2$ goes from $0.83$, in the $100$ elements mesh, to $0.64$, in the $25600$ elements mesh. Elements in corners have norms near $1$, which means the sensitivity error is pratically unbounded; elements in free edges have norms below $1$, but still relatively high; elements inside the structure have norms below $0.7$; and elements in the clamped edge have the lowest norms, around $0.5$. This is related to the potential each element has to disconnect a load from the constrained part of the structure. A corner element can always produce a singularity: if there is a load in the corner and the element is removed the displacement of such a node will go towards infinity. When the mesh is refined, the ratio between elements in the interfaces and elements inside the structure is reduced, so the overall behavior of the objective function is improved. It should be noted that, even though every interface element has higher upper bounds for the sensitivity error, this does not mean the error will necessarily be high.

In Fig.~\ref{fig:refining_sens}, sensitivity maps are presented for the fully solid topology in the $100$ elements mesh and in the $25600$ elements mesh, obtained through the FOCI, CGM-$2$J ($2$ steps with Jacobi preconditioning) and Woodbury approaches. The CGM initial conditions were the same: $\mathbf{u_0} = \mathbf{\bar{u}}$ and $\mathbf{d_0} = \pm\mathbf{M}^{-1}\,\mathbf{K_i}\,\mathbf{\bar{u}}$. Since the exact sensitivity absolute values of the loaded corner elements go towards infinity, these elements were not considered in this sensitivity analysis. It can be seen that the highly non-linear behavior for elements in the vertical tie vanished in the refined mesh. For this mesh, the FOCI approach was able to produce coherent sensitivity values and the CGM approach produced very satisfactory results. Evidently, the sensitivity numbers assume smaller absolute values for finer meshes, since the influence of each element alone is reduced when the elements are smaller.

% [ Figure 14 -- original position ]
\setcounter{figure}{14}

\begin{figure}[h!]
\centering
\includegraphics[width=0.5\textwidth]{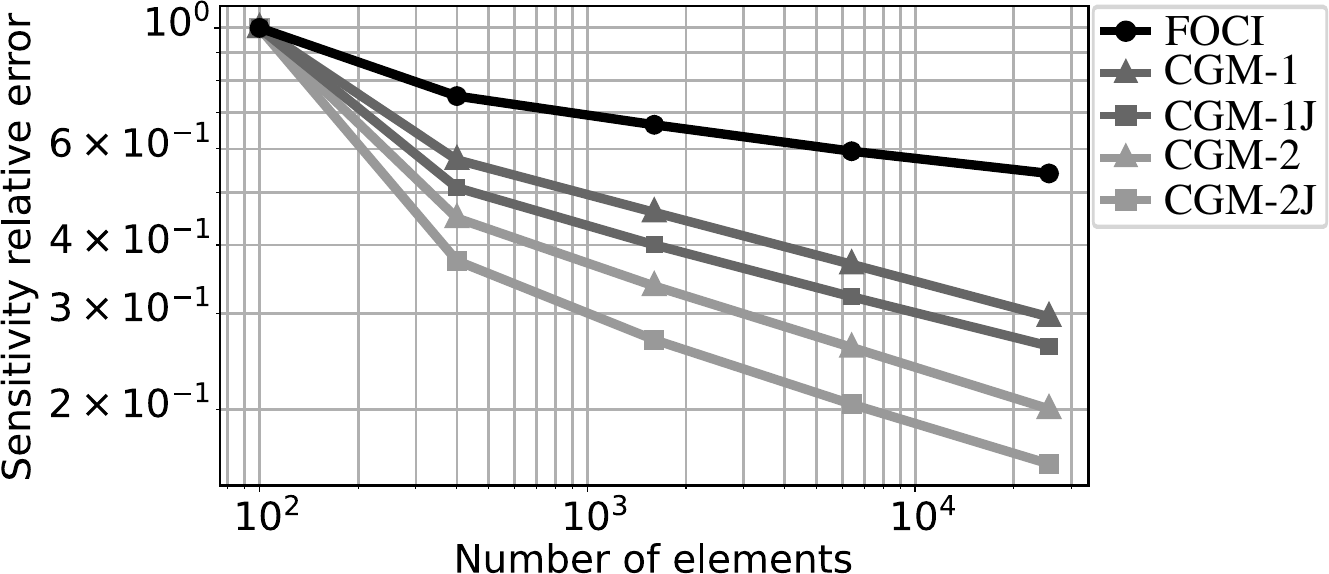}
\caption{Sensitivity relative $l^2$-error with respect to mesh refinement.}
\label{fig:refining_error}
\end{figure}

For fully solid topologies, Fig.~\ref{fig:refining_error} presents the progression of the relative $l^2$-error, given by the $l^2$-norm of the error vector divided by the $l^2$-norm of exact sensitivity vector, with respect to the number of elements in the mesh. The CGM with $1$ and $2$ steps were considered, with and without Jacobi preconditioning. The FOCI results were also presented for comparison. In the coarse mesh, the relative $l^2$-error is around $100\%$ for all cases, because of the very high error for the elements in the vertical tie. From the second mesh ($400$ elements) to the last one ($25600$ elements), it seems that the sensitivity errors are reduced following fixed power laws, since their plots in logarithmic scale result in nearly straight lines.

In all cases, including the FOCI approach, the overall sensitivity error was reduced with the mesh refinement. This means that, as the mesh is refined, the objective function becomes ``more linear'' (its linear approximation becomes more accurate) with respect to each elemental density. Moreover, by using the proposed CGM sensitivity analysis, there was a substantial improvement of the error values and in how fast they drop with mesh refinement. For the most refined mesh, even when using a single CGM step without preconditioning, this measure of error was reduced from more than $50\%$ to around $30\%$, when compared with the FOCI result.

\subsection{Cantilever beam}

\begin{figure}[h!]
\centering
\hspace*{1.4em}
\begin{subfigure}{0.40\textwidth}
    \includegraphics[width=\linewidth]{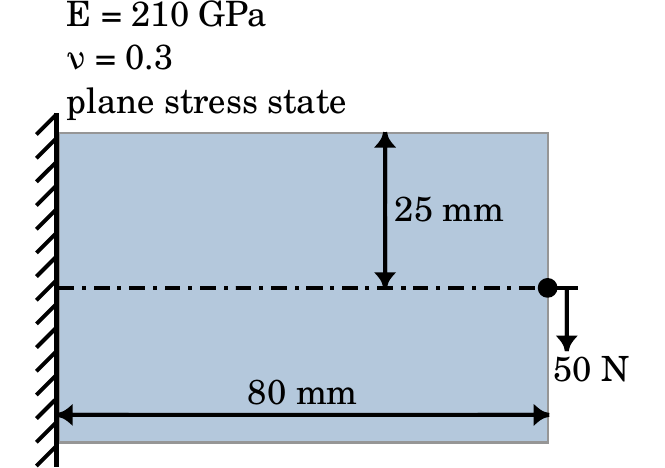}
    \caption{Design domain.}
\end{subfigure}
\begin{subfigure}{0.40\textwidth}
    \includegraphics[width=\linewidth]{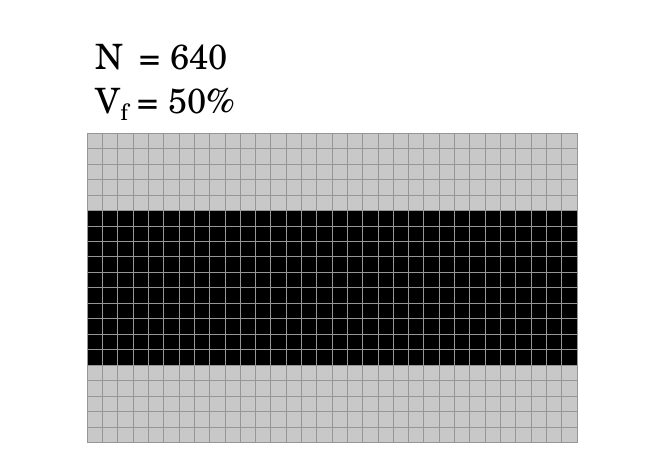}
    \caption{Initial topology.}
\end{subfigure}
\caption{Cantilever beam.}
\label{fig:cantilever_beam}
\end{figure}

Fig.~\ref{fig:cantilever_beam} presents the design domain and the initial topology considered for the optimization of a cantilever beam. A mesh with $32 \times 20$ elements of dimensions $2.5 \times 2.5\,\text{mm}$ was considered.

\begin{figure*}[h!]
\centering
\includegraphics[width=0.80\textwidth]{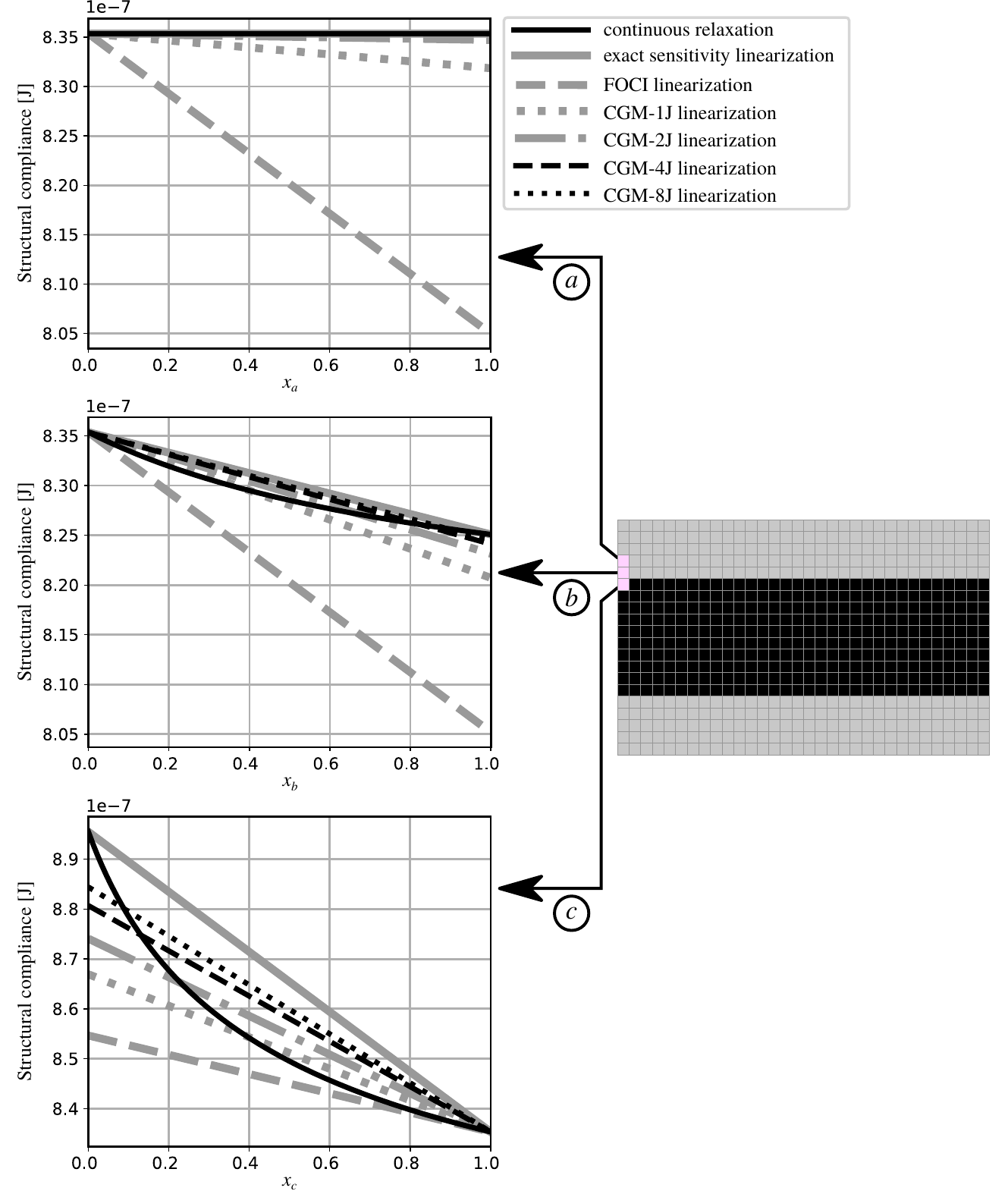}
\caption{FVSA linearizations for the cantilever beam.}
\label{fig:cant_linearizations}
\end{figure*}

In Fig.~\ref{fig:cant_linearizations}, the FVSA linearizations produced with the FOCI and CGM approaches are presented for three elements in the clamped extremity: the first solid element from top (element $c$); the void element right above it (element $b$); and the next void element above (element $a$). The relaxed function behavior for a linear interpolation is also presented. As before, $\mathbf{u_0} = \mathbf{\bar{u}}$ and $\mathbf{d_0} = \pm\mathbf{M}^{-1}\,\mathbf{K_i}\,\mathbf{\bar{u}}$, with Jacobi preconditioning.

It can be noted that the three relaxed functions are well behaved and a small number of CGM steps is needed to achieve accurate sensitivity values. The FOCI sensitivity analysis results in very inaccurate values for the disconnected void elements. This reinforces the argument that it may be better to always disregard the sensitivity values (set them to $0$) of void elements, which can be done without any changes to the implementation, by using $\varepsilon_v=0$.

\begin{figure}[h!]
\centering
\includegraphics[width=0.45\textwidth]{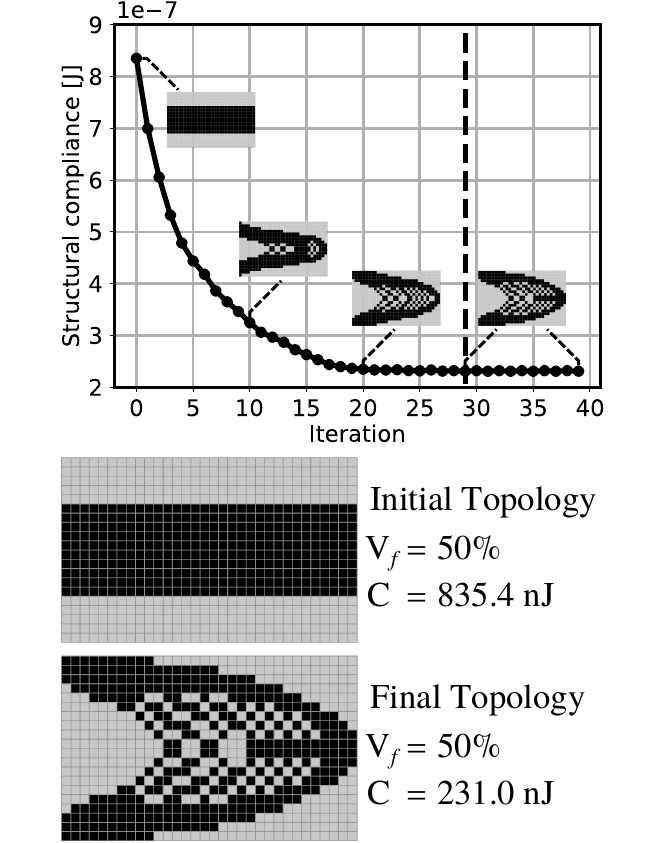}
\caption{Cantilever beam optimization with a fixed $V_f(\mathbf{\bar{x}}^{(j)})=50\%$.}
\label{fig:cant_opt}
\end{figure}

Fig.~\ref{fig:cant_opt} presents the topology optimization of the presented cantilever beam, using exact sensitivity values. It was performed with a fixed volume fraction of $V_f(\mathbf{\bar{x}}^{(j)})=V_f^*=50\%$. The constraints for all iterations were $V\!V^{(j)} = 0$ ($0.0 \%$) and $T\!V_{\max}^{(j)} = 28$ ($\approx 4.4 \%$), which corresponds to the BESO method with $E\!R = 0.0\%$ and $A\!R_{\max}=2.2\%$. The structure of minimal compliance was obtained at the $29$th iteration, after that, the topologies oscillate until the patience criterion is achieved. No filter was used so the sensitivity numbers truly correspond to objective function variations.

% [ Figure 18 -- original position ]

The influence of the initial condition used in the CGM approach was evaluated  for the sequence of topologies produced in this optimization process. Three cases were considered: in the first case, $\mathbf{u_0} = \mathbf{0}$ and $\mathbf{d_0} = \mathbf{M}^{-1}\,\mathbf{f}$; in the second case, $\mathbf{u_0} = \mathbf{\bar{u}}$ and $\mathbf{d_0} = \pm\mathbf{M}^{-1}\,\mathbf{K_i}\,\mathbf{\bar{u}}$; in the third case, $\mathbf{u_0} = \mathbf{0}$ and $\mathbf{d_0} = \mathbf{\bar{u}}$.

\begin{figure}[H]
\centering
  \begin{subfigure}{0.35\textwidth}
    \includegraphics[width=\textwidth]{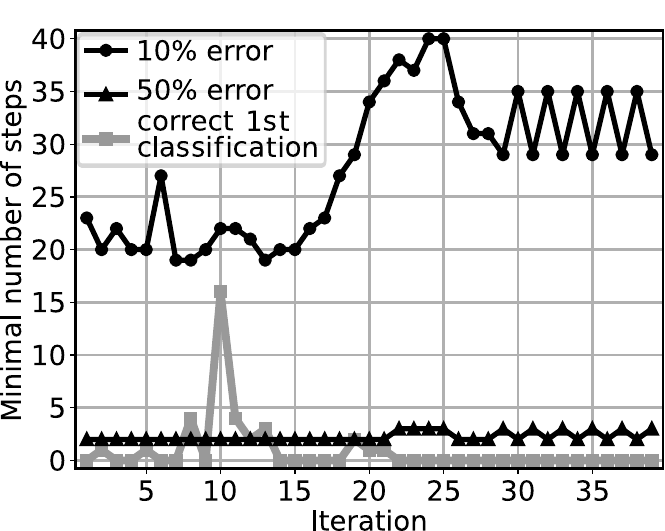}
    \caption{Second case without preconditioning.}
  \end{subfigure}\\
  \begin{subfigure}{0.35\textwidth}
    \includegraphics[width=\textwidth]{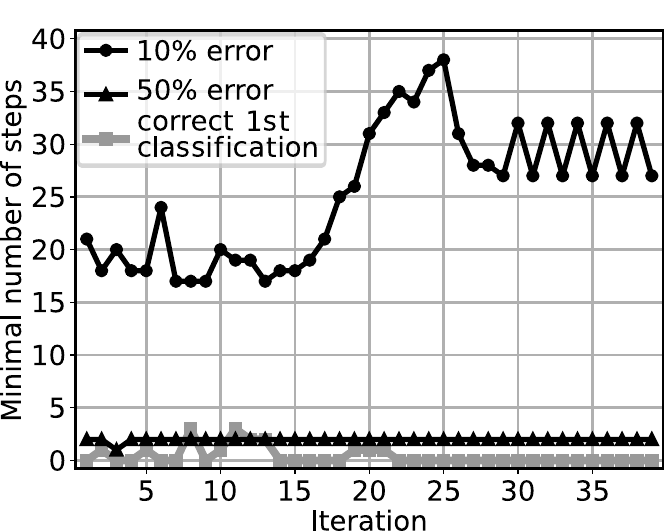}
    \caption{Second case with preconditioning.}
  \end{subfigure}\\
  \begin{subfigure}{0.35\textwidth}
    \includegraphics[width=\textwidth]{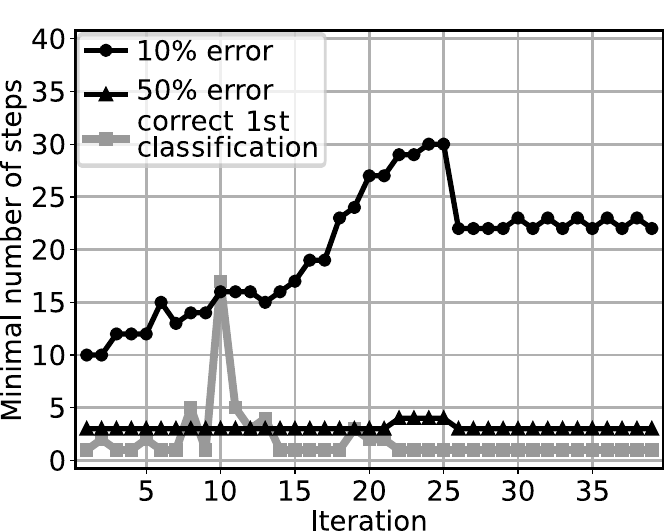}
    \caption{Third case without preconditioning.}
  \end{subfigure}\\
  \begin{subfigure}{0.35\textwidth}
    \includegraphics[width=\textwidth]{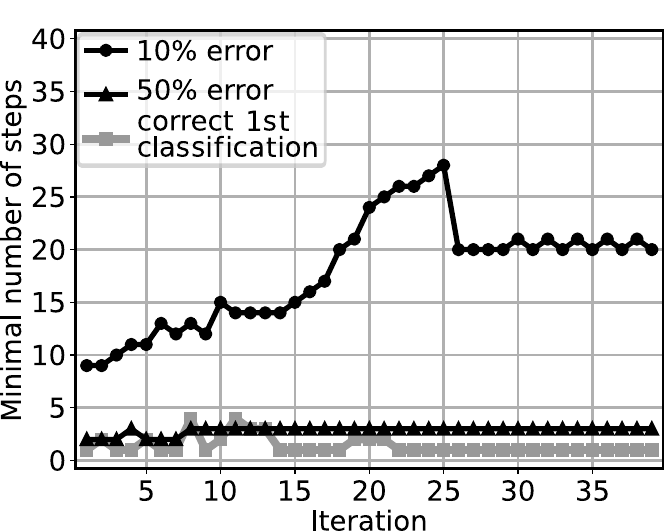}
    \caption{Third case with preconditioning.}
  \end{subfigure}
\caption{Number of CGM steps to achieve different criteria.}
\label{fig:cant_cgm_steps}
\end{figure}

For the second and third cases, with and without Jacobi preconditioning, Fig.~\ref{fig:cant_cgm_steps} presents the minimal number of steps to achieve different criteria for the topology of each iteration: the number of steps so that the relative $l^2$-error of the sensitivity vector is below $10\%$; the number of steps so that the relative $l^2$-error is below $50\%$; and the number of steps so that the solid element with the lowest sensitivity absolute value is correctly classified. Only solid elements were considered in this analysis.

% [ Figure 19 -- original position ]

It can be seen that the preconditioning consistently reduced the number of steps needed to achieve the criteria in both cases. For a small number of steps, the performance was better for the second case. This was expected since its initial guess is closer to the solution. For a large number of steps, the performance was better for the third case. This result indicates that there may be better initializations than the most intuitive one, given in the second case, which starts at the last equilibrium point and moves in the direction of steepest descent of the preconditioned problem.

In the second case without preconditioning, when the error was below $50\%$, at least $1.3\%$ of the solid elements were correctly classified; when the error is below $10\%$, at least $30.8\%$ of them were correctly classified. With Jacobi preconditioning those rates of correct classification were $4.4\%$ and $32.1\%$. For the third case, the results were similar for $50\%$ of error, but the error below $10\%$ resulted only to minimal rates of correct classification around $22\%$.

% [ Figure 18 -- repositioned so it appears in the right place ]

The relative $l^2$-error is a good measure to evaluate if the overall behavior of the sensitivity vector is being well predicted. However, it should be noted that, even if such an error is small, elements with similar sensitivity values may still be incorrectly classified, when compared to the exact sensitivity classification.

The results for the first case were substantially worse. For the case with preconditioning, the smallest number of steps to achieve one of the criteria was $82$. The poor performance for the first case is because it does not take advantage of the knowledge that was already obtained about the structural behavior, stored in $\mathbf{\bar{u}}$.

\begin{figure}[h!]
\centering
\begin{subfigure}{0.45\textwidth}
\includegraphics[width=\textwidth]{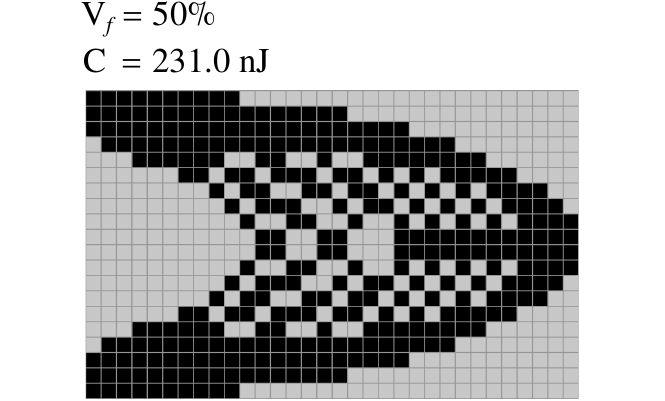}
\caption{Without sensitivity filter.}
\end{subfigure}
\begin{subfigure}{0.45\textwidth}
\includegraphics[width=\textwidth]{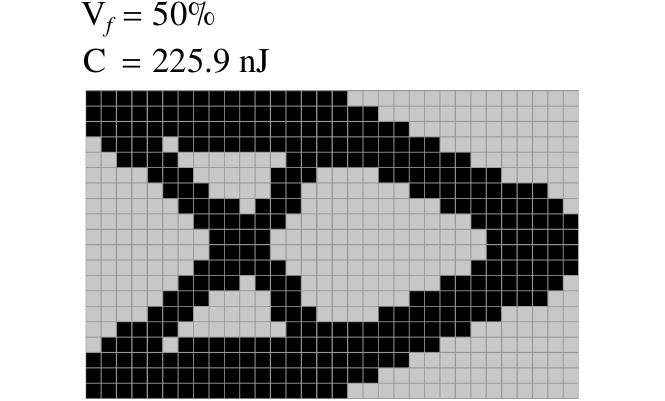}
\caption{With sensitivity filter.}
\end{subfigure}
\caption{Optimized cantilever beams.}
\label{fig:cant_filtered}
\end{figure}

In this section, a topology optimization without filtering was considered and non-filtered sensitivity values were compared. Evidently, to obtain proper optimized structures, without the checkerboard problem, the filtering procedure is essential. When a filter with radius of $3\,\text{mm}$ was included to the optimization algorithm, the topology shown in Fig.~\ref{fig:cant_filtered} was obtained.

% [ Figure 19 -- repositioned so it appears in the right place ]
\vfill \phantom{.}

\subsection{MBB Beam}

\begin{figure}[h!]
\centering
\includegraphics[width=0.45\textwidth]{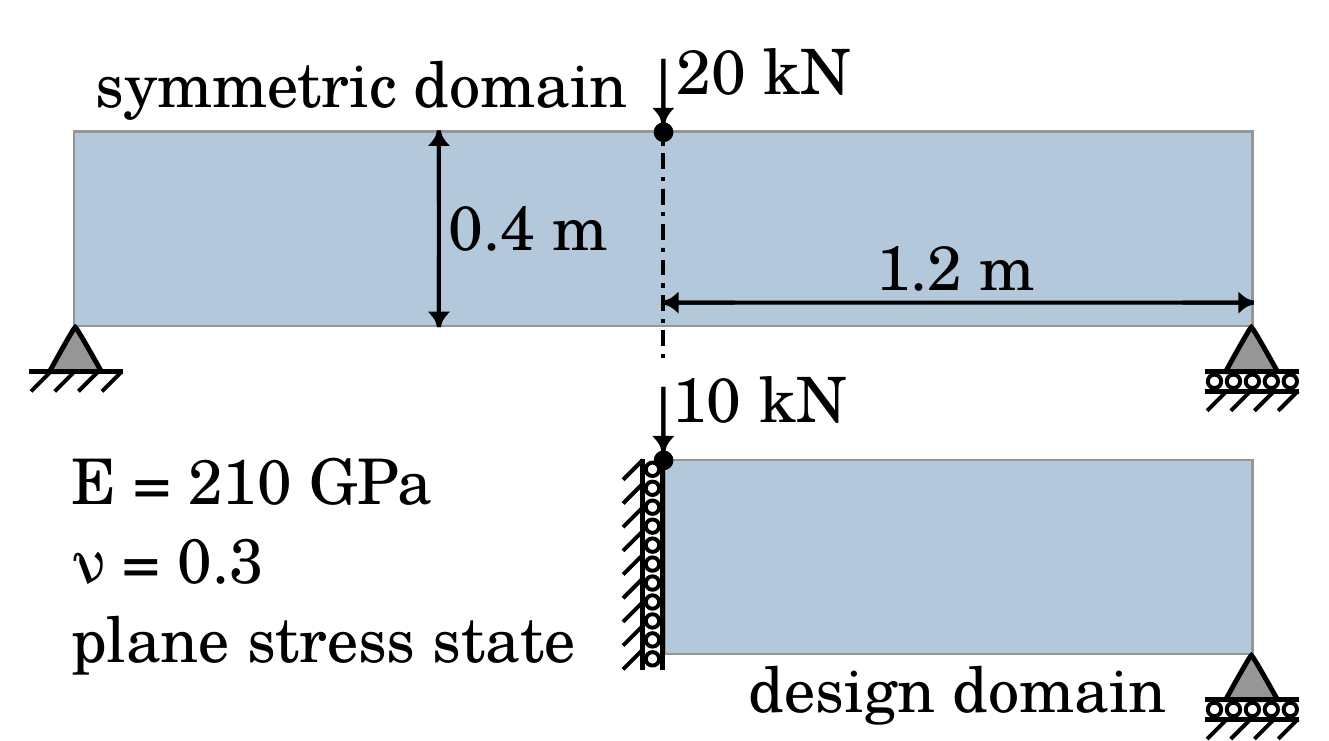}
\caption{MBB beam.}
\label{fig:mbb_beam}
\end{figure}

Fig.~\ref{fig:mbb_beam} presents the design domain considered for the optimization of a MBB beam. A mesh with $300 \times 100$ elements of dimensions $4 \times 4\,\text{mm}$ was considered. A sensitivity filter with radius of $40\,mm$ was used, as well as the presented momentum strategy.

The optimization was performed for six different sensitivity analyses: FOCI with $\varepsilon_v = 10^{-6}$ (simply denoted by FOCI); FOCI with $\varepsilon_v = 0$ (denoted by FOCI-s); CGM with $1$ step, without preconditioning, for void and solid elements (denoted by CGM-$1$); CGM with $1$ step, without preconditioning, for solid elements and void sensitivity values assigned as $0$ (denoted by CGM-$1$s); CGM with $2$ steps, with Jacobi preconditioning, for void and solid elements (denoted by CGM-$2$J); and CGM with $2$ steps, with Jacobi preconditioning, for solid elements and void sensitivity values assigned as $0$ (denoted by CGM-$2$Js). The considered initial condition for the CGM was $\mathbf{u_0} = \mathbf{\bar{u}}$ and $\mathbf{d_0} = \pm\mathbf{M}^{-1}\,\mathbf{K_i}\,\mathbf{\bar{u}}$.

Firstly, a fully solid initial topology was used, so $V_f(\mathbf{\bar{x}}^{(0)}) = 100\%$. For a target volume fraction of $V_f^*=50\%$, four sets of constraints were considered:

\begin{itemize}
 \item $V\!V^{(j)} = -300$ ($1\%$) and $T\!V_{\max}^{(j)} = 1500$ ($5\%$) until the target volume fraction is achieved, then $V\!V^{(j)} = 0$ ($0\%$) and $T\!V_{\max}^{(j)} = 1200$ ($4\%$), which corresponds to the BESO method with $E\!R = 1\%$ and $A\!R_{\max}=2\%$;
 \item no constraint over $T\!V(\mathbf{y})$, $V\!V^{(j)} = -300$ ($1\%$) until the target volume fraction is achieved, then $V\!V^{(j)} = 0$ ($0\%$), which corresponds to the BESO method with $E\!R=1\%$ and $A\!R_{\max}=100\%$;
 \item $V\!V^{(j)} = -3000$ ($10\%$) and $T\!V_{\max}^{(j)} = 4200$ ($14\%$) until the target volume fraction is achieved, then $V\!V^{(j)} = 0$ ($0\%$) and $T\!V_{\max}^{(j)} = 1200$ ($4\%$), which corresponds to the BESO method with $E\!R=10\%$ and $A\!R_{\max}=2\%$;
 \item no constraint over $T\!V(\mathbf{y})$, $V\!V^{(j)} = -3000$ ($10\%$) until the target volume fraction is achieved, then $V\!V^{(j)} = 0$ ($0\%$), which corresponds to the BESO method with $E\!R=10\%$ and $A\!R_{\max}=100\%$.
\end{itemize}

Table~\ref{tab:mbb_100} presents, for each case, the minimized compliance and the number of iterations needed to achieve it.

\begin{table*}[h!]
\caption{Results for different MBB optimizations with $V_f(\mathbf{\bar{x}}^{(0)})=100\%$ and $V_f^*=50\%$.}
\centering
\begin{tabular}{p{0.20\textwidth} p{0.15\textwidth} p{0.15\textwidth} p{0.20\textwidth} p{0.20\textwidth}}
\toprule
Sensitivity Analysis & $\phantom{}V\!V^{(j)}$ & $T\!V_{\max}^{(j)}$ & Number of Iterations & Compliance [$mJ$] \\
\midrule
FOCI & $-300\phantom{0} \;|\; 0$ & $1500\phantom{0} \;|\; 1200$ & $\mathbf{131} \;\; (52)$ & $\mathbf{45.74} \;\; (45.12)$\\
FOCI & $-300\phantom{0} \;|\; 0$ & $30000 \;|\; 30000$ & $\mathbf{147} \;\; (64)$ & $\mathbf{45.71} \;\; (45.10)$\\
FOCI & $-3000 \;|\; 0$ & $4200\phantom{0} \;|\; 1200$ & $\mathbf{189} \;\; (100)$ & $\mathbf{45.62} \;\; (45.54)$\\
FOCI & $-3000 \;|\; 0$ & $30000 \;|\; 30000$ & $\mathbf{285}$ & $\mathbf{45.08}$\\
\addlinespace
FOCI-s & $-300\phantom{0} \;|\; 0$ & $1500\phantom{0} \;|\; 1200$ & $\mathbf{131} \;\; (52)$ & $\mathbf{45.74} \;\; (45.12)$\\
FOCI-s & $-300\phantom{0} \;|\; 0$ & $30000 \;|\; 30000$ & $\mathbf{131} \;\; (52)$ & $\mathbf{45.74} \;\; (45.11)$\\
FOCI-s & $-3000 \;|\; 0$ & $4200\phantom{0} \;|\; 1200$ & $\mathbf{186}$ & $\mathbf{45.59}$\\
FOCI-s & $-3000 \;|\; 0$ & $30000 \;|\; 30000$ & $\mathbf{294}$ & $\mathbf{45.08}$\\
\addlinespace
CGM-$1$ & $-300\phantom{0} \;|\; 0$ & $1500\phantom{0} \;|\; 1200$ & ----- & \textbf{degenerated}\\
CGM-$1$ & $-300\phantom{0} \;|\; 0$ & $30000 \;|\; 30000$ & ----- & \textbf{degenerated}\\
CGM-$1$ & $-3000 \;|\; 0$ & $4200\phantom{0} \;|\; 1200$ & ----- & \textbf{degenerated}\\
CGM-$1$ & $-3000 \;|\; 0$ & $30000 \;|\; 30000$ & ----- & \textbf{degenerated}\\
\addlinespace
CGM-$1$s & $-300\phantom{0} \;|\; 0$ & $1500\phantom{0} \;|\; 1200$ & $\mathbf{152} \;\; (53)$ & $\mathbf{45.65} \;\; (45.29)$\\
CGM-$1$s & $-300\phantom{0} \;|\; 0$ & $30000 \;|\; 30000$ & $\mathbf{164} \;\; (53)$ & $\mathbf{45.61} \;\; (45.29)$\\
CGM-$1$s & $-3000 \;|\; 0$ & $4200\phantom{0} \;|\; 1200$ & $\textbf{243}$ & $\textbf{44.94}$\\
CGM-$1$s & $-3000 \;|\; 0$ & $30000 \;|\; 30000$ & $\mathbf{291}$ & $\mathbf{44.92}$\\
\addlinespace
CGM-$2$J & $-300\phantom{0} \;|\; 0$ & $1500\phantom{0} \;|\; 1200$ & $\mathbf{163}$ & $\mathbf{44.78}$\\
CGM-$2$J & $-300\phantom{0} \;|\; 0$ & $30000 \;|\; 30000$ & $\mathbf{142}$ & $\mathbf{44.78}$\\
CGM-$2$J & $-3000 \;|\; 0$ & $4200\phantom{0} \;|\; 1200$ & $\mathbf{109}$ & $\mathbf{45.32}$\\
CGM-$2$J & $-3000 \;|\; 0$ & $30000 \;|\; 30000$ & ----- & \textbf{degenerated} \\
\addlinespace
CGM-$2$Js & $-300\phantom{0} \;|\; 0$ & $1500\phantom{0} \;|\; 1200$ & $\mathbf{122}$ & $\mathbf{44.87}$\\
CGM-$2$Js & $-300\phantom{0} \;|\; 0$ & $30000 \;|\; 30000$ & $\mathbf{131}$ & $\mathbf{44.87}$\\
CGM-$2$Js & $-3000 \;|\; 0$ & $4200\phantom{0} \;|\; 1200$ & $\textbf{168}$ & $\textbf{45.76}$\\
CGM-$2$Js & $-3000 \;|\; 0$ & $30000 \;|\; 30000$ & $\mathbf{284}$ & $\mathbf{44.88}$\\
\bottomrule
\end{tabular}
\label{tab:mbb_100}
\end{table*}

In some cases, an early convergence occurred, resulting in a structure with thin components. These early results were presented within parentheses. In such cases, the actual results (without thin components) were obtained after performing a new optimization process starting from the solutions with thin components. The structural compliances of the whole MBB beams are twice the presented values, since these were computed for the design domain (half of the symmetric structure). The notation $\cdot\,|\,\cdot$ was used to express the constraints until the target volume was achieved (number on the left) and the constraints for the iterations at a constant volume fraction (number on the right).

When using CGM-$1$, the process was unstable and resulted in degenerated structures. This was due the inaccurate sensitivity values assigned to disconnected void elements. However, except for the case with $V\!V^{(j)}=-3000|0$ and unconstrained $T\!V(\mathbf{y})$, CGM-$2$J was able to stabilize the process. FOCI was stable because a small value was used to penalize the sensitivity numbers of void elements. As can be seen, FOCI and FOCI-s produced very similar results, so there was no negative effect when the sensitivity values of void elements were simply assigned as $0$.

Counterintuitively, when greater topological variations were performed in each iteration, most cases needed more iterations to converge. This may happen because more unstable procedures are more susceptible to produce oscillations between iterations, delaying the convergence.

\begin{figure}[h!]
\centering
\includegraphics[width=0.45\textwidth]{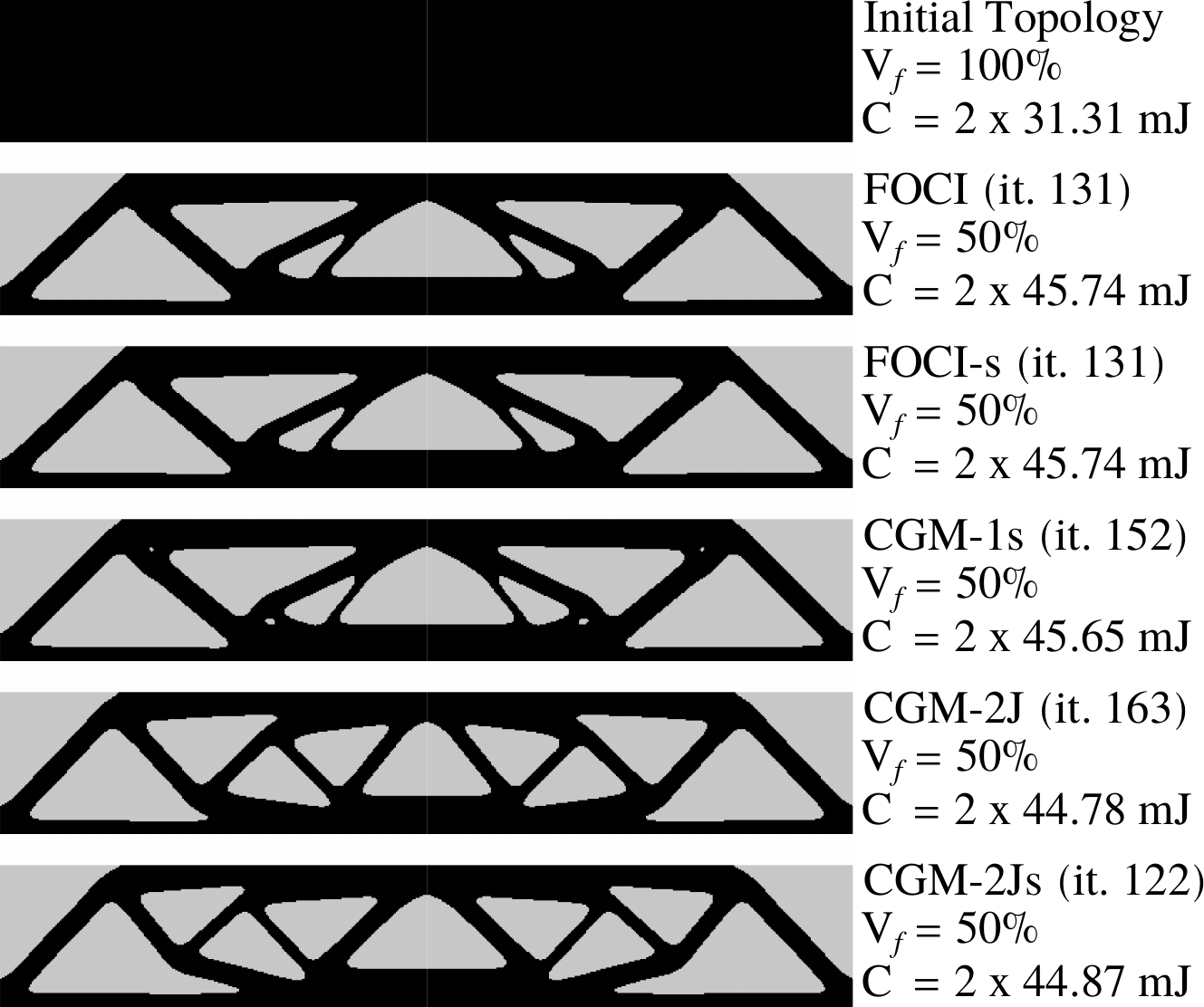}
\caption{Optimized MBB beams for $V\!V^{(j)} = -300|0$ and $T\!V_{\max}^{(j)} = 4200|1200$.}
\label{fig:mbb_beam_results100}
\end{figure}

Fig.~\ref{fig:mbb_beam_results100} presents the results for the most reasonable constraints, $V\!V^{(j)} =-300|0$ and $T\!V_{\max}^{(j)} = 4200|1200$. The whole MBB beams are presented, in meshes with $600 \times 100$ elements. When using CGM-$2$J or CGM-$2$Js, more efficient topologies were obtained.

\begin{figure}[h!]
\centering
\includegraphics[width=0.45\textwidth]{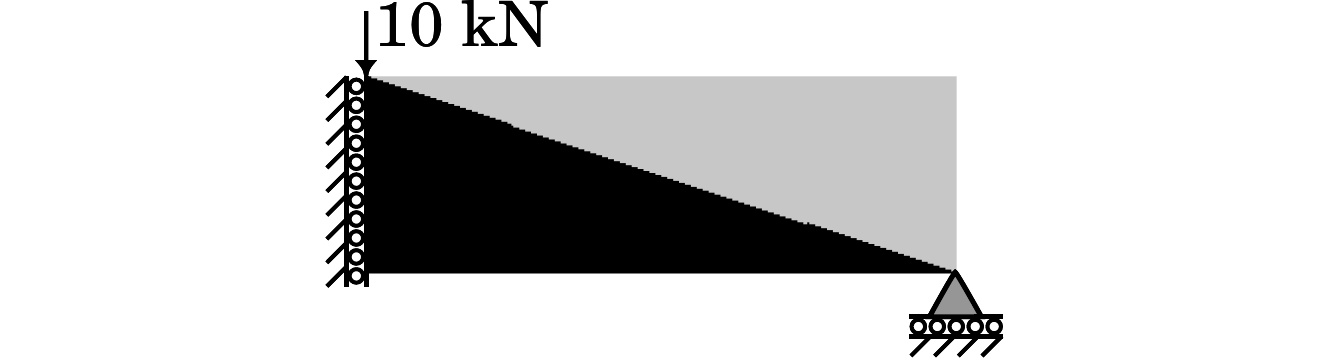}
\caption{Initial Topology for MBB beam with $V_f(\mathbf{\bar{x}}^{(0)})=50\%$.}
\label{fig:mbb_beam_initial50}
\end{figure}

Next, optimizations were performed for a fixed volume fraction of $V_f(\mathbf{\bar{x}}^{(j)})=V_f^*=50\%$, from the initial topology presented in Fig.~\ref{fig:mbb_beam_initial50}. Two sets of constraints were considered:

\begin{itemize}
 \item $V\!V^{(j)} = 0$ ($0\%$) and $T\!V_{\max}^{(j)} = 1200$ ($4\%$), which corresponds to the BESO method with $E\!R=0\%$ and $A\!R_{\max}=2\%$;
 \item no constraint over $T\!V(\mathbf{y})$ and $V\!V^{(j)} = 0$ ($0\%$), which corresponds to the BESO method with $E\!R=0\%$ and $A\!R_{\max}=100\%$.
\end{itemize}

Table~\ref{tab:mbb_50} presents, for each case, the minimized compliance and the number of iterations needed to achieve it. Again, the structural compliances of the whole MBB beams are twice the presented values.

\begin{table*}[h!]
\caption{Results for different MBB optimizations with a fixed $V_f(\mathbf{\bar{x}}^{(j)})=50\%$.}
\centering
\begin{tabular}{p{0.20\textwidth} p{0.15\textwidth} p{0.15\textwidth} p{0.20\textwidth} p{0.20\textwidth}}
\toprule
Sensitivity Analysis & $V\!V^{(j)}$ & $T\!V_{\max}^{(j)}$ & Number of Iterations & Compliance [$mJ$] \\
\midrule
FOCI & $0$ & $1200$ & $\mathbf{89}$ & $\mathbf{45.74}$\\
FOCI & $0$ & $30000$ & $\mathbf{120}$ & $\mathbf{45.65}$\\
\addlinespace
FOCI-s & $0$ & $1200$ & $\mathbf{92}$ & $\mathbf{45.74}$\\
FOCI-s & $0$ & $30000$ & $\mathbf{114}$ & $\mathbf{45.66}$\\
\addlinespace
CGM-$1$ & $0$ & $1200$ & ----- & \textbf{degenerated}\\
CGM-$1$ & $0$ & $30000$ & ----- & \textbf{degenerated}\\
\addlinespace
CGM-$1$s & $0$ & $1200$ & $\mathbf{112}$ & $\mathbf{45.60}$\\
CGM-$1$s & $0$ & $30000$ & $\mathbf{179}$ & $\mathbf{45.36}$\\
\addlinespace
CGM-$2$J & $0$ & $1200$ & $\mathbf{86}$ & $\mathbf{45.84}$\\
CGM-$2$J & $0$ & $30000$ & $\mathbf{67}$ & $\mathbf{45.85}$\\
\addlinespace
CGM-$2$Js & $0$ & $1200$ & $\mathbf{90}$ & $\mathbf{45.66}$\\
CGM-$2$Js & $0$ & $30000$  & $\mathbf{161}$ & $\mathbf{45.34}$\\
\bottomrule
\end{tabular}
\label{tab:mbb_50}
\end{table*}

\begin{figure}[h!]
\centering
\includegraphics[width=0.45\textwidth]{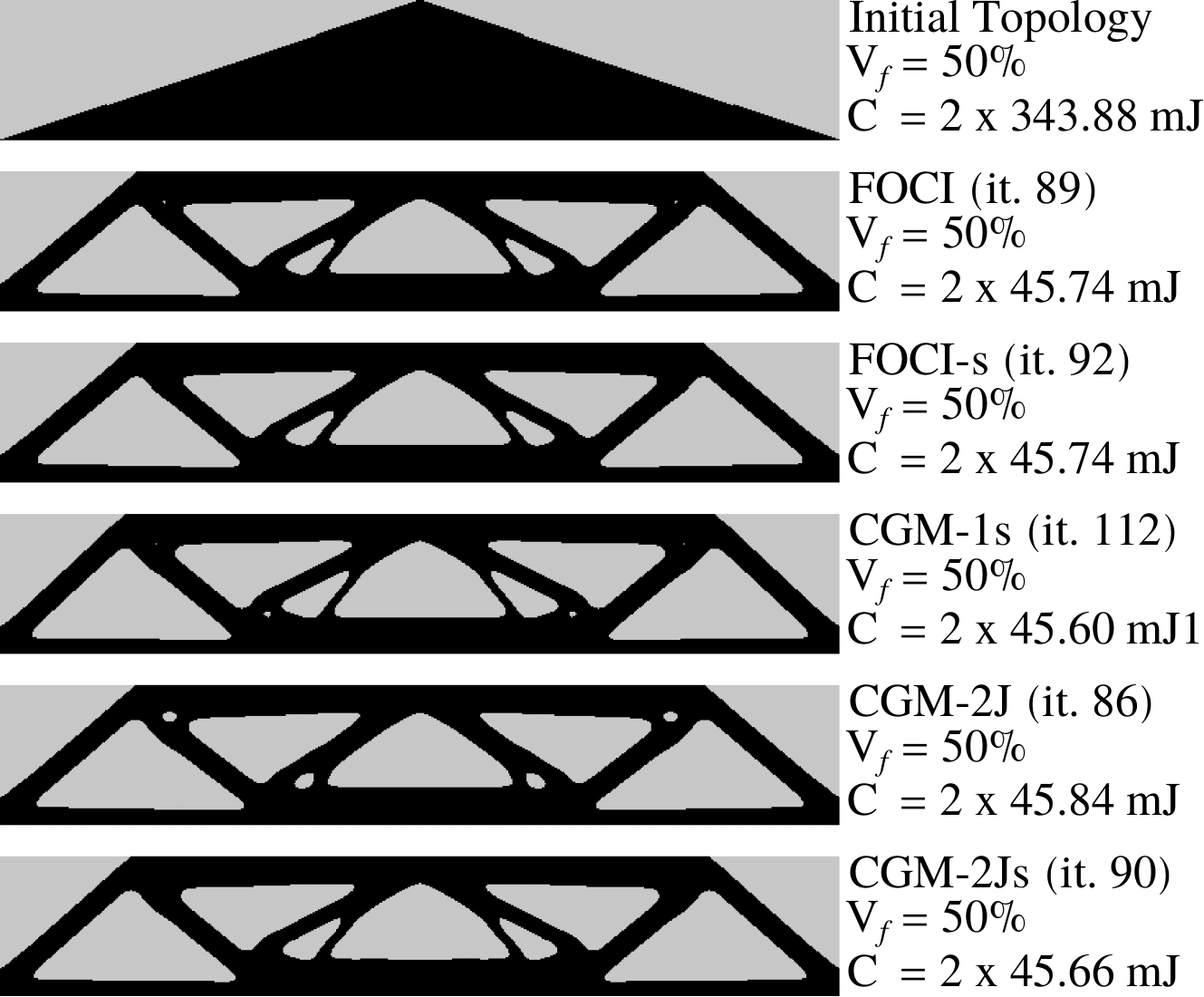}
\caption{Optimized MBB beams for $V\!V^{(j)}=0$ and $T\!V_{\max}^{(j)} = 1200$.}
\label{fig:mbb_beam_results50}
\end{figure}

Fig.~\ref{fig:mbb_beam_results50} presents the results for the most reasonable constraints, $V\!V^{(j)}=0$ and $T\!V_{\max}^{(j)} = 1200$. The whole MBB beams are presented, in meshes with $600 \times 100$ elements. All strategies resulted in similar topologies.

\section{Conclusions}
\label{sec:conclusions}

In this work, a more formal description of the BESO method was presented, in which the sensitivity analysis consists in estimating finite variations of the objective function. In addition to the naive approach, four ways to perform the sensitivity analysis were presented: FOCI, HOCI, Woodbury and CGM approaches. For the problem of structural compliance minimization with volume constraint, these analyses were developed, compared through numerical examples and discussed.

The standard FOCI approach was reformulated to have simpler parameters. Instead of allowing the interpolation function to assume an arbitrary form, it is fixed as a linear function and a parameter $\varepsilon_v$ is used to penalize the sensitivity values of void elements. In the compliance minimization problem, it was shown that $\varepsilon_v$ should assume non-negative values below $1$ to improve the accuracy of the sensitivity analysis.

Next, the convergence conditions for the Taylor series considered in the HOCI approach were discussed. In \ref{ap_eigenvalues_A} and \ref{ap_counterexample}, it was shown that the series is always convergent for solid elements and that it may be divergent for void elements. The Woodbury approach was proposed as an alternative over the HOCI approach, providing a closed-form expression for the exact sensitivity values, for both solid and void elements. Although it is an improvement over the HOCI approach, it may still be a computationally prohibitive strategy, because a selective inverse of the global stiffness matrix would have to be known in each iteration to perform this sensitivity analysis. Nonetheless, if the selective inverse is known at the first iteration, it would not be necessary to compute it again from scratch, since it can be updated after each topological variation instead, as shown in \ref{ap_selective}. For a conclusive viability evaluation, specific algorithms for selective inversion should be properly implemented.

The CGM approach was then derived. In contrast to the previous approach, it is computationally viable, providing an estimated sensitivity vector that is guaranteed to be more accurate than the one obtained through the FOCI approach, when $\mathbf{u_0}~=~\mathbf{\bar{u}}$ and $\mathbf{d_0}~=~\pm\mathbf{M}^{-1}\,\mathbf{K_i}\,\mathbf{\bar{u}}$. Different CGM initial conditions were also explored and explicit formulations can be found in \ref{ap_cgm_explicit}.

With the FVSA approaches formulated, the bounds of the errors for certain approaches were investigated. Upper bounds, that are independent from the applied load, were presented for the element-wise sensitivity error. For void elements, the bounds were presented for the error of the FOCI expression; for solid elements, they were presented for the error of the HOCI expression.

A coarsely meshed cantilever tie-beam problem was then considered. It was shown that the use of accurate sensitivity values may prevent highly non-optimal topological variations. General properties of the optimization method were discussed, such as the need of a minimally refined mesh and the dependence on the initial topology. It was shown that the upper bounds for the sensitivity errors are reduced with mesh refinement. Moreover, for finer meshes, the CGM approach produced substantially more accurate sensitivity values than the FOCI approach.

Next, a coarsely meshed cantilever beam was optimized with exact sensitivity numbers and different CGM sensitivity expressions were compared for each topology of the optimization procedure. The results were better when the information of the displacements vector was used to define the initial conditions of the CGM. Jacobi preconditioning consistently reduced the number of steps needed to achieve a chosen precision.

Lastly, a finely meshed MMB beam was optimized with FOCI and CGM approaches, using different parameters. The best results were obtained by CGM approach, with two steps and Jacobi preconditioning. It was shown that unpenalized sensitivity values for void elements may produce unstable behavior. The results suggest that it is advantageous to perform the sensitivity analysis only for solid elements, and to assign the void sensitivity numbers as $0$ before the filtering procedure.

From the results, it can be seen that the FOCI sensitivity analysis performs well when the mesh is sufficiently refined and reasonable parameters are used for the optimization method. The HOCI and Woodbury analyses provide useful expressions to understand the problem. For appropriate initial conditions, the CGM analysis provides a more accurate expression than the one obtained through FOCI approach, which may result in more stable processes and in more efficient optimized structures.

Although the FOCI approach seems to be enough for the compliance minimization problem, it may not be true when different objective functions and constraints are considered. Therefore, the use of such approach in BESO-type methods must be justified and it must be shown that there are conditions in which it can reasonably estimate the finite variations of the objective function.

To summarize, this paper presented new methods to perform the sensitivity analysis for discrete topology optimization problems. Before defining the sensitivity expression to be used in a new problem, the behavior of the objective function with respect to finite variations of the design variables should be carefully evaluated. Special conditions which may inform beforehand the exact sensitivity values of some elements should be identified, e.g., disconnected or externally loaded elements. Closed-form expressions should be developed for the sensitivity vector. The Taylor series of the relaxed objective function should be constructed and its convergence conditions should be evaluated. CGM estimated sensitivity expressions should be developed for appropriate initial conditions. Moreover, the viability and accuracy of the different approaches should be compared for different mesh refinements.

The presented contributions can also be used to produce improved linearizations for integer LP approaches. Furthermore, the HOCI, Woodbury and CGM approaches can all be used to predict the effects of switching the state of multiple elements simultaneously, which may lead to the development of new discrete optimization methods.

\renewcommand{\thesection}{Appendix A}
\renewcommand{\theequation}{A.\arabic{equation}}
\renewcommand{\thefigure}{A.\arabic{figure}}
\renewcommand{\thetable}{A.\arabic{table}}
\setcounter{equation}{0}
\setcounter{figure}{0}
\setcounter{table}{0}
\section{Upper bound of $\|\mathbf{\overline{A}_i}\|_2$ for solid elements}
\label{ap_eigenvalues_A}

Considering that the $i$th element is solid ($x_i = 1$), $\mathbf{\bar{K}}$ can be separated as

\begin{equation}
 \mathbf{\bar{K}} = \mathbf{\bar{R}_i} + \mathbf{K_i} \,,
 \label{eq:separated_K}
\end{equation}

\noindent
in which $\mathbf{\bar{R}_i}$ is a positive definite matrix. Thus, $\mathbf{\bar{A}_i}$ can be written as

\begin{equation}
 \mathbf{\bar{A}_i} = \sqrt{\mathbf{K_i}} \left[\mathbf{\bar{R}_i} + \mathbf{K_i}\right]^{-1} \sqrt{\mathbf{K_i}} \,.
 \label{eq:separated_A}
\end{equation}

Through Woodbury identity, it becomes

\begin{equation}
\begin{aligned}
\mathbf{\bar{A}_i} = {}& \sqrt{\mathbf{K_i}}\,\Bigg[\mathbf{\bar{R}_i}^{-1} -  \mathbf{\bar{R}_i}^{-1}\,\sqrt{\mathbf{K_i}}\,\Big[\,\mathbf{I} \\& + \sqrt{\mathbf{K_i}}\,\mathbf{\bar{R}_i}^{-1}\,\sqrt{\mathbf{K_i}}\,\Big]^{-1} \sqrt{\mathbf{K_i}}\,\mathbf{\bar{R}_i}^{-1}\Bigg]\,\sqrt{\mathbf{K_i}} \,.
\end{aligned}
 \label{eq:separated_A_woodbury}
\end{equation}

Let $\mathbf{\bar{B}_i}$ be defined by

 \begin{equation}
 \mathbf{\bar{B}_i} = \sqrt{\mathbf{K_i}} \, \mathbf{\bar{R}_i}^{-1} \sqrt{\mathbf{K_i}} \,,
 \label{eq:B_matrix}
\end{equation}

\noindent
since it is symmetric and positive semi-definite, it has an orthogonal eigenvectors matrix $\mathbf{F}$ and a diagonal eigenvalues matrix $\mathbf{L}$, composed by non-negative values. Its diagonalization is given by

\begin{equation}
 \mathbf{\bar{B}_i} = \mathbf{F}\,\mathbf{L}\,\mathbf{F}^T \,.
 \label{eq:diagonalization_R}
\end{equation}

So Eq.~(\ref{eq:separated_A_woodbury}) can be rewritten as

\begin{equation}
\begin{aligned}
\mathbf{\bar{A}_i} {}& = \mathbf{F} \left[ \mathbf{L} - \mathbf{L}\left[\mathbf{I}+\mathbf{L}\right]^{-1}\mathbf{L}\right] \mathbf{F}^T \\& = \mathbf{F} \left[\mathbf{L}\left[\mathbf{I} - \left[\mathbf{I}+\mathbf{L}\right]^{-1}\mathbf{L}\right]\right] \mathbf{F}^T \,.
\end{aligned}
 \label{eq:simplified_A}
\end{equation}

By simplifying it, $\mathbf{\bar{A}_i}$ is obtained as

\begin{equation}
 \mathbf{\bar{A}_i} = \mathbf{F} \left[\mathbf{L}\left[\mathbf{I}+\mathbf{L}\right]^{-1}\right] \mathbf{F}^T \,,
 \label{eq:super_simplified_A}
\end{equation}

\noindent
which means that $\mathbf{F}$ is also an eigenvectors matrix of $\mathbf{\bar{A}_i}$ and $\mathbf{L}\left[\mathbf{I}+\mathbf{L}\right]^{-1}$ is the corresponding eigenvalues matrix. Therefore, each eigenvalue $\lambda_k$ of $\mathbf{\bar{A}_i}$ can be written with respect to its corresponding term of $\mathbf{L}$:

\begin{equation}
 \lambda_k = \frac{L_k}{1 + L_k} \in [0,1[ \,.
 \label{eq:eigenvalues_A}
\end{equation}

This means that, when $x_i=1$, all eigenvalues of $\mathbf{\bar{A}_i}$ must be non-negative and strictly less than $1$. Thus,

\begin{equation}
 \boxed{x_i = 1 \Rightarrow \|\mathbf{\bar{A}_i}\|_2  < 1 \,.}
 \label{eq:upper_A_norm}
\end{equation}

\renewcommand{\thesection}{Appendix B}
\renewcommand{\theequation}{B.\arabic{equation}}
\renewcommand{\thefigure}{B.\arabic{figure}}
\renewcommand{\thetable}{B.\arabic{table}}
\setcounter{equation}{0}
\setcounter{figure}{0}
\setcounter{table}{0}
\section{Uncertainty about $\|\mathbf{\overline{A}_i}\|_2$ for void elements}
\label{ap_counterexample}

Considering that the $i$th element is void ($x_i = 0$), $\mathbf{\bar{K}}$ can be separated as

\begin{equation}
 \mathbf{\bar{K}} = \mathbf{\bar{R}_i} - \mathbf{K_i} \,,
 \label{eq:separated_K2}
\end{equation}

\noindent
in which $\mathbf{\bar{R}_i}$ is a positive definite matrix. Thus, $\mathbf{\bar{A}_i}$ can be written as

\begin{equation}
 \mathbf{\bar{A}_i} = \sqrt{\mathbf{K_i}} \left[\mathbf{\bar{R}_i} - \mathbf{K_i}\right]^{-1} \sqrt{\mathbf{K_i}} \,.
 \label{eq:separated_A2}
\end{equation}

Through Woodbury identity, it becomes

\begin{equation}
\begin{aligned}
\mathbf{\bar{A}_i} = {}& \sqrt{\mathbf{K_i}}\,\Bigg[\mathbf{\bar{R}_i}^{-1} + \mathbf{\bar{R}_i}^{-1}\,\sqrt{\mathbf{K_i}} \,\Big[\,\mathbf{I} \\& - \sqrt{\mathbf{K_i}}\,\mathbf{\bar{R}_i}^{-1}\,\sqrt{\mathbf{K_i}}\,\Big]^{-1} \sqrt{\mathbf{K_i}}\,\mathbf{\bar{R}_i}^{-1}\Bigg]\, \sqrt{\mathbf{K_i}} \,.
\end{aligned}
 \label{eq:separated_A_woodbury2}
\end{equation}

Let $\mathbf{\bar{B}_i}$ be defined by

 \begin{equation}
 \mathbf{\bar{B}_i} = \sqrt{\mathbf{K_i}} \, \mathbf{\bar{R}_i}^{-1} \sqrt{\mathbf{K_i}} \,,
 \label{eq:B_matrix2}
\end{equation}

\noindent
since it is symmetric and positive semi-definite, it has an orthogonal eigenvectors matrix $\mathbf{F}$ and a diagonal eigenvalues matrix $\mathbf{L}$, composed by non-negative values. Its diagonalization is given by

\begin{equation}
 \mathbf{\bar{B}_i} = \mathbf{F}\,\mathbf{L}\,\mathbf{F}^T \,.
 \label{eq:diagonalization_R2}
\end{equation}

So Eq.~(\ref{eq:separated_A_woodbury2}) can be rewritten as

\begin{equation}
\begin{aligned}
\mathbf{\bar{A}_i} {}& = \mathbf{F} \left[ \mathbf{L} + \mathbf{L}\left[\mathbf{I}-\mathbf{L}\right]^{-1}\mathbf{L}\right] \mathbf{F}^T \\& = \mathbf{F} \left[\mathbf{L}\left[\mathbf{I} + \left[\mathbf{I}-\mathbf{L}\right]^{-1}\mathbf{L}\right]\right] \mathbf{F}^T \,.
\end{aligned}
 \label{eq:simplified_A2}
\end{equation}

By simplifying it, $\mathbf{\bar{A}_i}$ is obtained as

\begin{equation}
 \mathbf{\bar{A}_i} = \mathbf{F} \left[\mathbf{L}\left[\mathbf{I}-\mathbf{L}\right]^{-1}\right] \mathbf{F}^T \,,
 \label{eq:super_simplified_A2}
\end{equation}

\noindent
which means that $\mathbf{F}$ is also an eigenvectors matrix of $\mathbf{\bar{A}_i}$ and $\mathbf{L}\left[\mathbf{I}-\mathbf{L}\right]^{-1}$ is the corresponding eigenvalues matrix. Therefore, each eigenvalue $\lambda_k$ of $\mathbf{\bar{A}_i}$ can be written with respect to its corresponding term of $\mathbf{L}$:

\begin{equation}
 \lambda_k = \frac{L_k}{1 - L_k} \,.
 \label{eq:eigenvalues_A2}
\end{equation}

Since $\mathbf{\bar{A}_i}$ is a semi-positive matrix with finite eigenvalues, Eq.~(\ref{eq:eigenvalues_A2}) establishes an upper bound for $L_k$: $L_k \in [0,1[$. As for $\lambda_k$, it will only be lesser than $1$ when $L_k$ is lesser than $1/2$:

\begin{equation}
 \begin{aligned}
    &L_k \in \left[0\,,\frac{1}{2}\right[ \;\Rightarrow \; \lambda_k \in [0,1[ \;;\\
    &L_k \in \left[\frac{1}{2}\,, 1\right[ \;\Rightarrow \; \lambda_k \in [1,\infty[ \;.
 \end{aligned}
\label{eq:bounds}
\end{equation}

Thus,

\begin{equation}
 \boxed{
 \begin{aligned}
  & x_i = 0 \text{ and } \|\mathbf{\bar{B}_i}\|_2 < \frac{1}{2} \Rightarrow \|\mathbf{\bar{A}_i}\|_2<1 \,,\\
  & x_i = 0 \text{ and } \|\mathbf{\bar{B}_i}\|_2 \geq \frac{1}{2} \Rightarrow \|\mathbf{\bar{A}_i}\|_2 \geq 1 \,.
 \end{aligned}}
 \label{eq:bounds_A_norm}
\end{equation}

In the counterexample below, it is shown that $\|\mathbf{\bar{A}_i}\|_2$ may indeed assume values greater than $1$ when $x_i=0$.

The value of $\|\mathbf{\bar{A}_i}\|_2$ highly depends on the current topology and on the soft-kill parameter $\varepsilon_k$. The presence of nodes connected only to void elements, especially for a small $\varepsilon_k$, favors higher values of $\|\mathbf{\bar{A}_i}\|_2$.

\begin{figure}[h!]
\centering
\includegraphics[width=0.45\textwidth]{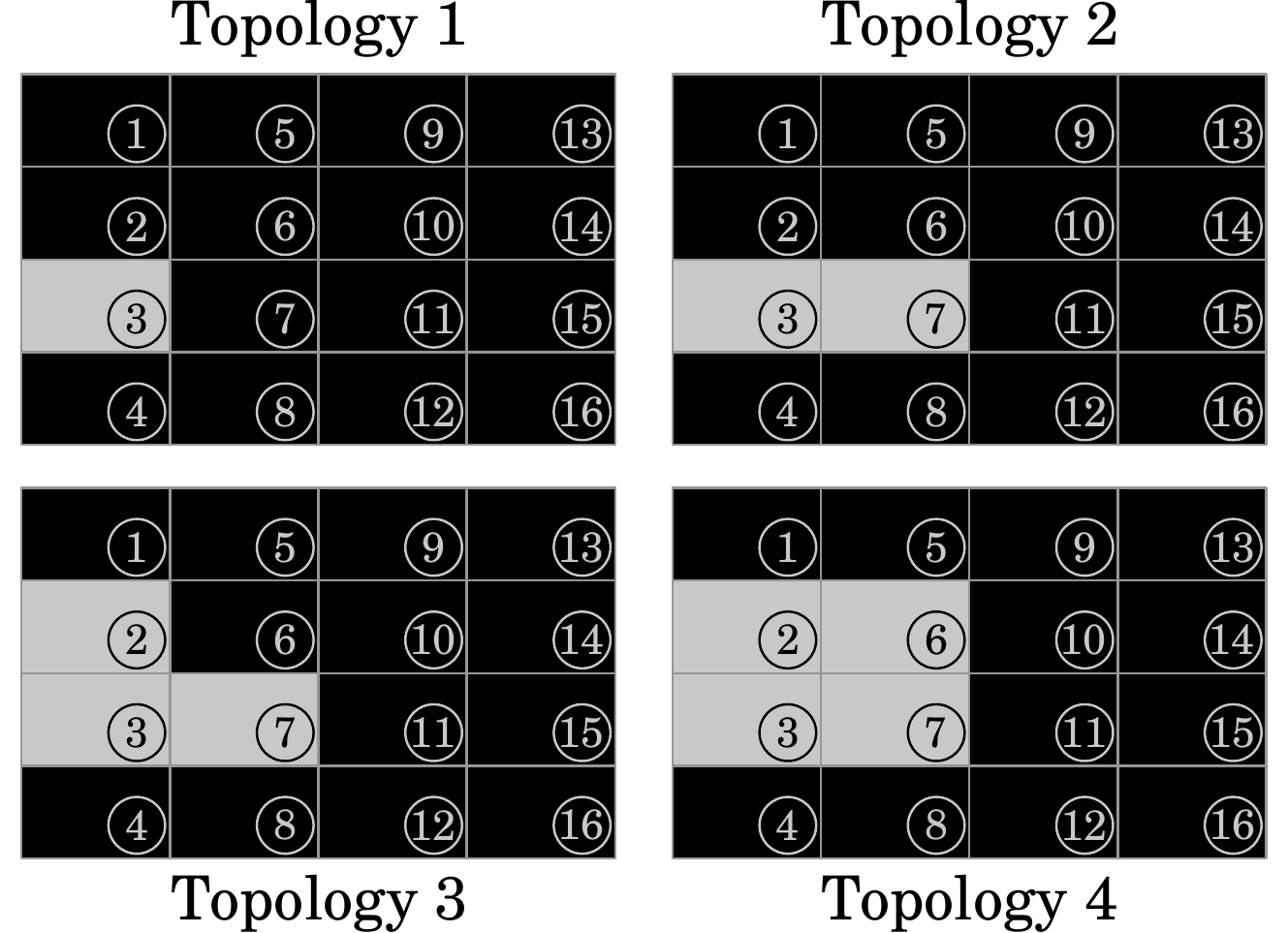}
\caption{Different topologies for a $4 \times 4$ mesh.}
\label{fig:divergent_series}
\end{figure}

In Fig.~\ref{fig:divergent_series}, four topologies are presented for a cantilever beam of dimensions $80 \times 50\,\text{mm}$, clamped on its left end. Solid elements are represented in black and void elements are represented in gray, they are numbered from top to bottom, starting at the leftmost column. Four-nodes bilinear quadrilateral elements of dimensions $20.0 \times 12.5\,\text{mm}$ were considered, in plane stress state. The material is homogeneous and isotropic, with a Young's modulus of $210\,\text{GPa}$ and a Poisson's ratio of $0.3$.

For $\varepsilon_k = 0.1$, Table~\ref{tab:bounds_AB} presents the values of $\|\mathbf{\bar{A}_i}\|_2$ and $\|\mathbf{\bar{B}_i}\|_2$ for each void element in these topologies. It can be seen that Eq.~(\ref{eq:eigenvalues_A2}) is satisfied and that $\|\mathbf{\bar{A}_i}\|_2$ is less than $1$ when $\|\mathbf{\bar{B}_i}\|_2$ is less than $1/2$.

\begin{table*}[h!]
\caption{Values of $\|\mathbf{\bar{A}_i}\|_2$ and $\|\mathbf{\bar{B}_i}\|_2$ for void elements.}
\begin{tabular}{p{0.15\textwidth} p{0.08\textwidth} p{0.08\textwidth} p{0.08\textwidth} p{0.08\textwidth} p{0.08\textwidth} p{0.08\textwidth} p{0.08\textwidth} p{0.08\textwidth}}
\toprule 
\textbf{} & $\|\mathbf{\bar{A}_3}\|_2$ & $\|\mathbf{\bar{A}_7}\|_2$ & $\|\mathbf{\bar{A}_2}\|_2$ & $\|\mathbf{\bar{A}_6}\|_2$ & $\|\mathbf{\bar{B}_3}\|_2$ & $\|\mathbf{\bar{B}_7}\|_2$ & $\|\mathbf{\bar{B}_2}\|_2$ & $\|\mathbf{\bar{B}_6}\|_2$ \\
\midrule
\textbf{Topology 1} & $0.900$ & ----- & ----- & ----- & $0.474$ & ----- & ----- & ----- \\
\textbf{Topology 2} & $1.794$ & $2.007$ & ----- & ----- & $0.642$ & $0.667$ & ----- & ----- \\
\textbf{Topology 3} & $2.046$ & $2.598$ & $1.663$ & ----- & $0.672$ & $0.722$ & $0.624$ & ----- \\
\textbf{Topology 4} & $3.427$ & $3.632$ & $3.427$ & $3.632$ & $0.774$ & $0.784$ & $0.774$ & $0.784$ \\
\bottomrule
\end{tabular}
\label{tab:bounds_AB}
\end{table*}

Even for such unusually high $\varepsilon_k$, except for the void element in Topology 1, all values of $\|\mathbf{\bar{A}_i}\|_2$ were greater than $1$. This suggests that, in a practical case, $\|\mathbf{\bar{A}_i}\|_2$ will hardly be less than $1$ for any void element.

\renewcommand{\thesection}{Appendix C}
\renewcommand{\theequation}{C.\arabic{equation}}
\renewcommand{\thefigure}{C.\arabic{figure}}
\renewcommand{\thetable}{C.\arabic{table}}
\setcounter{equation}{0}
\setcounter{figure}{0}
\setcounter{table}{0}
\section{Selective inverse update}
\label{ap_selective}

Considering that the selective inverse of $\mathbf{\bar{K}}$ is known, it is desired to obtain its new value after a variation $\mathbf{\Delta K}$. The matrix $\mathbf{I_\Delta}$ matches the identity matrix for entries corresponding to valued terms of $\mathbf{\Delta K}$, and it is zero-valued elsewhere. The updated complete inverse can be written through Woodbury identity as

\begin{equation}
\begin{aligned}
\left[\mathbf{\bar{K}} + \mathbf{\Delta K}\right]^{-1} = {}& \mathbf{\bar{K}}^{-1} - \mathbf{\bar{K}}^{-1}\,\mathbf{I_\Delta}\,\mathbf{\Delta K} \Big[\,\mathbf{I} \\& + \mathbf{I_\Delta}\,\mathbf{\bar{K}}^{-1}\,\mathbf{I_\Delta}\,\mathbf{\Delta K}\Big]^{-1} \mathbf{I_\Delta}\,\mathbf{\bar{K}}^{-1} \,.
\end{aligned}
\label{eq:woodbury_selective}
\end{equation}

By defining $\mathbf{Q}$ and $\mathbf{T}$ as

\begin{equation}
 \mathbf{Q} = \mathbf{\Delta K} \left[\mathbf{I} + \mathbf{I_\Delta}\,\mathbf{\bar{K}}^{-1}\,\mathbf{I_\Delta}\,\mathbf{\Delta K}\right]^{-1} \mathbf{I_\Delta}
 \label{eq:woodbury_Q}
\end{equation}

\noindent
and

\begin{equation}
 \mathbf{T} = \mathbf{I_\Delta}\,\mathbf{\bar{K}}^{-1} \,,
 \label{eq:woodbury_T}
\end{equation}

\noindent
the updated complete inverse can be written as shown below:

\begin{equation}
 \left[\mathbf{\bar{K}} + \mathbf{\Delta K}\right]^{-1} = \mathbf{\bar{K}}^{-1} -\mathbf{T}^T\,\mathbf{Q}\,\mathbf{T} \,.
 \label{eq:woodbury_selective_var}
\end{equation}

Let $g_{\Delta}$ be the dimension of the valued part of $\mathbf{\Delta K}$; and let $G$ be the dimension of the whole matrix. After each alteration, $g_{\Delta}$ linear systems of dimension $G$ must be solved to compute $\mathbf{T}$, and a matrix of dimension $g_{\Delta}$ must be inverted to compute $\mathbf{Q}$.

All columns $\mathbf{t_i}$ of $\mathbf{T}$ have only $g_{\Delta}$ valued terms. The dimension of the valued submatrix of $\mathbf{Q}$ is also $g_{\Delta}$. Therefore, after obtaining $\mathbf{T}$ and $\mathbf{Q}$, each updated entry, given by

\begin{equation}
 \boxed{\left[\mathbf{\bar{K}} + \mathbf{\Delta K}\right]^{-1}_{i_1 i_2}  = \mathbf{\bar{K}}^{-1}_{i_1 i_2} -\mathbf{t_{i_1}}^T\,\mathbf{Q}\,\mathbf{t_{i_2}} \,,}
 \label{eq:woodbury_quadratic}
\end{equation}

\noindent
can be obtained by computing a bilinear expression of dimension $g_{\Delta}$.

Even for a small topological variation, which means a small $g_{\Delta}$, this update procedure would still be costly, since it involves solving linear systems of dimension $G$. Nevertheless, it should be advantageous over obtaining the selective inverse from scratch in each iteration of the topology optimization algorithm.

\renewcommand{\thesection}{Appendix D}
\renewcommand{\theequation}{D.\arabic{equation}}
\renewcommand{\thefigure}{D.\arabic{figure}}
\renewcommand{\thetable}{D.\arabic{table}}
\setcounter{equation}{0}
\setcounter{figure}{0}
\setcounter{table}{0}
\section{Explicit expressions for the CGM approach}
\label{ap_cgm_explicit}

\subsection{Definitions and notation}

For the $i$th element, let the matrix $\mathbf{\bar{\bar{K}}}$ be defined as

\begin{equation}
 \mathbf{\bar{\bar{K}}} = \mathbf{\bar{K}} + \mathbf{\Delta K} \,,
 \label{eq:Kbarbar}
\end{equation}

\noindent
where $\mathbf{\Delta K}$ is given by

\begin{equation}
 \mathbf{\Delta K} = \left\{\begin{aligned}
 &\phantom{.-}\mathbf{K_i} \;\;,\; \text{if } x_i = 0\,,\\
 &-\mathbf{K_i} \;\;,\; \text{if } x_i = 1\,.
 \end{aligned}\right.
 \label{eq:deltaK}
\end{equation}

Let $\bar{C}$, $C_i$, $C_{\Delta}$ and $C_T$ be the scalars defined as

\begin{equation}
 \bar{C} = \frac{1}{2}\,\mathbf{\bar{u}}^T \, \mathbf{f} \,,
 \label{eq:cgm_compliance}
\end{equation}

\begin{equation}
 C_i = \frac{1}{2}\,\mathbf{\bar{u}}^T \, \mathbf{K_i} \, \mathbf{\bar{u}} \,,
 \label{eq:cgm_compliance_i}
\end{equation}

\begin{equation}
 C_{\Delta} = \frac{1}{2}\,\mathbf{\bar{u}}^T \, \mathbf{\Delta K} \, \mathbf{\bar{u}} \,,
 \label{eq:cgm_compliance_delta}
\end{equation}

\noindent
and

\begin{equation}
 C_{T} = \bar{C} + C_{\Delta} \,.
 \label{eq:cgm_compliance_t}
\end{equation}

For a given preconditioner matrix $\mathbf{M}$, let the matrices $\mathbf{W_j}$ be defined as

\begin{equation}
\mathbf{W_j} = \mathbf{M}^{-1} \left[\mathbf{\bar{\bar{K}}}\,\mathbf{M}^{-1}\right]^{j} \,.
\label{eq:cgm_Wj}
\end{equation}

The following notation will be used in this appendix to represent inner products in a more compact way:

\begin{equation}
 \left\langle \mathbf{a_1} \,,\, \mathbf{a_2} \right\rangle_{j} = \left\langle \mathbf{a_1} \,,\, \mathbf{a_2} \right\rangle_{\mathbf{W_j}} = \mathbf{a_1}^T\,\mathbf{W_j}\,\mathbf{a_2} \,;
 \label{eq:notation1}
\end{equation}

\begin{equation}
 \left\langle \mathbf{a} \right\rangle_{j} = \left\langle \mathbf{a} \right\rangle_{\mathbf{W_j}} = \|\mathbf{a}\|_{\mathbf{W_j}}^2 = \mathbf{a}^T\,\mathbf{W_j}\,\mathbf{a} \,.
 \label{eq:notation2}
\end{equation}

Three initial conditions were considered: in the first case, $\mathbf{u_0} = \mathbf{0}$ and $\mathbf{d_0} = \mathbf{M}^{-1}\,\mathbf{f}$ (direction of steepest descent); in the second case, $\mathbf{u_0} = \mathbf{\bar{u}}$ and $\mathbf{d_0} = -\mathbf{M}^{-1}\,\mathbf{\Delta K}\,\mathbf{\bar{u}}$ (direction of steepest descent); and in the third case, $\mathbf{u_0} = \mathbf{0}$ and $\mathbf{d_0} = \mathbf{\bar{u}}$.

\subsection{Explicit expressions for the 1st case}

Considering $\mathbf{u_0} = \mathbf{0}$ and $\mathbf{d_0} = \mathbf{M}^{-1}\,\mathbf{f}$, the first step of the CGM results in the displacements vector

\begin{equation}
 \mathbf{u_1} = \left[\frac{\langle \mathbf{f} \rangle_0}{\langle \mathbf{f} \rangle_1}\right] \mathbf{W_0}\,\mathbf{f} \,.
 \label{eq:first_case_u1}
\end{equation}

The coefficients $\langle \mathbf{f} \rangle_0$ and $\langle \mathbf{f} \rangle_1$ are given by

\begin{equation}
 \langle \mathbf{f} \rangle_0 = \mathbf{f}^T\,\mathbf{v_M}
 \label{eq:first_case_coeff_0}
\end{equation}

\noindent
and

\begin{equation}
 \langle \mathbf{f} \rangle_1 = \mathbf{v_{K}}^T \, \mathbf{v_M} \,,
\label{eq:first_case_coeff_1}
\end{equation}

\noindent
where the vectors $\mathbf{v_M}$ and $\mathbf{v_K}$ correspond to

\begin{equation}
 \mathbf{v_M} = \mathbf{M}^{-1}\,\mathbf{f}
 \label{eq:first_case_vectors_vm}
\end{equation}

\noindent
and

\begin{equation}
 \mathbf{v_K} = \mathbf{\bar{K}}\,\mathbf{v_M} + \mathbf{\Delta K}\,\mathbf{v_M} \,.
 \label{eq:first_case_vectors_vk}
\end{equation}

From Eq.~(\ref{eq:sensitivity_cgm_u0df}), the sensitivity expression is obtained as

\begin{equation}
 \boxed{\alpha^{\langle 1 \rangle}_i =
 \left\{\begin{aligned}
  & - \left[\bar{C} - \frac{{\langle \mathbf{f} \rangle_0}^2}{2\,\langle \mathbf{f} \rangle_1}\right]  \;, & \text{if } x_i = 0 \,,\\
  &  - \left[\frac{{\langle \mathbf{f} \rangle_0}^2}{2\,\langle \mathbf{f} \rangle_1} - \bar{C}\right]  \;, & \text{if } x_i = 1 \,.
 \end{aligned}\right.}
 \label{eq:first_case_sensM_u1}
\end{equation}

For a preconditioner matrix $\mathbf{M} = \mathbf{\bar{K}}$, $\mathbf{v_M} = \mathbf{\bar{u}}$ and the sensitivity expression becomes

\begin{equation}
 \boxed{\alpha^{\langle 1 \rangle}_i = - \left[\frac{\bar{C}}{C_{T}}\right] C_i  \;\;,\;\;  x_i \in \{0,1\} \,.}
 \label{eq:first_case_sens_u1}
\end{equation}

The second step of the CGM results in the displacements vector

\begin{equation}
\begin{aligned}
\mathbf{u_2} = {}& \left[\frac{\langle \mathbf{f} \rangle_0 \, \langle \mathbf{f} \rangle_3 - \langle \mathbf{f} \rangle_1 \, \langle \mathbf{f} \rangle_2}{\langle \mathbf{f} \rangle_1 \, \langle \mathbf{f} \rangle_3 - \langle \mathbf{f} \rangle_2 \, \langle \mathbf{f} \rangle_2}\right] \mathbf{W_0}\,\mathbf{f} \\& + \left[\frac{\langle \mathbf{f} \rangle_1 \, \langle \mathbf{f} \rangle_1 - \langle \mathbf{f} \rangle_0 \, \langle \mathbf{f} \rangle_2}{\langle \mathbf{f} \rangle_1 \, \langle \mathbf{f} \rangle_3 - \langle \mathbf{f} \rangle_2 \, \langle \mathbf{f} \rangle_2}\right] \mathbf{W_1}\,\mathbf{f} \,.
\end{aligned}
 \label{eq:first_case_u2}
\end{equation}

The coefficients $\langle \mathbf{f} \rangle_2$ and $\langle \mathbf{f} \rangle_3$ are given by

\begin{equation}
\langle \mathbf{f} \rangle_2 = \mathbf{v_K}^T \, \mathbf{v_L}
\label{eq:first_case_coeff_2}
\end{equation}

\noindent
and

\begin{equation}
\langle \mathbf{f} \rangle_3 = \mathbf{v_R}^T \, \mathbf{v_L} \,,
\label{eq:first_case_coeff_3}
\end{equation}

\noindent
where the vectors $\mathbf{v_L}$ and $\mathbf{v_R}$ correspond to

\begin{equation}
\mathbf{v_L} = \mathbf{M}^{-1}\,\mathbf{v_K}
\label{eq:first_case_vectors2_vl}
\end{equation}

\noindent
and

\begin{equation}
\mathbf{v_R} = \mathbf{\bar{K}}\,\mathbf{v_L} + \mathbf{\Delta K}\,\mathbf{v_L} \,.
\label{eq:first_case_vectors2_vr}
\end{equation}

From Eq.~(\ref{eq:sensitivity_cgm_u0df}), the sensitivity expression is obtained as

\begin{equation}
 \boxed{\alpha^{\langle 1 \rangle}_i =
 \left\{\begin{aligned}
  &\begin{aligned}
  -{}&\left[\bar{C} - \frac{{\langle \mathbf{f} \rangle_0}^2\,\langle \mathbf{f} \rangle_3 - 2\,\langle \mathbf{f} \rangle_0\,\langle \mathbf{f} \rangle_1\,\langle \mathbf{f} \rangle_2 + {\langle \mathbf{f} \rangle_1}^3}{2\,[\langle \mathbf{f} \rangle_1\,\langle \mathbf{f} \rangle_3 - {\langle \mathbf{f} \rangle_2}^2 ]}\right]
  \\& \text{if } x_i = 0\,,\end{aligned}\\
  &\begin{aligned}
  -{}&\left[\frac{{\langle \mathbf{f} \rangle_0}^2\,\langle \mathbf{f} \rangle_3 - 2\,\langle \mathbf{f} \rangle_0\,\langle \mathbf{f} \rangle_1\,\langle \mathbf{f} \rangle_2 + {\langle \mathbf{f} \rangle_1}^3}{2\,[\langle \mathbf{f} \rangle_1\,\langle \mathbf{f} \rangle_3 - {\langle \mathbf{f} \rangle_2}^2 ]} - \bar{C}\right]
  \\& \text{if } x_i = 1\,.\end{aligned}
 \end{aligned}\right.}
 \label{eq:first_case_sens_u2}
\end{equation}

For this and all the following sensitivity expressions, the selective inverse of $\mathbf{\bar{K}}$ would be needed to use $\mathbf{M} = \mathbf{\bar{K}}$. To reduce computational costs, a diagonal $\mathbf{M}$ is recommended (Jacobi preconditioning).

\subsection{Explicit expressions for the 2nd case}

Considering $\mathbf{u_0} = \mathbf{\bar{u}}$ and $\mathbf{d_0} = -\mathbf{M}^{-1}\,\mathbf{\Delta K}\,\mathbf{\bar{u}}$, the first step of the CGM results in the displacements vector

\begin{equation}
 \mathbf{u_1} = \mathbf{\bar{u}} + \left[\frac{\langle \mathbf{b} \rangle_0}{\langle \mathbf{b} \rangle_1}\right] \mathbf{W_0}\,\mathbf{b} \,,
 \label{eq:second_case_u1}
\end{equation}

\noindent
where the vector $\mathbf{b}$ is corresponds to

\begin{equation}
 \mathbf{b} = -\mathbf{\Delta K}\,\mathbf{\bar{u}} = \left\{
 \begin{aligned}
 &-\mathbf{K_i}\,\mathbf{\bar{u}} \;\;,\; \text{if } x_i = 0 \,,\\
 &\phantom{.-}\mathbf{K_i}\,\mathbf{\bar{u}} \;\;,\; \text{if } x_i = 1 \,.
 \end{aligned}\right.
 \label{eq:second_case_b}
\end{equation}

The coefficients $\langle \mathbf{b} \rangle_0$ and $\langle \mathbf{b} \rangle_1$ are given by

\begin{equation}
\langle \mathbf{b} \rangle_0 = \mathbf{b}^T\,\mathbf{v_M}
\label{eq:second_case_coeff_0}
\end{equation}

\noindent
and

\begin{equation}
\langle \mathbf{b} \rangle_1 = \mathbf{v_{K}}^T \, \mathbf{v_M} \,,
\label{eq:second_case_coeff_1}
\end{equation}

\noindent
where the vectors $\mathbf{v_M}$ and $\mathbf{v_K}$ were redefined as

\begin{equation}
\mathbf{v_M} = \mathbf{M}^{-1}\,\mathbf{b}
\label{eq:second_case_vectors_vm}
\end{equation}

\noindent
and

\begin{equation}
\mathbf{v_K} = \mathbf{\bar{K}}\,\mathbf{v_M} + \mathbf{\Delta K}\,\mathbf{v_M} \,.
\label{eq:second_case_vectors_vk}
\end{equation}

From Eq.~(\ref{eq:sensitivity_cgm_uudg}), the sensitivity expression is obtained as

\begin{equation}
 \boxed{\alpha^{\langle 1 \rangle}_i =
 \left\{\begin{aligned}
  & - \left[C_i - \frac{{\langle \mathbf{b} \rangle_0}^2}{2\,\langle \mathbf{b} \rangle_1}\right]  \;, & \text{if } x_i = 0 \,,\\
  &  - \left[C_i + \frac{{\langle \mathbf{b} \rangle_0}^2}{2\,\langle \mathbf{b} \rangle_1}\right]  \;, & \text{if } x_i = 1 \,.
 \end{aligned}\right.}
 \label{eq:second_case_sens_u1}
\end{equation}

The second step of the CGM results in the displacements vector

\begin{equation}
\begin{aligned}
\mathbf{u_2} ={}& \mathbf{\bar{u}} + \left[\frac{\langle \mathbf{b} \rangle_0 \, \langle \mathbf{b} \rangle_3 - \langle \mathbf{b} \rangle_1 \, \langle \mathbf{b} \rangle_2}{\langle \mathbf{b} \rangle_1 \, \langle \mathbf{b} \rangle_3 - \langle \mathbf{b} \rangle_2 \, \langle \mathbf{b} \rangle_2}\right] \mathbf{W_0}\,\mathbf{b} \\& + \left[\frac{\langle \mathbf{b} \rangle_1 \, \langle \mathbf{b} \rangle_1 - \langle \mathbf{b} \rangle_0 \, \langle \mathbf{b} \rangle_2}{\langle \mathbf{b} \rangle_1 \, \langle \mathbf{b} \rangle_3 - \langle \mathbf{b} \rangle_2 \, \langle \mathbf{b} \rangle_2}\right] \mathbf{W_1}\,\mathbf{b} \,.
\end{aligned}
 \label{eq:second_case_u2}
\end{equation}

The coefficients $\langle \mathbf{b} \rangle_2$ and $\langle \mathbf{b} \rangle_3$ are given by

\begin{equation}
\langle \mathbf{b} \rangle_2 = \mathbf{v_K}^T \, \mathbf{v_L}
\label{eq:second_case_coeff_2}
\end{equation}

\noindent
and

\begin{equation}
\langle \mathbf{b} \rangle_3 = \mathbf{v_R}^T \, \mathbf{v_L} \,,
\label{eq:second_case_coeff_3}
\end{equation}

\noindent
where the vectors $\mathbf{v_L}$ and $\mathbf{v_R}$ were redefined as

\begin{equation}
\mathbf{v_L} = \mathbf{M}^{-1}\,\mathbf{v_K}
\label{eq:second_case_vectors2_vl}
\end{equation}

\noindent
and

\begin{equation}
\mathbf{v_R} = \mathbf{\bar{K}}\,\mathbf{v_L} + \mathbf{\Delta K}\,\mathbf{v_L} \,.
\label{eq:second_case_vectors2_vr}
\end{equation}

From Eq.~(\ref{eq:sensitivity_cgm_uudg}), the sensitivity expression is obtained as

\begin{equation}
 \boxed{\alpha^{\langle 1 \rangle}_i =
 \left\{\begin{aligned}
  &\begin{aligned} -{}&\left[C_i - \frac{{\langle \mathbf{b} \rangle_0}^2\,\langle \mathbf{b} \rangle_3 - 2\,\langle \mathbf{b} \rangle_0\,\langle \mathbf{b} \rangle_1\,\langle \mathbf{b} \rangle_2 + {\langle \mathbf{b} \rangle_1}^3}{2\,[\langle \mathbf{b} \rangle_1\,\langle \mathbf{b} \rangle_3 - {\langle \mathbf{b} \rangle_2}^2 ]}\right]
  \\& \text{if } x_i = 0 \,,\end{aligned}\\
  &\begin{aligned} -{}&\left[C_i + \frac{{\langle \mathbf{b} \rangle_0}^2\,\langle \mathbf{b} \rangle_3 - 2\,\langle \mathbf{b} \rangle_0\,\langle \mathbf{b} \rangle_1\,\langle \mathbf{b} \rangle_2 + {\langle \mathbf{b} \rangle_1}^3}{2\,[\langle \mathbf{b} \rangle_1\,\langle \mathbf{b} \rangle_3 - {\langle \mathbf{b} \rangle_2}^2 ]}\right]
  \\& \text{if } x_i = 1 \,.\end{aligned}
 \end{aligned}\right.}
\label{eq:second_case_sens_u2}
\end{equation}

\subsection{Explicit Expressions for the 3rd Case}

Considering $\mathbf{u_0} = \mathbf{0}$ and $\mathbf{d_0} = \mathbf{\bar{u}}$, the first step of the CGM results in the displacements vector

\begin{equation}
 \mathbf{u_1} = \frac{\bar{C}}{C_{T}}\,\mathbf{\bar{u}} \,.
 \label{eq:third_case_u1}
\end{equation}

Both Eqs.~(\ref{eq:sensitivity_cgm_u0df}) and (\ref{eq:sensitivity_cgm_uudg}) produce the same sensitivity expression, given by

\begin{equation}
 \boxed{\alpha^{\langle 1 \rangle}_i = - \left[\frac{\bar{C}}{C_{T}}\right] C_i  \;\;,\;\;  x_i \in \{0,1\}} \;,
 \label{eq:third_case_sens_u1}
\end{equation}

\noindent
which is the same as Eq.~(\ref{eq:first_case_sens_u1}).

The second step of the CGM results in the displacements vector

\begin{equation}
\begin{aligned}
\mathbf{u_2} = {}&  \left[\frac{\bar{C}\,{\langle \mathbf{z} \,,\, \mathbf{g} \rangle_0}^2 - 2\,\bar{C}\,C_T\,\langle \mathbf{g} \rangle_1 - C_T\,\langle \mathbf{g} \rangle_0\,\langle \mathbf{z} \,,\, \mathbf{g}\rangle_0}{C_T\,{\langle \mathbf{z} \,,\, \mathbf{g} \rangle_0}^2 - 2\,{C_T}^2\,\langle \mathbf{g} \rangle_1}\right] \mathbf{\bar{u}} \\& +\left[\frac{2\,{C_T}^2\,\langle \mathbf{g} \rangle_0}{C_T\,{\langle \mathbf{z} \,,\, \mathbf{g} \rangle_0}^2 - 2\,{C_T}^2\,\langle \mathbf{g} \rangle_1}\right] \mathbf{W_0}\,\mathbf{g} \,,
\end{aligned}
 \label{eq:third_case_u2}
\end{equation}

\noindent
where the vectors $\mathbf{z}$ and $\mathbf{g}$ correspond to

\begin{equation}
 \mathbf{z} = \mathbf{f} - \mathbf{b}
 \label{eq:third_case_z}
\end{equation}

\noindent
and

\begin{equation}
 \mathbf{g} = \frac{\bar{C}}{C_T}\,\mathbf{z} -\mathbf{f} \,.
 \label{eq:third_case_g}
\end{equation}

The coefficients $\langle \mathbf{z} \,,\, \mathbf{g} \rangle_0$, $\langle \mathbf{g} \rangle_0$ and $\langle \mathbf{g} \rangle_1$ are given by

\begin{equation}
\langle \mathbf{z} \,,\, \mathbf{g} \rangle_0 = \mathbf{v_M}^T \, \mathbf{z} \,,
\label{eq:third_case_coeffs_zg0}
\end{equation}

\begin{equation}
\langle \mathbf{g} \rangle_0 = \mathbf{v_M}^T \, \mathbf{g}
\label{eq:third_case_coeffs_g0}
\end{equation}

\noindent
and

\begin{equation}
\langle \mathbf{g} \rangle_1 = \mathbf{v_M}^T \, \mathbf{v_K} \,,
\label{eq:third_case_coeffs_g1}
\end{equation}

\noindent
where the vectors $\mathbf{v_M}$ and $\mathbf{v_K}$ were redefined as

\begin{equation}
\mathbf{v_M} = \mathbf{M}^{-1}\,\mathbf{g}
\label{eq:third_case_vectors_vm}
\end{equation}

\noindent
and

\begin{equation}
\mathbf{v_K} = \mathbf{\bar{K}}\,\mathbf{v_M} + \mathbf{\Delta K}\,\mathbf{v_M} \,.
\label{eq:third_case_vectors_vk}
\end{equation}

Both Eqs.~(\ref{eq:sensitivity_cgm_u0df}) and (\ref{eq:sensitivity_cgm_uudg}) produce the same sensitivity expression, given by

\begin{equation}
 \boxed{\alpha^{\langle 1 \rangle}_i =
 \left\{\begin{aligned}
  &\begin{aligned} - {}& \left[\frac{\bar{C}}{C_{T}}\right] C_i +\left[\frac{{C_T}\,{\langle \mathbf{g} \rangle_0}^2}{2\,{C_T}\,\langle \mathbf{g} \rangle_1 - {\langle \mathbf{z} \,,\, \mathbf{g} \rangle_0}^2} \right]
  \\& \text{if } x_i = 0 \,,\end{aligned}\\
  &\begin{aligned} - {}& \left[\frac{\bar{C}}{C_{T}}\right] C_i -\left[\frac{{C_T}\,{\langle \mathbf{g} \rangle_0}^2}{2\,{C_T}\,\langle \mathbf{g} \rangle_1 - {\langle \mathbf{z} \,,\, \mathbf{g} \rangle_0}^2} \right]
  \\& \text{if } x_i = 1 \,.\end{aligned}
 \end{aligned}\right.}
\label{eq:third_case_sens_u2}
\end{equation}

\section*{Declarations}

\paragraph{Funding} This work was supported by the São Paulo Research Foundation (FAPESP), grant numbers:  \mbox{2013/08293-7}, \mbox{2019/05393-7}, \mbox{2019/19237-7} and \mbox{2020/07391-9}

\paragraph{Conflicts of interest} The authors declare that they have no conflict of interest.

\paragraph{Availability of data and material} Not applicable.

\paragraph{Code availability} Not applicable.

\paragraph{Ethics approval} Not applicable.

\paragraph{Consent to participate} Not applicable.

\paragraph{Consent for publication} Not applicable.

\section*{Replication of results}
All necessary information was provided in the numerical examples. Any researcher with an implemented BESO program for structural compliance minimization can easily adapt their sensitivity analysis with the presented expressions and replicate the results.

\bibliographystyle{spbasic}      % basic style, author-year citations
\bibliography{references}        % name your BibTeX data base

\begin{thebibliography}{51}
\providecommand{\natexlab}[1]{#1}
\providecommand{\url}[1]{{#1}}
\providecommand{\urlprefix}{URL }
\expandafter\ifx\csname urlstyle\endcsname\relax
  \providecommand{\doi}[1]{DOI~\discretionary{}{}{}#1}\else
  \providecommand{\doi}{DOI~\discretionary{}{}{}\begingroup
  \urlstyle{rm}\Url}\fi
\providecommand{\eprint}[2][]{\url{#2}}

\bibitem[{de~Almeida et~al.(2019)de~Almeida, Cunha, and
  Pavanello}]{de2019topology}
de~Almeida BV, Cunha DC, Pavanello R (2019) Topology optimization of bimorph
  piezoelectric energy harvesters considering variable electrode location.
  Smart Materials and Structures 28(8):085030

\bibitem[{Azevedo et~al.(2018)Azevedo, Moura, Vicente, Picelli, and
  Pavanello}]{azevedo2018topology}
Azevedo F, Moura M, Vicente W, Picelli R, Pavanello R (2018) Topology
  optimization of reactive acoustic mufflers using a bi-directional
  evolutionary optimization method. Structural and Multidisciplinary
  Optimization 58(5):2239--2252

\bibitem[{Beckers(1999)}]{beckers1999topology}
Beckers M (1999) Topology optimization using a dual method with discrete
  variables. Structural Optimization 17(1):14--24

\bibitem[{Bends{\o}e(1989)}]{bendsoe1989optimal}
Bends{\o}e MP (1989) Optimal shape design as a material distribution problem.
  Structural optimization 1(4):193--202

\bibitem[{Bojczuk and Mr{\'o}z(2009)}]{bojczuk2009topological}
Bojczuk D, Mr{\'o}z Z (2009) Topological sensitivity derivative and finite
  topology modifications: application to optimization of plates in bending.
  Structural and Multidisciplinary Optimization 39(1):1

\bibitem[{C{\'e}a et~al.(2000)C{\'e}a, Garreau, Guillaume, and
  Masmoudi}]{cea2000shape}
C{\'e}a J, Garreau S, Guillaume P, Masmoudi M (2000) The shape and topological
  optimizations connection. Computer methods in applied mechanics and
  engineering 188(4):713--726

\bibitem[{Cunha and Pavanello(2017)}]{cunha2017evolutionary}
Cunha DC, Pavanello R (2017) Evolutionary topology optimization for designing
  cellular fluid actuators. In: World Congress of Structural and
  Multidisciplinary Optimisation, Springer, pp 1484--1496

\bibitem[{van Dijk et~al.(2013)van Dijk, Maute, Langelaar, and
  Van~Keulen}]{van2013level}
van Dijk NP, Maute K, Langelaar M, Van~Keulen F (2013) Level-set methods for
  structural topology optimization: a review. Structural and Multidisciplinary
  Optimization 48(3):437--472

\bibitem[{Edwards et~al.(2007)Edwards, Kim, and Budd}]{edwards2007evaluative}
Edwards CS, Kim HA, Budd CJ (2007) An evaluative study on eso and simp for
  optimising a cantilever tie—beam. Structural and Multidisciplinary
  Optimization 34(5):403--414

\bibitem[{de~Faria et~al.(2007)de~Faria, Novotny, Feij{\'o}o, Taroco, and
  Padra}]{de2007second}
de~Faria JR, Novotny AA, Feij{\'o}o RA, Taroco E, Padra C (2007) Second order
  topological sensitivity analysis. International journal of solids and
  structures 44(14-15):4958--4977

\bibitem[{Fried(1973)}]{fried1973bounds}
Fried I (1973) Bounds on the spectral and maximum norms of the finite element
  stiffness, flexibility and mass matrices. International Journal of Solids and
  Structures 9(9):1013--1034

\bibitem[{Ghabraie(2015)}]{ghabraie2015eso}
Ghabraie K (2015) The eso method revisited. Structural and Multidisciplinary
  Optimization 51(6):1211--1222

\bibitem[{Groenwold and Etman(2010)}]{groenwold2010quadratic}
Groenwold AA, Etman LFP (2010) A quadratic approximation for structural
  topology optimization. International Journal for Numerical Methods in
  Engineering 82(4):505--524

\bibitem[{Guttman(1946)}]{guttman1946enlargement}
Guttman L (1946) Enlargement methods for computing the inverse matrix. The
  annals of mathematical statistics 17(3):336--343

\bibitem[{Hager(1989)}]{hager1989updating}
Hager WW (1989) Updating the inverse of a matrix. SIAM review 31(2):221--239

\bibitem[{Hassine and Khelifi(2016)}]{hassine2016higher}
Hassine M, Khelifi K (2016) Higher-order topological sensitivity analysis for
  the laplace operator. Comptes Rendus Mathematique 354(10):993--999

\bibitem[{Huang and Xie(2007)}]{huang2007convergent}
Huang X, Xie YM (2007) Convergent and mesh-independent solutions for the
  bi-directional evolutionary structural optimization method. Finite Elements
  in Analysis and Design 43(14):1039--1049

\bibitem[{Huang and Xie(2008)}]{huang2008new}
Huang X, Xie YM (2008) A new look at eso and beso optimization methods.
  Structural and Multidisciplinary Optimization 35(1):89--92

\bibitem[{Huang and Xie(2009)}]{huang2009bi}
Huang X, Xie YM (2009) Bi-directional evolutionary topology optimization of
  continuum structures with one or multiple materials. Computational Mechanics
  43(3):393

\bibitem[{Huang and Xie(2010)}]{huang2010further}
Huang X, Xie YM (2010) A further review of eso type methods for topology
  optimization. Structural and Multidisciplinary Optimization 41(5):671--683

\bibitem[{Jacquelin et~al.(2016)Jacquelin, Lin, and
  Yang}]{jacquelin2016pselinv}
Jacquelin M, Lin L, Yang C (2016) Pselinv—a distributed memory parallel
  algorithm for selected inversion: The symmetric case. ACM Transactions on
  Mathematical Software (TOMS) 43(3):1--28

\bibitem[{Liang and Cheng(2019)}]{liang2019topology}
Liang Y, Cheng G (2019) Topology optimization via sequential integer
  programming and canonical relaxation algorithm. Computer Methods in Applied
  Mechanics and Engineering 348:64--96

\bibitem[{Lin et~al.(2011)Lin, Yang, Meza, Lu, Ying, and E}]{lin2011selinv}
Lin L, Yang C, Meza JC, Lu J, Ying L, E W (2011) Selinv---an algorithm for
  selected inversion of a sparse symmetric matrix. ACM Transactions on
  Mathematical Software (TOMS) 37(4):1--19

\bibitem[{Lopes et~al.(2021)Lopes, Mahfoud, and Pavanello}]{lopes2021high}
Lopes HN, Mahfoud J, Pavanello R (2021) High natural frequency gap topology
  optimization of bi-material elastic structures and band gap analysis.
  Structural and Multidisciplinary Optimization pp 1--16

\bibitem[{Martinez(2005)}]{martinez2005note}
Martinez JM (2005) A note on the theoretical convergence properties of the simp
  method. Structural and Multidisciplinary Optimization 29(4):319--323

\bibitem[{Mr{\'o}z and Bojczuk(2003)}]{mroz2003finite}
Mr{\'o}z Z, Bojczuk D (2003) Finite topology variations in optimal design of
  structures. Structural and Multidisciplinary Optimization 25(3):153--173

\bibitem[{Norato et~al.(2007)Norato, Bends{\o}e, Haber, and
  Tortorelli}]{norato2007topological}
Norato JA, Bends{\o}e MP, Haber RB, Tortorelli DA (2007) A topological
  derivative method for topology optimization. Structural and Multidisciplinary
  Optimization 33(4-5):375--386

\bibitem[{Novotny et~al.(2003)Novotny, Feij{\'o}o, Taroco, and
  Padra}]{novotny2003topological}
Novotny AA, Feij{\'o}o RA, Taroco E, Padra C (2003) Topological sensitivity
  analysis. Computer methods in applied mechanics and engineering
  192(7-8):803--829

\bibitem[{Picelli et~al.(2015)Picelli, Vicente, and Pavanello}]{picelli2015bi}
Picelli R, Vicente WM, Pavanello R (2015) Bi-directional evolutionary
  structural optimization for design-dependent fluid pressure loading problems.
  Engineering Optimization 47(10):1324--1342

\bibitem[{Querin et~al.(1998)Querin, Steven, and Xie}]{querin1998evolutionary}
Querin OM, Steven GP, Xie YM (1998) Evolutionary structural optimisation (eso)
  using a bidirectional algorithm. Engineering computations 15(8):1031--1048

\bibitem[{Rietz(2001)}]{rietz2001sufficiency}
Rietz A (2001) Sufficiency of a finite exponent in simp (power law) methods.
  Structural and Multidisciplinary Optimization 21(2):159--163

\bibitem[{Rozvany(2009)}]{rozvany2009critical}
Rozvany GIN (2009) A critical review of established methods of structural
  topology optimization. Structural and multidisciplinary optimization
  37(3):217--237

\bibitem[{Rozvany and Querin(2002{\natexlab{a}})}]{rozvany2002combining}
Rozvany GIN, Querin OM (2002{\natexlab{a}}) Combining eso with rigorous
  optimality criteria. International journal of vehicle design 28(4):294--299

\bibitem[{Rozvany and Querin(2002{\natexlab{b}})}]{rozvany2002theoretical}
Rozvany GIN, Querin OM (2002{\natexlab{b}}) Theoretical foundations of
  sequential element rejections and admissions (sera) methods and their
  computational implementation in topology optimization. In: 9th AIAA/ISSMO
  symposium on multidisciplinary analysis and optimization, p 5521

\bibitem[{S{\'a} et~al.(2016)S{\'a}, Amigo, Novotny, and
  Silva}]{sa2016topological}
S{\'a} LFN, Amigo RCR, Novotny AA, Silva ECN (2016) Topological derivatives
  applied to fluid flow channel design optimization problems. Structural and
  Multidisciplinary Optimization 54(2):249--264

\bibitem[{Sigmund and Maute(2012)}]{sigmund2012sensitivity}
Sigmund O, Maute K (2012) Sensitivity filtering from a continuum mechanics
  perspective. Structural and Multidisciplinary Optimization 46(4):471--475

\bibitem[{Sigmund and Petersson(1998)}]{sigmund1998numerical}
Sigmund O, Petersson J (1998) Numerical instabilities in topology optimization:
  a survey on procedures dealing with checkerboards, mesh-dependencies and
  local minima. Structural optimization 16(1):68--75

\bibitem[{Sivapuram and Picelli(2018)}]{sivapuram2018topology}
Sivapuram R, Picelli R (2018) Topology optimization of binary structures using
  integer linear programming. Finite Elements in Analysis and Design 139:49--61

\bibitem[{Sivapuram et~al.(2018)Sivapuram, Picelli, and
  Xie}]{sivapuram2018topology2}
Sivapuram R, Picelli R, Xie YM (2018) Topology optimization of binary
  microstructures involving various non-volume constraints. Computational
  Materials Science 154:405--425

\bibitem[{Tanskanen(2002)}]{tanskanen2002evolutionary}
Tanskanen P (2002) The evolutionary structural optimization method: theoretical
  aspects. Computer methods in applied mechanics and engineering
  191(47-48):5485--5498

\bibitem[{Vicente et~al.(2016)Vicente, Zuo, Pavanello, Calixto, Picelli, and
  Xie}]{vicente2016concurrent}
Vicente W, Zuo Z, Pavanello R, Calixto T, Picelli R, Xie Y (2016) Concurrent
  topology optimization for minimizing frequency responses of two-level
  hierarchical structures. Computer Methods in Applied Mechanics and
  Engineering 301:116--136

\bibitem[{Wang(2009)}]{wang2009analysis}
Wang MY (2009) An analysis of the compliant mechanism models. In: 2009
  ASME/IFToMM International Conference on Reconfigurable Mechanisms and Robots,
  IEEE, pp 377--385

\bibitem[{Woodbury(1950)}]{woodbury1950inverting}
Woodbury MA (1950) Inverting modified matrices. Memorandum report 42(106):336

\bibitem[{Xia and Breitkopf(2017)}]{xia2017recent}
Xia L, Breitkopf P (2017) Recent advances on topology optimization of
  multiscale nonlinear structures. Archives of Computational Methods in
  Engineering 24(2):227--249

\bibitem[{Xia et~al.(2018{\natexlab{a}})Xia, Xia, Huang, and Xie}]{xia2018bi}
Xia L, Xia Q, Huang X, Xie YM (2018{\natexlab{a}}) Bi-directional evolutionary
  structural optimization on advanced structures and materials: a comprehensive
  review. Archives of Computational Methods in Engineering 25(2):437--478

\bibitem[{Xia et~al.(2018{\natexlab{b}})Xia, Zhang, Xia, and
  Shi}]{xia2018stress}
Xia L, Zhang L, Xia Q, Shi T (2018{\natexlab{b}}) Stress-based topology
  optimization using bi-directional evolutionary structural optimization
  method. Computer Methods in Applied Mechanics and Engineering 333:356--370

\bibitem[{Xie and Steven(1993)}]{xie1993simple}
Xie YM, Steven GP (1993) A simple evolutionary procedure for structural
  optimization. Computers \& structures 49(5):885--896

\bibitem[{Yan et~al.(2015)Yan, Huang, Sun, and Xie}]{yan2015two}
Yan X, Huang X, Sun G, Xie YM (2015) Two-scale optimal design of structures
  with thermal insulation materials. Composite Structures 120:358--365

\bibitem[{Yang et~al.(1999)Yang, Xie, Steven, and
  Querin}]{yang1999bidirectional}
Yang XY, Xie YM, Steven GP, Querin OM (1999) Bidirectional evolutionary method
  for stiffness optimization. AIAA journal 37(11):1483--1488

\bibitem[{Zhou and Rozvany(1991)}]{zhou1991coc}
Zhou M, Rozvany GIN (1991) The coc algorithm, part ii: Topological, geometrical
  and generalized shape optimization. Computer methods in applied mechanics and
  engineering 89(1--3):309--336

\bibitem[{Zhou and Rozvany(2001)}]{zhou2001validity}
Zhou M, Rozvany GIN (2001) On the validity of eso type methods in topology
  optimization. Structural and Multidisciplinary Optimization 21(1):80--83

\end{thebibliography}

\end{document}